\documentclass[twocolumn]{aastex631}
\usepackage{epsf}
\usepackage{graphicx}
\usepackage{float}
\usepackage{amsmath}
\usepackage{url}
\listofchanges

\defcitealias{O'Connell1951}{O51}
\defcitealias{Davidge1984}{D84}
\defcitealias{Wilsey2009}{W09}

\makeatletter
\DeclareRobustCommand\citetp
  {\begingroup
   \let\NAT@nmfmt\NAT@posfmt
   \NAT@swafalse\let\NAT@ctype\z@\NAT@partrue
   \@ifstar{\NAT@fulltrue\NAT@citetp}{\NAT@fullfalse\NAT@citetp}}
\let\NAT@orig@nmfmt\NAT@nmfmt
\def\NAT@posfmt#1{\NAT@orig@nmfmt{#1's}}
\makeatother

\begin{document}

\title{Characteristics of \emph{Kepler} Eclipsing Binaries Displaying a Significant O'Connell Effect}
\author[0000-0002-8540-739X]{Matthew F. Knote}
\affil{Florida Institute of Technology}
\author[0000-0002-8348-5191]{Saida M. Caballero-Nieves}
\affil{Florida Institute of Technology}
\author[0000-0001-8796-4686]{Vayujeet Gokhale}
\affil{Truman State University}
\author[0000-0002-9235-5807]{Kyle B. Johnston}
\affil{Florida Institute of Technology}
\affil{Booz Allen Hamilton}
\author[0000-0002-3099-1664]{Eric S. Perlman}
\affil{Florida Institute of Technology}

\begin{abstract}
The O'Connell effect -- the presence of unequal maxima in eclipsing binaries -- remains an unsolved riddle in the study of close binary systems. The \emph{Kepler} space telescope produced high precision photometry of nearly 3,000 eclipsing binary systems, providing a unique opportunity to study the O'Connell effect in a large sample and in greater detail than in previous studies. We have characterized the observational properties -- including temperature, luminosity, and eclipse depth -- of a set of 212 systems (7.3\% of \emph{Kepler} eclipsing binaries) that display a maxima flux difference of at least 1\%, representing the largest sample of O'Connell effect systems yet studied. We explored how these characteristics correlate with each other to help understand the O'Connell effect's underlying causes. We also describe some system classes with peculiar light curve features aside from the O'Connell effect ($\sim$24\% of our sample), including temporal variation and asymmetric minima. We found that the O'Connell effect size's correlations with period and temperature are inconsistent with \citeauthor{Kouzuma2019}'s starspot study. Up to 20\% of systems display the parabolic eclipse timing variation signal expected for binaries undergoing mass transfer. Most systems displaying the O'Connell effect have the brighter maximum following the primary eclipse, suggesting a fundamental link between which maximum is brighter and the O'Connell effect's physical causes. Most importantly, we find that the O'Connell effect occurs exclusively in systems where the components are close enough to significantly affect each other, suggesting that the interaction between the components is ultimately responsible for causing the O'Connell effect.
\end{abstract}

\section{Introduction\label{sec:Introduction}}

\begin{figure}
\begin{centering}
\includegraphics[width=\columnwidth]{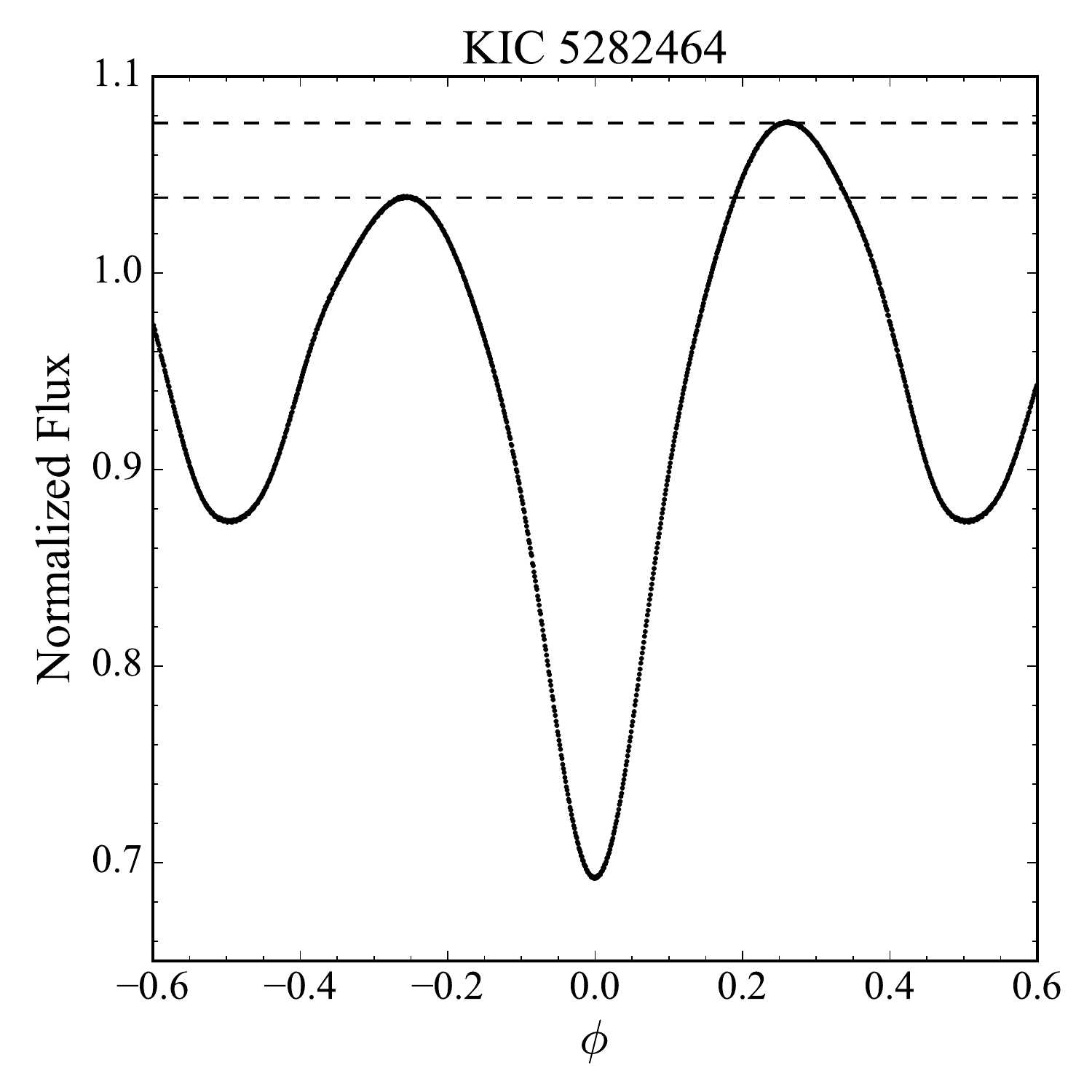}
\par\end{centering}
\caption{Averaged light curve of KIC~5282464 showing the O'Connell effect. The maximum following the primary eclipse (indicated with the upper dashed line) is noticeably brighter than the maximum preceding it (indicated by the lower dashed line).
\label{fig:O'Connell-effect-example}}
\end{figure}

The O'Connell effect \citep{O'Connell1951, Milone1968} is a poorly understood asymmetry seen in some eclipsing binaries where the maxima between eclipses are not equal in brightness (see Figure~\ref{fig:O'Connell-effect-example}). At the quadrature phases ($\phi = 0.25,~0.75$ for circular orbits), we see the stars side-by-side, so a difference in brightness implies that one hemisphere of a component emits a different amount of radiation than the other hemisphere. \citet{Roberts1906} first discussed the O'Connell effect, where he attributed it to stars in an eccentric orbit becoming tidally distorted near periapsis. His theory did not explain the presence of the O'Connell effect in systems with circular orbits, however, and is no longer considered a likely cause of the O'Connell effect in most systems. \citet{Wilsey2009} outlines a few possible explanations for the O'Connell effect: chromospheric spots on one or both components, a hotspot caused by mass transfer, or circumbinary material impacting the stars as they orbit \citep{Liu2003}. \citet{Wilsey2009} notes that none of these explains the O'Connell effect in all cases, however, and \citet{Papageorgiou2014} states that ``the O'Connell effect is still one of the most perplexing challenges in binary studies.''

\subsection{Background\label{subsec:Background}}

\citet[hereafter \citetalias{O'Connell1951}]{O'Connell1951} found strong correlations between the O'Connell effect size (OES, also commonly known as $\Delta m$) and the following parameters: the difference in OES at different wavelengths, the ellipticity of the component stars, the ratio of stellar radii, and the ratio of stellar densities. Most notably, he found that the OES becomes larger at shorter wavelengths as the OES increases, which is to say OES\textsubscript{blue} -- OES\textsubscript{red} increases as OES increases. \citetalias{O'Connell1951} restricted his study to systems with a constant OES and excluded W Ursae Majoris systems.

\citet[hereafter \citetalias{Davidge1984}]{Davidge1984} built upon the work of \citetalias{O'Connell1951} by introducing more systems displaying the O'Connell effect and using a second, non-parametric form of analysis less prone to outliers in addition to the parametric analysis used by \citetalias{O'Connell1951}. Like \citetalias{O'Connell1951}, \citetalias{Davidge1984} restricted their sample to systems with a constant OES and excluded overcontact systems. The higher precision photometry available to \citetalias{Davidge1984} meant that many systems included in \citetalias{O'Connell1951} were found to have a variable OES and so were excluded. \citetalias{Davidge1984} found essentially the same strong correlations as \citetalias{O'Connell1951}. The correlation between the OES and the size at different wavelengths was the strongest correlation in \citetalias{O'Connell1951} and \citetalias{Davidge1984}. However, \citetalias{Davidge1984} found that the OES becomes larger at \emph{longer} wavelengths as the OES increases, the opposite of \citetalias{O'Connell1951}'s correlation. They attributed this discrepancy to the sample selection difference between \citetalias{O'Connell1951} and \citetalias{Davidge1984}, as there are only six systems in common between the two studies. Note that both \citetalias{O'Connell1951} and \citetalias{Davidge1984} used ground-based data taken in the era before CCDs.

\citet[hereafter \citetalias{Wilsey2009}]{Wilsey2009} is the most recent major work on the O'Connell effect. \citetalias{Wilsey2009} examines the current theoretical models used to explain the O'Connell effect: starspots, gas stream impacts, circumstellar material, and asymmetric circumfluence due to Coriolis forces. They state that starspots are the most commonly used explanation for the O'Connell effect. These spot models can be used to explain both chromospheric starspots like those found on our Sun and starspots arising from the impact of a matter stream. However, \citetalias{Wilsey2009} states that the spots need to be unrealistically large to explain the observed O'Connell effect in many systems, and \citet{Maceroni1993} states that the photometric light curves used to build these models are often insufficient to justify the existence of starspots. Furthermore, starspots introduce four new parameters per spot to the parameter space: spot latitude, longitude, radius, and temperature factor. These additional parameters increase the uniqueness problem \citep{Kallrath2009}, making finding the parameter space's global minimum (and thus the best model of the system) nearly impossible. Next, \citetalias{Wilsey2009} discusses how a hotspot created by a matter stream can create a maximum difference if it is offset from the central axis connecting the centers of both stars. They also identify two candidate systems where this matter stream hypothesis has explained the O'Connell effect: V361 Lyrae and GR Tauri. Finally, \citetalias{Wilsey2009} discusses two papers (\citealt{Liu2003} and \citealt{Zhou1990}) that propose models for the O'Connell effect. \citet{Liu2003} proposed that the impact of material surrounding both stars heats the leading hemispheres, causing a difference in brightness and temperature. However, \citetalias{Wilsey2009} notes that their model's assumptions are unrealistic. \citet{Zhou1990} proposed that the deflection of circulating material by Coriolis forces can explain the O'Connell effect in overcontact systems.

Recent surveys such as the \emph{Kepler} \citep{Borucki2010} and Transiting Exoplanet Survey Satellite \citep[TESS;][]{Ricker2014} missions and future surveys such as the Legacy Survey of Space and Time \citep[LSST;][]{Ivezic2019} conducted by the Vera C. Rubin Observatory provide an enormous amount of data on the variable universe. These surveys uncover and provide complete samples of sources that display poorly understood phenomena (including the O'Connell effect), allowing us to completely characterize these systems for the first time. The first step to understanding these phenomena is determining the characteristics of sources exhibiting them. This characterization allows us to anticipate which newly observed sources might show them and provides clues to the physical processes underlying the phenomenon in question.

\subsection{Our Study\label{subsec:Our-Study}}

Our project aims to characterize the observational properties (including period, color, OES, and eclipse depth) of systems displaying a significant O'Connell effect, which serves as a prelude to understanding the O'Connell effect's physical causes. We have selected 258 eclipsing binaries observed by \emph{Kepler}, of which 212 constitute a core sample showing an average $|$OES$|$ larger than 1\% of their normalized flux. Our core sample represents the largest sample of O'Connell effect binaries studied to date. This paper describes our characterization of the core sample. In addition to this characterization, we determined how the characteristics correlate with each other. These correlations provide important insights into the processes underlying the O'Connell effect. This paper also details our use of two statistical measures introduced by \citet{McCartney1999} and refined by \citetalias{Wilsey2009}. Our project serves as a continuation of the studies presented in \citetalias{O'Connell1951} and \citetalias{Davidge1984} using modern observational techniques like space-based CCD photometry. However, our lack of bulk physical characteristics (mass, density, surface gravity, etc.)\ due to lacking radial velocity data forces us to probe a different parameter space than those works.

Our project seeks to answer questions about the O'Connell effect using a sample selected in a consistent and unbiased manner. We primarily see the O'Connell effect in short-period eclipsing binaries, where the stars are close enough to interact with each other. Is this a selection bias, or is there a genuine link between binary interaction and the O'Connell effect? \citetalias{O'Connell1951} found that the brighter maximum almost uniformly followed the primary eclipse rather than preceding it. \citetalias{Davidge1984}, by contrast, found a more even distribution of maxima preference, with about 60\% of their sample having the brighter maximum follow the primary eclipse. Which result better describes the maxima distribution of the underlying population? Do the characteristics of these systems correlate in the manner \citet{Kouzuma2019} suggests spotted stars should? Or do we instead see the eclipse timing variations characteristic of mass transfer in a majority of systems? Is the OES the best way to characterize the O'Connell effect, or are the O'Connell Effect Ratio and Light Curve Asymmetry statistical measures introduced by \citet{McCartney1999} better? Finally, how common is the O'Connell effect?

We now define the light curve classifications (Algol-, $\beta$ Lyrae-, and W Ursae Majoris-type systems) and morphological classifications (detached, semi-detached, and overcontact) we will adopt in this paper. Algol-type systems have sharp, well-defined eclipses and minimal out-of-eclipse variations, $\beta$ Lyrae-type systems have continuously variable light curves and minima of significantly unequal depth, and W Ursae Majoris-type systems have continuously variable light curves and minima of equal or nearly equal depth. We quantify our light curve classification in Section~\ref{subsec:Morphology-Parameter}. Our light curve classification does not consider the binary components' spectral type or evolutionary state (e.g.\ we would classify V382 Cygni as W Ursae Majoris-type rather than $\beta$ Lyrae-type as \citealt{Landolt1975} did). Meanwhile, detached systems have neither star filling their Roche lobe, semi-detached systems have one star (either the primary or secondary) exactly filling their Roche lobe, and overcontact systems have both stars overflowing their Roche lobes. We define the primary star as the component eclipsed at the deeper minimum without regard to the component masses.

Section~\ref{sec:Target-Selection} describes how we determined our sample and the biases that are inherent to it. Section~\ref{sec:Methodology} discusses our methodology in analyzing the \emph{Kepler} data. Section~\ref{sec:Results} gives the characteristics of our core sample, including their light curve properties, information obtained from other sources, and properties derived from available data. Section~\ref{sec:Results} also describes four system classes with peculiar features in their light curves aside from the O'Connell effect, including temporal variation of their light curves and asymmetric minima. Section~\ref{sec:Analysis-and-Discussion} discusses the observed trends found in our study and the implications of these trends. Finally, Section~\ref{sec:Conclusion} summarizes our major results and details future work on this project.

\section{Target Selection\label{sec:Target-Selection}}

We drew our sample from the \emph{Kepler} Eclipsing Binary Catalog (KEBC), a compilation of eclipsing binary systems observed by the \emph{Kepler} space telescope. We defined a single criterion to define our sample: $|$OES$| \geq 0.01$, where OES is in units of normalized flux and $|$OES$|$ is the absolute value of the OES that folds the negative OES values over the positive ones. We created a code to determine the OES, which we used to select a core sample of 212 system based on their long-cadence \emph{Kepler} data.

\subsection{\emph{Kepler} Space Telescope\label{subsec:Kepler}}

The \emph{Kepler} space telescope was launched in 2009 with the purpose of observing transiting planets. The spacecraft imaged over 150,000 objects every thirty minutes for four years, observing numerous variable stars in the process. \emph{Kepler}'s photometry is accurate within 29 parts per million (ppm) for a star with an apparent magnitude of 12 and 80.7 ppm at an apparent magnitude of 14.5 \citep{Gilliland2011}. For comparison, \citet{Tregloan-Reed2013} cites a precision of 258 ppm \citep[of a V = 11.45 star using the 4-meter Mayall telescope;][]{Gilliland1993} as the most precise ground-based observations known. Therefore, it is clear that \emph{Kepler}'s photometry is much more precise than ground-based data, even for fainter stars, making features such as kinks or subtle changes over time more apparent. As a result of this precision, we are more confident that any peculiar features observed in a light curve are real structures rather than a result of statistical noise. Some targets have short-cadence data (sampling every minute) in addition to the long-cadence data (sampling every 30 minutes) that all targets have, allowing us to study short-timescale variations in these systems. However, our analysis in this paper is based exclusively on the long-cadence data.

\emph{Kepler} observed most systems nearly continuously for over three years, providing an opportunity to see how each system changes over that time. Such changes can be attributed to various factors, including starspot evolution, accretion disk instabilities, or a change in temperature due to thermal equilibration. The long observation span and high precision also maximizes our sensitivity to transient effects like flares, allowing us to characterize the prevalence of these events in our sample. Additionally, variations in the eclipse timing can indicate the presence of a third body in the system or an actual change in the period due to mass transfer. Finally, observing targets for such a long, continuous time minimizes the risk of artifacts due to poor sampling or transient effects such as flares. The effects of such artifacts are unpredictable, with one such effect being an oscillatory signal introduced in systems with a period that is a near-integer multiple of the \emph{Kepler} cadence.

Using \emph{Kepler} as a basis for our study introduces the same biases inherent to the \emph{Kepler} mission. The observed targets were drawn from the \emph{Kepler} Input Catalog \citep[KIC;][]{Brown2011}, which estimated each star's spectral type and luminosity class based on photometrically determined colors. \citet{Batalha2010} describes the \emph{Kepler} target prioritization and statistics for the operational target list. They assigned the highest priority to systems with a \emph{Kepler} magnitude $K_p < 14$, and where an Earth-sized planet in the habitable zone would produce at least three transits in the 3.5-year mission with a signal-to-noise ratio greater than $7.1\sigma$. They chose the first condition to facilitate follow-up high-precision spectroscopy to confirm planet detections. The second condition, meanwhile, excluded most O- and B-type stars. As a result of this prioritization, there are more G- and F-type stars than K- or M-type stars among \emph{Kepler} targets, and \emph{Kepler} only observed $\sim$3,000 M-type stars and fewer than 200 O- and B-type stars. The \emph{Kepler} target list also has a bias against giants and subgiants because planetary transits are harder to detect due to the smaller ratio between the star and planet radii. Comparing Tables~2 and 3 of \citet{Batalha2010} shows that the \emph{Kepler} target list included fewer than 10\% of giants located within the \emph{Kepler} field of view. Because of these selection biases, our study undersamples the giants and subgiants as well as the low- and high-mass stellar population, a fact reinforced by the results we present in Section~\ref{subsec:Physical-Characteristics}. Consequently, our study provides less information on the O'Connell effect in systems containing these stars.

\citet{Bryson2020} and \citet{Wolniewicz2021} further investigated the \emph{Kepler} completeness using \emph{Gaia} DR2 data \citep{Gaia2018}, the former for exoplanet occurrence rates and the latter for stellar populations. \citet{Wolniewicz2021} looked specifically at \emph{Kepler}'s selection function, finding that the \emph{Kepler} sample is nearly unbiased for stars brighter than $K_p = 14$. For fainter stars ($K_p > 14$), they found a bias toward main-sequence and subgiant stars with late-F to early-M spectral types and a bias against cool giants. They state that the bias toward subgiants was because the KIC misidentified them as main-sequence stars. They also found a bias against binaries by analyzing the re-normalized unit weight error (RUWE), which is the normalized $\chi^2$ obtained from fitting the point-spread function of \emph{Gaia} sources re-normalized to correct for color-dependent biases \citep{Lindegren2018}. \citet{Wolniewicz2021} considered a system to be binary when its RUWE was greater than 1.2 (Kraus et al.\ 2020, in prep.). For $K_p > 14$, they found that \emph{Kepler}'s completeness was 8\% lower for main-sequence binaries than for solitary main-sequence stars. It is unclear how significantly this impacts the KEBC, however, because \citet{Wolniewicz2021} considered systems with stellar separations of order tens of AU and above. Even the most widely separated systems in the KEBC ($P \sim 1,\!000$~d) have separations of only a few AU\@. \emph{Gaia} EDR3 data \citep{Gaia2021} shows that some of the KEBC systems (24.5\% of the 2,861 systems with \emph{Gaia} parallaxes) have an RUWE above the 1.2 limit \citet{Wolniewicz2021} used. Based on their results, the fact that nearly a quarter of the KEBC shows a RUWE above this limit means that the \emph{Kepler} target list likely excluded some close eclipsing binaries.

\subsection{\emph{Kepler} Eclipsing Binary Catalog\label{subsec:KEBC}}

\citet{Kirk2016} compiled \emph{Kepler}'s data on eclipsing binaries into the KEBC, from which we drew our sample. The KEBC contains \emph{Kepler} observations on 2,907 unique identified eclipsing binary systems in 2,920 entries. Each system has an extracted light curve, and the catalog can be searched based on several parameters, including period and eclipse depth. The KEBC provides an excellent source for our study for three primary reasons: the large number of systems observed, the data's photometric precision, and the observation span. The 2,907 systems in the catalog represent a large, complete sample of eclipsing binaries. For comparison, the All-Sky Automated Survey (ASAS) found 11,099 eclipsing binaries south of declination +28$^{\circ}$ \citep{Obu2013}, while the Optical Gravitational Lensing Experiment (OGLE) survey found over 425,000 eclipsing binaries toward the galactic bulge \citep{Soszynski2016}. The KEBC also contains eclipsing binaries with many different characteristics and allows us to study a complete, statistically significant sample. \citet{Kirk2016} states that the KEBC completeness is 89.1\% for eclipsing binaries, with essentially 100\% completeness for systems with periods of order one day or less. However, \citet{Bienias2021} recently found 547 short-period ($P \lesssim 0.5$~d) eclipsing binaries in the \emph{Kepler} field not included in the KEBC\@. These new eclipsing binaries are fainter than the KEBC systems, with an average magnitude of 18.2. Nevertheless, the KEBC's high completeness for brighter ($K_p \lesssim 16$) systems means it is well-representative of the true eclipsing binary population, increasing its suitability for our study.

We compare our core sample to the entire KEBC throughout this paper. However, we use three separate subsets of the KEBC in this paper, each more restrictive than the last. The first subset -- the comparative subset -- removed four systems, one due to data corruption and three because they are not in the KEBC (note that the three non-KEBC systems are not included in the system count given in the previous paragraph). We use the comparative subset in the histograms in Sections~\ref{sec:Results} and \ref{sec:Analysis-and-Discussion} and the color-magnitude diagram in Section~\ref{sec:Results}. The second subset -- the trend subset -- removed a further 228 systems due to missing certain parameters or having multiple KEBC entries. We use the trend subset in the plots in Sections~\ref{subsec:Characteristic-Trends} and \ref{subsec:Statistical-Analysis-Discussion}. Finally, the third subset -- the analysis subset -- removed a still further 1,309 systems to select only systems similar in period and light curve shape to our sample. We use the analysis subset in our statistical analysis explained in Section~\ref{subsec:Statistical-Analysis} and discussed in Section~\ref{subsec:Statistical-Analysis-Discussion}. We discuss these subsets further in Sections~\ref{sec:Results}, \ref{subsec:Characteristic-Trends}, and \ref{subsec:Statistical-Analysis-Discussion}, respectively.

\subsection{Selection Criterion}

\begin{figure}
\begin{centering}
\includegraphics[width=\columnwidth]{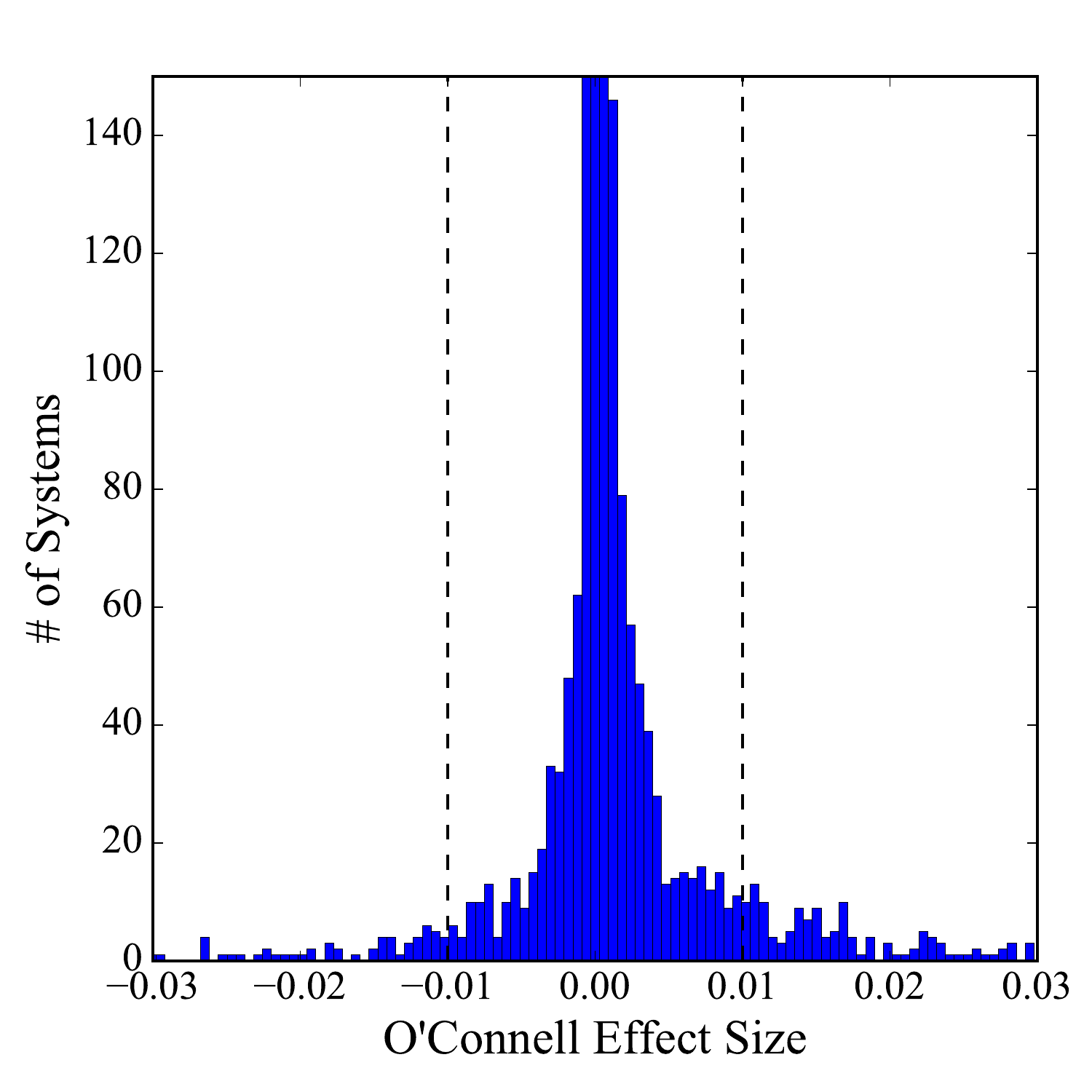}
\par\end{centering}
\caption{Histogram showing the OES distribution for the KEBC\@. The central peak at an OES of 0.00 rises to a value of 1,367 systems and is truncated for clarity. There are 26 additional systems with $|$OES$| > 0.03$ not included in this figure for clarity. The dashed lines indicate the $|$OES$| = 0.01$ cutoff defining our sample.
\label{fig:O'Connell-size-histogram}}
\end{figure}

To ensure that the sample only includes systems with a significant O'Connell effect, we instituted a cutoff in the absolute value of the normalized flux difference between the two maxima of 0.01 (1\% of the normalized flux). We applied this cutoff to averaged light curves (defined in Section~\ref{subsec:O'Connell-Effect-Size-Determination}) similar to the one Figure~\ref{fig:O'Connell-effect-example} shows. We determined the median, mean, and standard deviation ($\sigma$) of the KEBC's OES by applying the definitions of these measures to the KEBC systems' OESs. This method shows that the KEBC's median OES is 0.002\%, the mean value is 0.075\%, and its $\sigma$ is 0.837\%. The corresponding values for our core sample are a median of 1.303\%, a mean value of 0.071\%, and a $\sigma$ of 2.886\%. Figure~\ref{fig:O'Connell-size-histogram} shows a histogram of the OES distribution for the KEBC with the 1\% cutoff indicated by dashed lines. We determined that our core sample does not include 45 systems with an $|$OES$|$ larger than the 1$\sigma$ value but below our 1\% cutoff (equivalent to 1.19$\sigma$). Our criterion is therefore more restrictive than the 1$\sigma$ value. Unlike \citetalias{O'Connell1951} and \citetalias{Davidge1984}, we do not exclude W Ursae Majoris-type or overcontact systems, respectively, nor do we exclude systems with a variable OES\@.

\subsection{Selection Method: O'Connell Effect Size Determination\label{subsec:O'Connell-Effect-Size-Determination}}

We tried several methods to produce our sample. We first tried manually selecting our sample based on visually inspecting KEBC light curves, but this method is inherently subjective and non-reproducible. We then tried using the automated detector \citet{Johnston2019c} describes, as they made this detector to create our sample. However, due in part to the light curve scaling \citet{Johnston2019c} used, it found over 70 systems that did not meet our criterion. Additionally, the detector did not find several other systems that we knew from visual inspection should be in our sample. Therefore, we developed a code that directly calculated the OES, which we used to select our sample.

First, we downloaded the short- and long-cadence \emph{Kepler} data for each eclipsing binary from the KEBC website. Our code split each cadence's data into 1,001 equally-spaced bins in phase space for $\phi \in [-0.5, 0.5]$, where the bins for $\phi = -0.5$ and 0.5 contain the same data. It then calculated the weighted average flux $F$ for each bin using inverse-variance weighting \citep{Hartung2008}:
\begin{equation}
    F = \frac{\sum_i{f_i / \sigma_{f_i}^2}}{\sum_i{1 / \sigma_{f_i}^2}}
    \label{eq:Bin-average}
\end{equation}
along with the error:
\begin{equation}
    \sigma_F = \sqrt{\frac{1}{\sum_i{1 / \sigma_{f_i}^2}}}
    \label{eq:Bin-error}
\end{equation}
where $\sigma_{f_i}$ is the error of the $i$\textsuperscript{th} data point in the bin. We estimated the flux value for any phase bins without data by linearly interpolating between the nearest bins with data. The code introduced a phase shift setting the phase bin with the lowest average flux to $\phi=0$. The resultant light curve represents an averaged light curve for the system, an example of which is shown in Figure~\ref{fig:O'Connell-effect-example}. Note that, throughout this work, we removed all data from KIC~9164694 with BJD in the ranges [2455309.2870355, 2455336.8939506] and [2455432.2188883, 2455552.5485149] to correct an apparent data processing issue wherein the flux values during these intervals were systematically reduced by $\sim$2\%. We also removed a single outlier datum from KIC~8029708's data. Finally, 160 KEBC systems have the flag QAM, which indicates that these systems' amplitudes differ between \emph{Kepler} quarters. Eighteen of these systems are in our sample. The KEBC corrected for this amplitude mismatch by rescaling the data to match between quarters, and we use this rescaled flux in our analysis. However, in at least one case (KIC~4474637), this rescaling did not correct the issue, and so this system exhibits significantly increased scatter in its data and its averaged light curve.

The code produced a smoothed version of the light curve by convolving each point in the averaged light curve with the immediately adjacent points. The convolution reduces the effect of data scatter and oscillations on the light curve, giving a more accurate measure of the OES\@. Since convolution in time space is equivalent to multiplication in frequency space \citep{Press2007}, we performed the convolution in frequency space after applying a discrete Fourier transform to the phased data. Our code multiplied the transformed data by the transformed convolution kernel 50 times in frequency space, equivalent to convolving the data 50 times in time space. It then transformed the data back into time space using the inverse discrete Fourier transform and measured the maximum value of this convolved curve on either side of the primary minimum. The difference between these two maxima is the OES\@. The OES is positive when the maximum after the primary eclipse is brighter than the one preceding it, and we say that a system with a positive OES has a positive O'Connell effect. Conversely, we say that a system where the maximum before the primary eclipse is brighter than the one following it has a negative O'Connell effect. We found the error of each bin in the convolved curve by convolving the errors given by Equation~\ref{eq:Bin-error} in the same fashion as we convolved the averaged data. The error in the OES is then given by propagating the error of the bins used to determine the OES\@. The maxima difference data from this code provided the basis for the histogram in Figure~\ref{fig:O'Connell-size-histogram}. Because we determine the OES from the convolved curve, which is derived from the averaged light curve, this OES represents the average size of the O'Connell effect over the duration of \emph{Kepler}'s observations.

This method for finding O'Connell effect binaries is viable only because we applied it to a set of known eclipsing binaries. If applied to a general population of stars, this method would find many non-eclipsing systems, such as pulsating variables and spotted stars. The KEBC therefore acts as a filter that allows this method to find O'Connell effect binaries. This method is also quite sensitive, and several systems near the $|$OES$|$ cutoff of 0.01 would cross that threshold when we would change our method. For instance, when we began using Equation~\ref{eq:Bin-average} instead of the standard definition of the mean, three systems with $|$OES$|$ above the criterion threshold decreased to under it, while three systems under it increased to above it. This sensitivity to the precise method of determining the OES makes it more challenging to accurately reproduce our sample. It also explains why this method initially added systems to our sample that ultimately did not meet the criterion. To estimate the sensitivity of our method, we varied the cutoff by a small amount and determined the number of systems meeting the new cutoff. We found that varying the cutoff value by 0.0002 in either direction changed the sample size by about $\pm5$ ($\pm2.3\%$), while varying it by 0.001 changed the sample size by about $\pm28$ ($\pm13.2\%$).

\begin{figure*}
\begin{centering}
\includegraphics[width=\columnwidth]{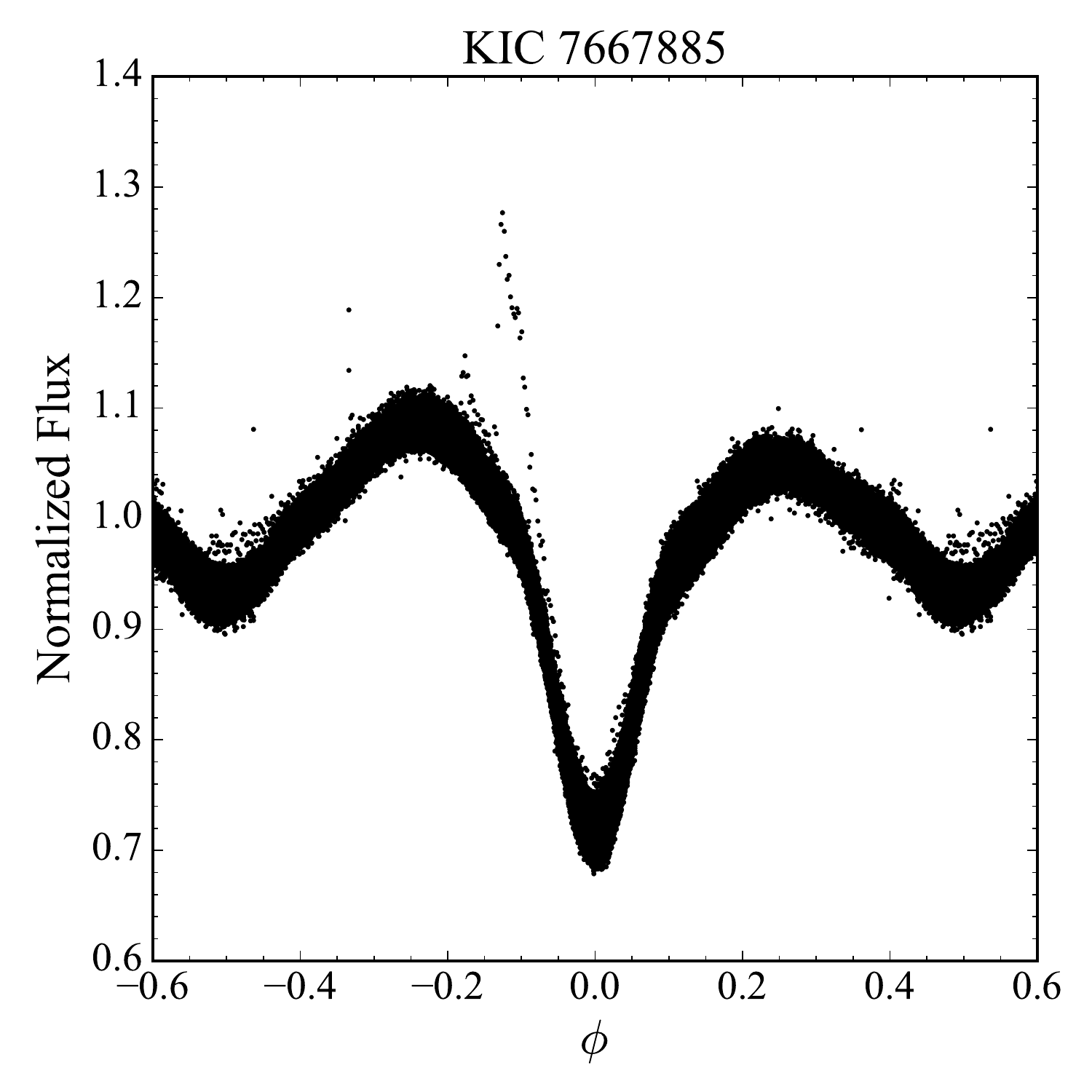}\unskip
\includegraphics[width=\columnwidth]{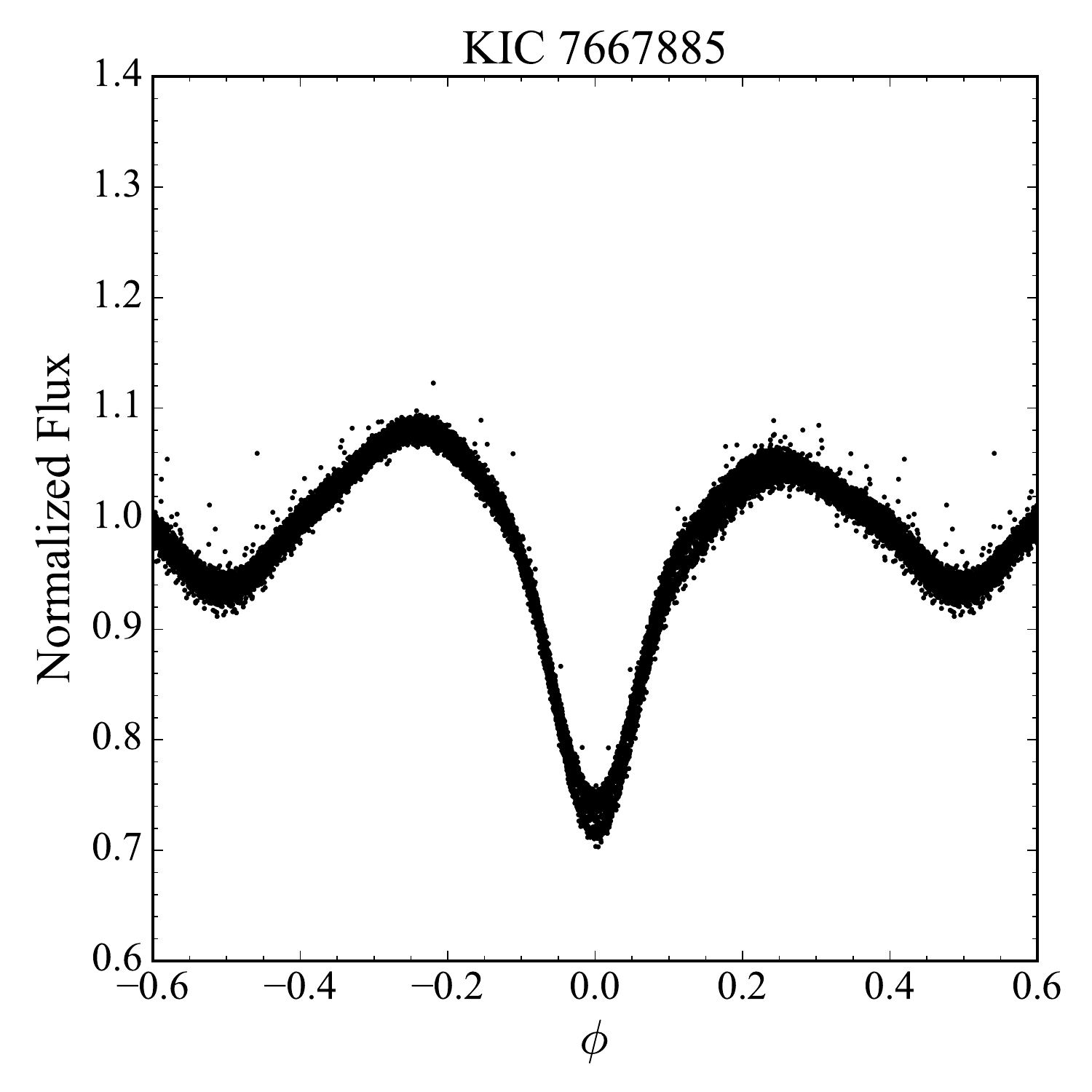}
\par\end{centering}
\caption{\emph{Kepler} short-cadence (left) and long-cadence (right) light curves of KIC~7667885. The system displays a negative O'Connell effect and at least one significant flare event.
\label{fig:KIC-7667885-light-curves}}
\end{figure*}

It is also important to keep in mind that our averaging process removes information about the temporally varying properties of the system. The non-averaged OES changes in magnitude and even sign in several systems, as does the overall shape of the light curve. Removing this information simplifies the analysis of these systems but reduces the utility of the averaged light curve in studying temporal variation. We will analyze the temporal variation separately by looking at the eclipse timing variation (discussed further in Section~\ref{subsec:ETV}).

\subsection{Target Sample\label{subsec:Target-Sample}}

Our final, complete sample consists of 211 systems found by our final OES determination code, 1 system (KIC~7667885) found in \citet{Ramsay2014}, and 46 systems found using earlier methods, for a total of 258 systems. Our OES determination code showed that three additional systems met our criterion. Of these, KIC~7950964 is a duplicate entry of a system already in the sample (KIC~7950962), and KIC~9137819 appears to be an RR Lyrae variable and not an eclipsing binary. These two systems have since been removed from the KEBC (A. Pr\v{s}a 2021, private comm.). The third system, KIC~8456774, is a heartbeat star \citep{Thompson2012}, which is an eccentric binary that becomes tidally distorted at periapsis. KIC~8456774, like many heartbeat stars, does not appear to be eclipsing, and we do not consider it to have an O'Connell effect due to the fundamentally different origin of its asymmetry. Additionally, it is inappropriate to apply the phase offset we introduced to place the primary minimum at $\phi = 0$ to non-eclipsing systems like KIC~8456774, and the OES without this phase offset is negligible. We do not include these three systems in our sample. We also do not include systems like KIC~11560447 that meet our criterion with their short-cadence data but not with their long-cadence data.

\setcounter{footnote}{3}

\emph{Kepler} observed KIC~7667885 during quarters 14-17, but the KEBC does not include the system. Inspection of KIC~7667885's short-cadence data from MAST\footnote{Mikulski Archive for Space Telescopes;\\ \url{http://archive.stsci.edu/kepler}} indicated that it met our criterion. We detrended the single-aperture photometry (SAP) flux by fitting Legendre polynomials up to order 150 to continuous data blocks (defined as a set of data with no gaps longer than a day) and then dividing the flux by the fitted trend. We then normalized the data by dividing the flux by the median flux value of each data block. This method is similar to the one outlined in Section~4.2 of \citet{Slawson2011}, which they used to produce the KEBC data. We determined a period of $0.314840 \pm 0.000004$~d using the analysis of variance \citep{Schwarzenberg-Czerny1989, Schwarzenberg-Czerny1996} method with the software Peranso. Figure~\ref{fig:KIC-7667885-light-curves} shows the resultant phased short- and long-cadence light curves. We did not determine the morphology parameter of KIC~7667885, nor did we perform an analysis of its ETV, but we otherwise fully incorporated the system into our sample.

One system in our sample, KIC~7879399, lies only 4'' from another \emph{Kepler} system, KIC~7879404. As a result of their proximity, these two systems' data are blended, making it difficult to determine which one is the eclipsing binary. The first two releases of the KEBC (\citealt{Prsa2011} and \citealt{Slawson2011}) identified KIC~7879404 as the binary. The third release (\citealt{Kirk2016} and \citealt{Abdul-Masih2016}), however, identified KIC~7879399 as the binary. C. Cynamon of the American Association of Variable Star Observers (AAVSO) observed these stars for us (C. Cynamon 2021, private comm.), and our analysis of his data conclusively confirms KIC~7879399 as the eclipsing binary.

Of the 258 systems in our sample, 46 selected before we finalized the selection method do not meet our criterion. We will describe these 46 marginal sample systems in a separate paper. The remaining 212 systems (the 211 KEBC systems found by our OES determination code and KIC~7667885, which our code also found when we added its processed data to the KEBC's) constitute our core sample, which represents 7.3\% of the systems in the KEBC\@. Table~\ref{tab:Target-List} lists all 258 systems in our complete sample.

\section{Methodology\label{sec:Methodology}}

We used several methods to characterize the systems in our sample: determinations of the eclipse depth, the O'Connell Effect Ratio and Light Curve Asymmetry, the KEBC's morphology parameter, and statistical analyses of the sample characteristics. The majority of these methods involved analyzing the phased \emph{Kepler} data taken from the KEBC using our in-house Python codes.

\subsection{Eclipse Depth Determination\label{subsec:Eclipse-Depth-Determination}}

\begin{figure}
\begin{centering}
\includegraphics[width=\columnwidth]{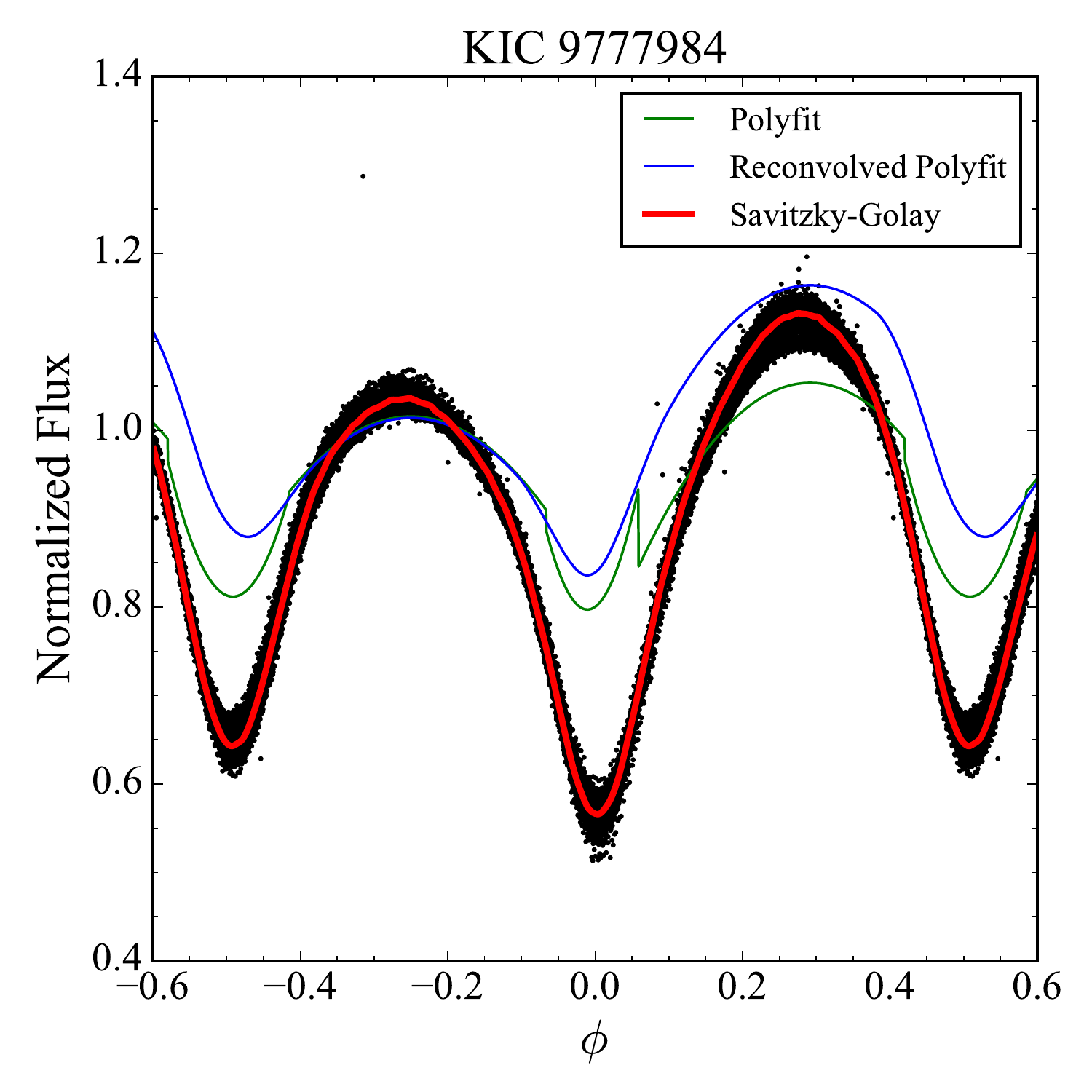}
\par\end{centering}
\caption{\emph{Kepler} light curve of KIC~9777984 showing three fits to the system: the KEBC's original polyfit (green), the KEBC's reconvolved polyfit (blue) that corrects for \emph{Kepler}'s non-zero integration time, and our fit using the Savitzky--Golay filter (red). The Savitzky--Golay curve better fits KIC~9777984's light curve and provides a more accurate measure of the eclipse depth for this system.
\label{fig:KIC-9777984-polyfit}}
\end{figure}

The KEBC provides a value for the eclipse depth determined using the polyfit method described in detail in Appendix A of \citet{Prsa2008}. In brief, the polyfit uses a chain of piecewise smooth polynomials connected at dynamically determined knots to fit the light curve, therefore providing a smoothed version of the light curve. This procedure works for the majority of systems in the KEBC\@. Unfortunately, it fails to accurately measure the eclipse depth for a small number of systems in our sample, as shown in Figure~\ref{fig:KIC-9777984-polyfit} with KIC~9777984's polyfits (note that \citealt{Kirk2016} and the KEBC identify this system as KIC~9777987, while \citealt{Abdul-Masih2016} identifies it as KIC~9777984). Furthermore, \citet{Kirk2016} states that the provided eclipse depths are approximate and only as accurate as the polyfits used to produce them.

We therefore needed to define our own method for determining eclipse depth. The convolved curve used to determine the OES consistently underestimates the depth of the eclipse, so we instead applied a Savitzky--Golay filter \citep{Savitzky1964} of polynomial order 4 and window size 7 to each averaged light curve. We applied the filter in the same way we convolved the data in Section~\ref{subsec:O'Connell-Effect-Size-Determination}, the sole difference being in the kernel. We used SciPy's \texttt{savgol\_coeffs} function to determine the kernel values for the Savitzky--Golay curve. This procedure produced a curve that approximates the eclipse depth much better than the polyfits for KIC 9777984 -- as Figure~\ref{fig:KIC-9777984-polyfit} shows -- despite retaining more of a oscillatory pattern than the convolved curve. We found the Savitzky--Golay curve's errors similarly to how we found the convolved curve's errors. We take the smallest value of this curve to be the point of maximum eclipse, and the error of the corresponding bin is the eclipse depth error. A limitation of this method is that Algol-type systems with narrow eclipses can have their eclipse depths underestimated by 50\% or more. Nevertheless, the Savitzky--Golay curve measures the eclipse depth more accurately than the Fourier curve (Section~\ref{subsec:OER-and-LCA-Determination}), particularly for some Algol-type systems like KIC~5700330 that lie outside of our sample. The Savitzky--Golay curve is also more accurate than the convolved curve in all but a few edge cases involving totally-eclipsing Algol-type systems.

Establishing a proper reference point from which to measure the depth is a non-trivial matter. We rejected using either maximum as the reference point as it is uncertain which maximum (if either) is unaffected by the process causing the O'Connell effect. Communication with the KEBC team revealed that they face the same uncertainty about a proper reference point and noted that modeling the light curve provides the only robust way of determining eclipse depth (A. Pr\v{s}a 2020, private comm.), a procedure which is beyond the scope of this paper. Therefore, we used the KEBC's normalized data and adopted a normalized flux value of unity to establish each system's baseline flux. We then subtracted the flux value at the eclipse minimum from this baseline to find the eclipse depth. \citet{Slawson2011} details the normalization scheme used for the KEBC data, and it is ``reasonable to assume that, for the most part, the data are median-normalized'' (A. Pr\v{s}a 2020, private comm.).

\subsection{O'Connell Effect Ratio and Light Curve Asymmetry\label{subsec:OER-and-LCA-Determination}}

The O'Connell Effect Ratio (OER) and Light Curve Asymmetry (LCA) are two statistical measures introduced by \citet{McCartney1999}, who used them with phased light curves broken into $n$ equally-spaced phase bins. While originally introduced to characterize the O'Connell effect in W Ursae Majoris-type systems, OER and LCA are valid measures for all light curve classes. The original definitions given by \citet{McCartney1999} are:
\begin{equation}
    \text{OER} = \frac{\sum_{i=1}^{n/2}{I_i - I_0}}{\sum_{i=(n/2)+1}^{n}{I_i - I_0}}
\end{equation}
and:
\begin{equation}
    \text{LCA} = \sqrt{\sum_{i=1}^{n/2}{\left(\frac{I_i - I_{n + 1 - i}}{I_i}\right)}^2}
\end{equation}
where $n$ is the number of phase bins, $I_i$ is the average intensity of the $i$\textsuperscript{th} bin, and $I_0$ is the minimum average intensity of the light curve. Note that the LCA should contain a normalization factor to account for the number of bins, i.e.:
\begin{equation}
    \text{LCA} = \sqrt{\frac{1}{n}\sum_{i=1}^{n/2}{\left(\frac{I_i - I_{n + 1 - i}}{I_i}\right)}^2}
\end{equation}
as in \citetalias{Wilsey2009}. \citetalias{Wilsey2009} transformed these summations into integrals (taking $n \to \infty$) such that:
\begin{equation}
    \text{OER} = \frac{\int_{0}^{1/2}{[I(\phi) - I(0)]~d\phi}}{\int_{1/2}^{1}{[I(\phi) - I(0)]~d\phi}}
    \label{eq:OER}
\end{equation}
and:
\begin{equation}
    \text{LCA} = \sqrt{\int_{0}^{1/2}{\left(\frac{I(\phi) - I(1 - \phi)}{I(\phi)}\right)^2}~d\phi}
    \label{eq:LCA}
\end{equation}
where $I(\phi)$ is the $N$-term Fourier series representation of the light curve:
\begin{equation}
    I(\phi) = \frac{a_0}{2} + \sum_{i = 1}^{N}{[a_i \cos{(2\pi i\phi)} + b_i \sin{(2\pi i\phi)}]}
\end{equation}
We use the integral formulation in this paper. Both OER and LCA assume that the secondary eclipse occurs at $\phi = 0.5$, so systems with eccentric orbits have inaccurate OER values. The LCA, by contrast, interprets the eccentricity as an asymmetry in the light curve.

\begin{figure}
\begin{centering}
\includegraphics[width=\columnwidth]{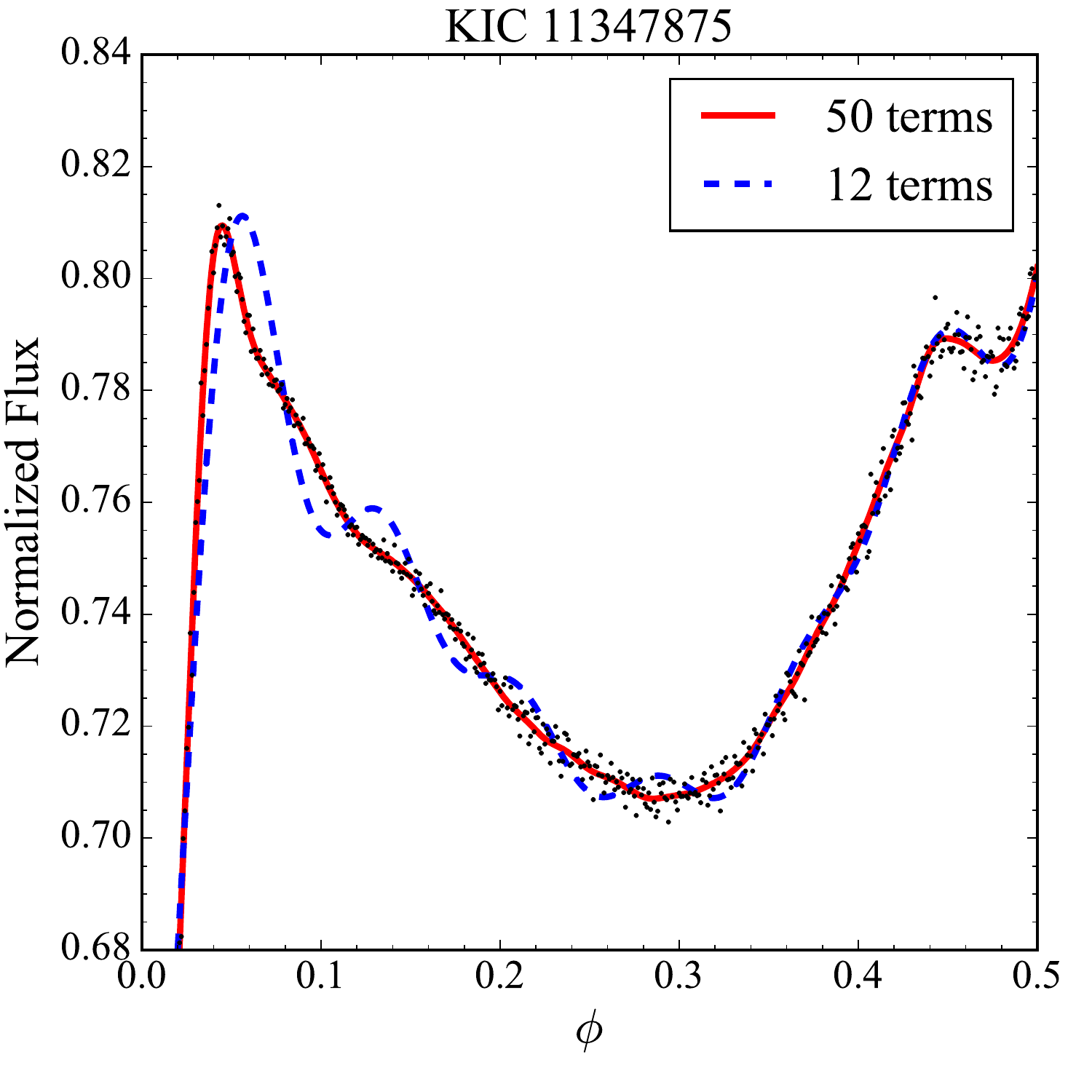}
\par\end{centering}
\caption{Averaged light curve of KIC~11347875 showing two Fourier series approximations of the system: the 50-term series used in this paper (solid red) and a 12-term series (dashed blue) similar to one \citet{Akiba2019} would use. The 50-term series follows the averaged data more closely than the 12-term series while not capturing the statistical scatter.
\label{fig:KIC-11347875-Fourier-comparison}}
\end{figure}

We chose to use a 50-term Fourier series based on a $\chi^2$ analysis of $N$ for our complete sample of 258 systems. Our code calculated the $\chi^2$ statistic by comparing the Fourier curve with $N$ coefficients to the averaged light curve of the system, where $N \in [2, 500]$. This analysis indicated a median optimal $N$ of 59. We adopted an $N$ of 50 because we found that varying $N$ in the range $[40, 60]$ changes the results little. Figure~\ref{fig:KIC-11347875-Fourier-comparison} compares our 50-term Fourier series of the core sample system KIC~11347875 to a 12-term Fourier series like those used in \citet{Akiba2019} and \citet{Hahs2020}. Figure~\ref{fig:KIC-11347875-Fourier-comparison} demonstrates that 50 terms are sufficient to capture most large-scale features in the light curve without having so many terms that the series approximates small-scale statistical scatter.

We obtained our Fourier series by applying a discrete Fourier transform to the averaged light curve described in Section~\ref{subsec:O'Connell-Effect-Size-Determination}. Our code converted the resultant complex-valued coefficients $c_i$ to the Fourier series coefficients $a_i$ and $b_i$ using the equations:
\begin{subequations}
\begin{align}
    a_i & = \frac{2\text{Re}(c_i)}{n}
    \label{eq:a-coefficient}\\
    b_i & = -\frac{2\text{Im}(c_i)}{n}
    \label{eq:b-coefficient}
\end{align}
\end{subequations}
We use the resulting Fourier series to calculate the OER and LCA using Equations~\ref{eq:OER} and \ref{eq:LCA}, respectively. We also use the $a_i$ coefficients to determine each system's light curve class, as we describe in Section~\ref{subsec:Morphology-Parameter}. The 50-term Fourier series is an excellent approximation of $\beta$ Lyrae- and W Ursae Majoris-type systems. However, it struggles to accurately represent several Algol-type systems, particularly those with sharp, narrow eclipses. Such systems dominate the long-period ($P > 10$~d) population of the KEBC but are rare in our sample.

We based our error analysis on the method described in \citet{Akiba2019}, but there were some differences caused by our use of \texttt{fft}, most prominently that all $a_i$ and $b_i$ coefficients with $i > 0$ have the same error. This error is given by:
\begin{equation}
    \sigma_X = \frac{\sqrt{2}}{n}\sqrt{\sum_{j = 1}^{n}{\displaystyle\sigma_{x_j}^2}}
    \label{eq:Coefficient-error}
\end{equation}
where $\sigma_{x_j}$ are the errors of each bin given by Equation~\ref{eq:Bin-error}. The error for $a_0$ is larger by a factor of $\sqrt{2}$. Our errors for OER and LCA are:
\begin{equation}
\begin{split}
    &\sigma_\text{OER} = \text{OER}\frac{\sigma_X}{2}\\
    &~~~\times~\sqrt{\frac{N + 2\sum_{i = 1}^{N}{\left(\frac{1}{\pi i}\right)^2}}{\left[H\left(\frac{1}{2}\right) - H(0)\right]^2} + \frac{5N + 2\sum_{i = 1}^{N}{\left(\frac{1}{\pi i}\right)^2}}{\left[H(1) - H\left(\frac{1}{2}\right)\right]^2}}
    \label{eq:OER-error}
\end{split}
\end{equation}
and:
\begin{equation}
\begin{split}
    \sigma_\text{LCA} & \approx \frac{\sigma_X}{\text{LCA}}\int_{0}^{1/2}{}\left[\left(\frac{I(\phi) - I(1 - \phi)}{I(\phi)}\right)^2\right.\\
    &~~~~~\left.\times~\sqrt{\frac{2\left(N + \frac{1}{2}\right)}{\left[I(\phi) - I(1 - \phi)\right]^2} + \frac{N + \frac{1}{2}}{\displaystyle I(\phi)^2}}\right]~d\phi
    \label{eq:LCA-error}
\end{split}
\end{equation}
where:
\begin{equation}
\begin{split}
    H(\phi) & = \sum_{i = 1}^{N}{\left[\frac{a_i}{2\pi i} \sin{(2\pi i\phi)} - \frac{b_i}{2\pi i} \cos{(2\pi i\phi)} \right.}\\
    &~~~~~~~~~~~\left. \vphantom{\frac{b_i}{2\pi i} \cos{(2\pi i\phi)}} {}-{} a_i\phi\right]
    \label{eq:H-equation}
\end{split}
\end{equation}
is the antiderivative of $I(\phi) - I(0)$.

\subsection{Morphology Parameter and Light Curve Classification\label{subsec:Morphology-Parameter}}

\citet{Matijevic2012} introduced the morphology parameter (which we represent with the symbol $\mu$) to determine the morphology class of KEBC systems. They used local linear embedding \citep{Roweis2000} to reduce the light curve's polyfit representation \citep{Prsa2008} to a single, one-dimensional parameter. This parameter proved to correlate very well with their manually identified morphology class for most KEBC systems. They determined that $\mu \leq 0.5$ implied detached systems, $0.5 < \mu \leq 0.7$ implied semi-detached systems, $0.7 < \mu \leq 0.8$ implied overcontact systems, and $\mu > 0.8$ implied ellipsoidal variables (non-eclipsing systems where the variation is due to the changing aspect of tidally distorted stars). They note, however, that the morphology parameter is a ``best guess'' at the morphology class and that system modeling can more accurately determine the morphology class. Nevertheless, we use the morphology parameter to estimate a system's morphology class throughout this paper, and Table~\ref{tab:Target-List} lists each system's morphology parameter.

Based on our own visual classification of the systems in our sample, we estimated that Algol-type systems have $\mu < 0.6$, $\beta$ Lyrae-type systems have $0.6 \leq \mu < 0.75$, and W Ursae Majoris-type systems have $\mu \geq 0.75$. These boundaries agree with those in \citet{Matijevic2012}, as Algol-type systems are generally detached or semi-detached, $\beta$ Lyrae-type systems can be of any morphological class, and W Ursae Majoris-type systems are predominantly overcontact (although detached and semi-detached eclipsing binaries, as well as ellipsoidal variables, may also appear as W Ursae Majoris-type systems).

In addition, we quantified the light curve classification using the Fourier coefficients obtained from Equation~\ref{eq:a-coefficient}. Our procedure is a modification of the one described in Section~3.2 of \citet{Akiba2019} and uses the following criteria to classify systems:
\begin{enumerate}
\item If $a_4 < a_2 (0.125 - a_2)$, the system is classified as an Algol-type system.

\item If $a_4 \geq a_2 (0.125 - a_2)$ and $|a_1 / a_2| > 0.25$, the system is classified as a $\beta$ Lyrae-type system.

\item If $a_4 \geq a_2 (0.125 - a_2)$ and $|a_1 / a_2| \leq 0.25$, the system is classified as a W Ursae Majoris-type system.
\end{enumerate}
Criterion 1 comes from \citet{Rucinski1997} and is unchanged from \citet{Akiba2019}, as are the first parts of Criteria~2 and 3. However, the second parts of Criteria~2 and 3 differ from \citet{Akiba2019}, who instead used $|a_1| > 0.05$ to define $\beta$ Lyrae-type systems and $|a_1| < 0.05$ to define W Ursae Majoris-type systems. These criteria arise from \citetalias{Wilsey2009}'s observation that $a_1$ is proportional to the difference in eclipse depth. Unfortunately, $a_1$ does not contain any information about the eclipse depth \emph{ratio}, which is the parameter related to the component temperature ratio \citep{Kallrath2009}. We introduced $a_2$ (which \citetalias{Wilsey2009} notes is proportional to the light curve amplitude) to quantify the eclipse depth ratio.

We note that our sample likely contains several ellipsoidal variables. It is difficult to distinguish an ellipsoidal variable from an overcontact system due to their similar light curves. Preliminary modeling of four systems in our sample with PHOEBE v0.31a \citep{Prsa2005} shows that the morphology parameter is not wholly effective at identifying ellipsoidal variables. As an example, KIC~8285349 has $\mu = 0.90$ but clearly shows an eclipse. Meanwhile, KICs~10815379 and 10979669 have $\mu = 0.79$ and 0.83, respectively, but are non-eclipsing according to our models. Identifying and removing ellipsoidal variables would therefore require large-scale modeling of the sample systems, which is beyond the scope of this paper.

\subsection{Statistical Analysis\label{subsec:Statistical-Analysis}}

To determine if our sample population was similar to the KEBC population as a whole, we performed the Kolmogorov--Smirnov test \citep[K--S test;][]{Kolmogorov1933, Smirnov1948} on the observational characteristics we studied. If the test statistic differs for a given characteristic, it suggests a potential link between that characteristic and a significant O'Connell effect. We expect the OES and $|$OES$|$ populations to differ due to how we defined our sample. We do not expect the distance populations to differ because the O'Connell effect should be independent of distance.

We determined correlations between characteristics using Spearman's $\rho$ coefficient \citep{Spearman1904}, as in \citetalias{Davidge1984}. We also used Kendall's $\tau$ coefficient \citep{Kendall1938} as a second test of correlation. Our decision to use both coefficients stemmed from research showing that the two measures complement each other \citep{Xu2013}. Unlike \citetalias{O'Connell1951} and \citetalias{Davidge1984}, we did not use Pearson's r coefficient due to its sensitivity to outliers and assumption of a linear correlation. We tested six functional forms for each correlation using SciPy's implementation of orthogonal distance regression \citep[ODR;][]{Boggs1987} to determine the best fit to the correlation. The functional forms we tested were linear, quadratic, exponential, logarithmic, power law, and inverse.

In order to test the robustness of our correlation analysis results, we performed a bootstrapping procedure on our sample. Our code randomly selected 20 subsets of 40 systems from our core sample (and 400 systems from the KEBC) and computed the Spearman's $\rho$ coefficients for each subset. We considered a correlation robust if at least 19 of the 20 subsets recovered the correlation.

\section{Results\label{sec:Results}}

We now present a detailed summary of various characteristics of the core sample, specifically the light curve characteristics (distributions of eclipse depth, OES, and positive and negative OESs) and physical characteristics (spatial, luminosity, temperature, and period distributions). Determining the characteristic distributions of the underlying population allows us to make inferences about the physical processes underlying the O'Connell effect. We also discuss the results of our analysis of the ETV, OER, and LCA\@. Finally, we present several system classes in our sample that show peculiar features in their light curves aside from the O'Connell effect. These features include changes in the light curve over time, asymmetric minima, and concave-up regions. In the figures in Sections~\ref{sec:Results} and \ref{sec:Analysis-and-Discussion}, the distribution or data set labeled ``Core Targets'' includes all 212 core sample systems, while the distribution or data set labeled ``KEBC Systems'' includes the core sample (except KIC~7667885) and the rest of the KEBC\@. In this section, we use the comparative subset of the KEBC discussed in Section~\ref{subsec:KEBC}, which excluded four systems: KIC~5217781, a long-period system with severe data corruption, KIC~7667885, due to it not being in the KEBC, and KICs~7950964 and 9137819, due to their removal from the KEBC (as explained in Section~\ref{subsec:Target-Sample}).

\subsection{Light Curve Characteristics\label{subsec:Light-Curve-Characteristics}}

We determined several characteristics from analyzing the \emph{Kepler} photometry of the KEBC systems: light curve class, OES, the proportion of positive and negative O'Connell effects (defined in Section~\ref{subsec:O'Connell-Effect-Size-Determination}), primary eclipse depth, and the ratio between $|$OES$|$ and the primary eclipse depth. This section discusses the range and distribution of these characteristics. Using the method described in Section~\ref{subsec:Morphology-Parameter}, we classified 54 systems (25\% of our sample) as Algol-type, 40 (19\%) as $\beta$ Lyrae-type, and 118 (56\%) as W Ursae Majoris-type. This distribution strongly differs from the KEBC's, wherein 1,990 (68\%) are Algol-type, 250 (9\%) are $\beta$ Lyrae-type, and 679 (23\%) are W Ursae Majoris-type.

The largest OES in our sample is KIC~11347875, which has an OES of $-0.265$ in units of normalized flux. However, KIC~1134785 is an unusual system in that one of its inter-eclipse maxima is instead a minimum. Therefore, it lacks an O'Connell effect in the traditional sense described in Section~\ref{sec:Introduction}. We discuss KIC~11347875 further in Section~\ref{subsubsec:Concave-Up-Systems}. The largest OES among systems with a more traditional O'Connell effect is KIC~9935311, with an OES of 0.115. A clear majority of systems (147, or 69\% of the sample) show a positive O'Connell effect, where the brighter maximum is the one following the primary eclipse. We discuss the implications of this preference for a positive O'Connell effect in Section~\ref{subsubsec:O'Connell-Effect-Size-Correlations}.

\begin{figure}
\begin{centering}
\includegraphics[width=\columnwidth]{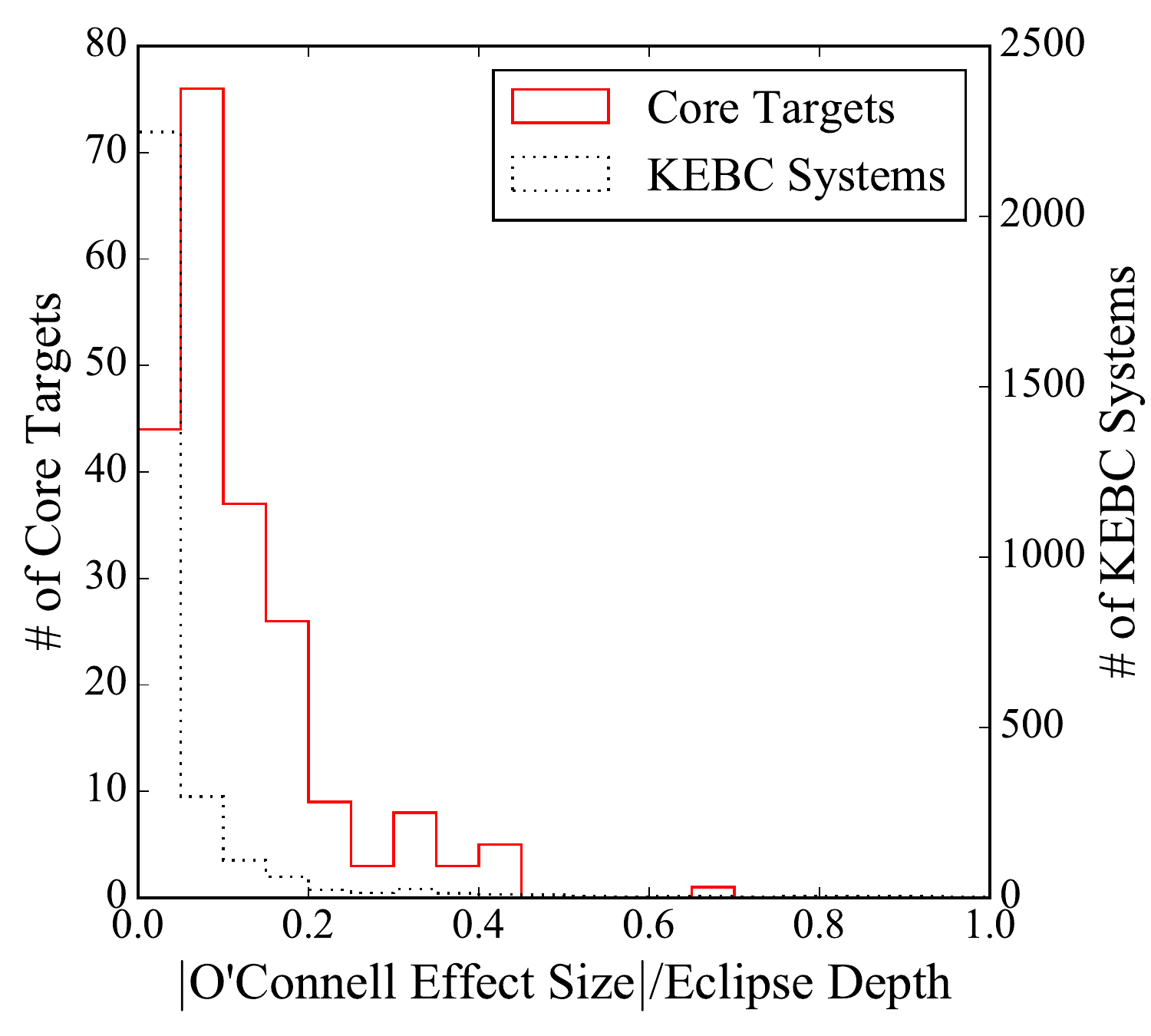}
\par\end{centering}
\caption{Histogram comparing the ratio $|$OES$|$/primary eclipse depth for our core sample (solid red) and all entries in the KEBC (dashed grey). Systems with a ratio above 1 are not displayed. Core sample systems have a larger ratio on average than other KEBC systems.
\label{fig:OES-Eclipse-Depth-Ratio-Histogram}}
\end{figure}

The primary eclipse depth ranges from 0.025 (KIC 8190491) to 0.916 (KIC~9101279) in units of normalized flux. Figure~\ref{fig:OES-Eclipse-Depth-Ratio-Histogram} shows the ratio between $|$OES$|$ and the primary eclipse depth for the core sample and the KEBC\@. We have determined that the overwhelming majority of systems with a ratio above 0.5 are heartbeat stars, a type of binary discussed in Section~\ref{subsec:Target-Sample}. The only system in our sample among such high ratio systems is the aforementioned non-heartbeat star KIC~11347875, which has a ratio of 0.658.

\subsection{Physical Characteristics\label{subsec:Physical-Characteristics}}

\begin{figure}
\begin{centering}
\includegraphics[width=\columnwidth]{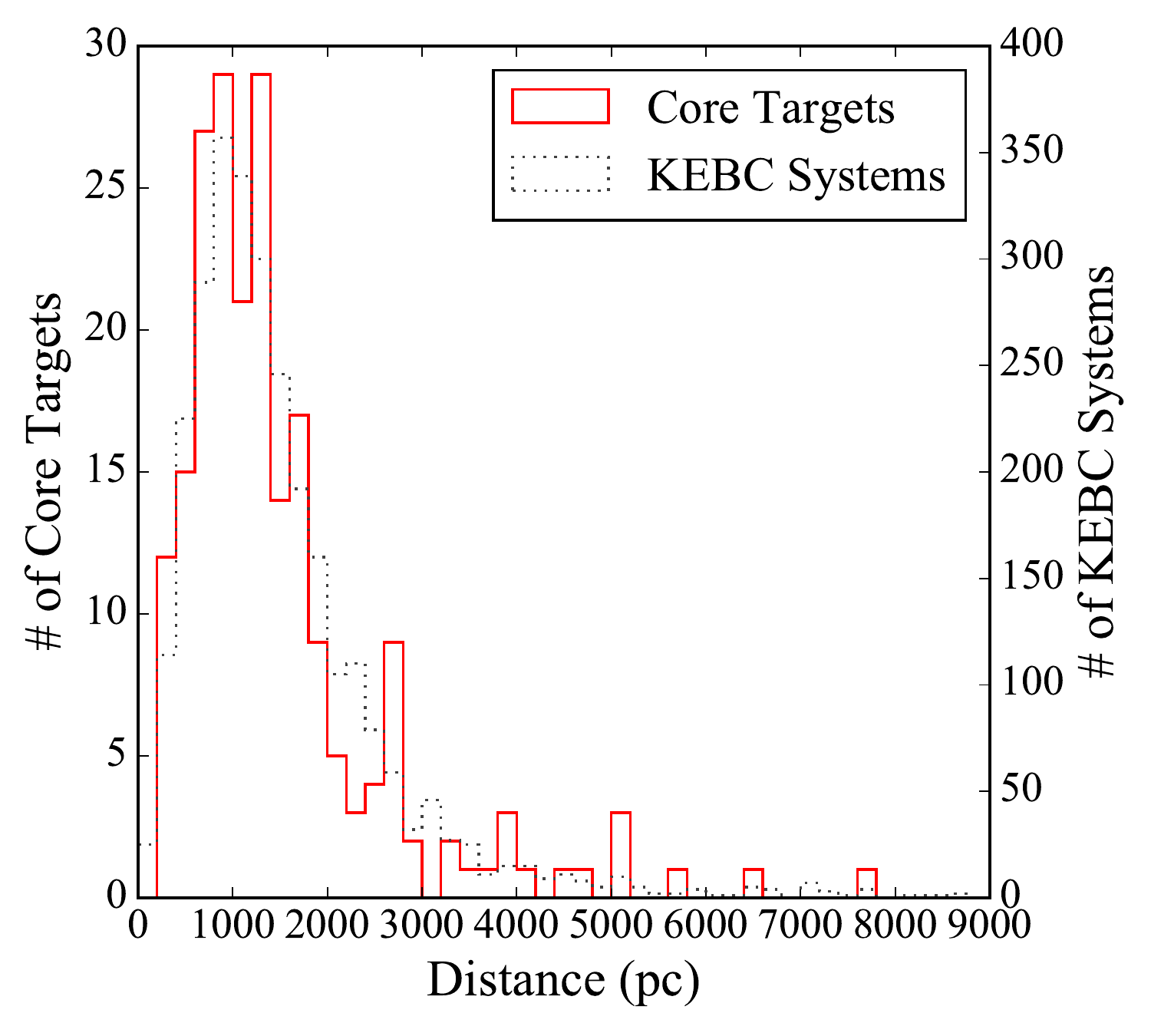}
\par\end{centering}
\caption{Histogram comparing the implied distances for the 212 core sample targets (solid red) and 2,860 of 2,862 entries in the KEBC (dashed grey) with \emph{Gaia} EDR3 parallaxes (excluding two systems with $\text{d} > 9,\!000$~pc). Our core sample's distances range from 210~pc (KIC~7671594) to 7,743~pc (KIC~4474637). The two distributions are similar.
\label{fig:Distance-Histogram}}
\end{figure}

\begin{figure}
\begin{centering}
\includegraphics[width=\columnwidth]{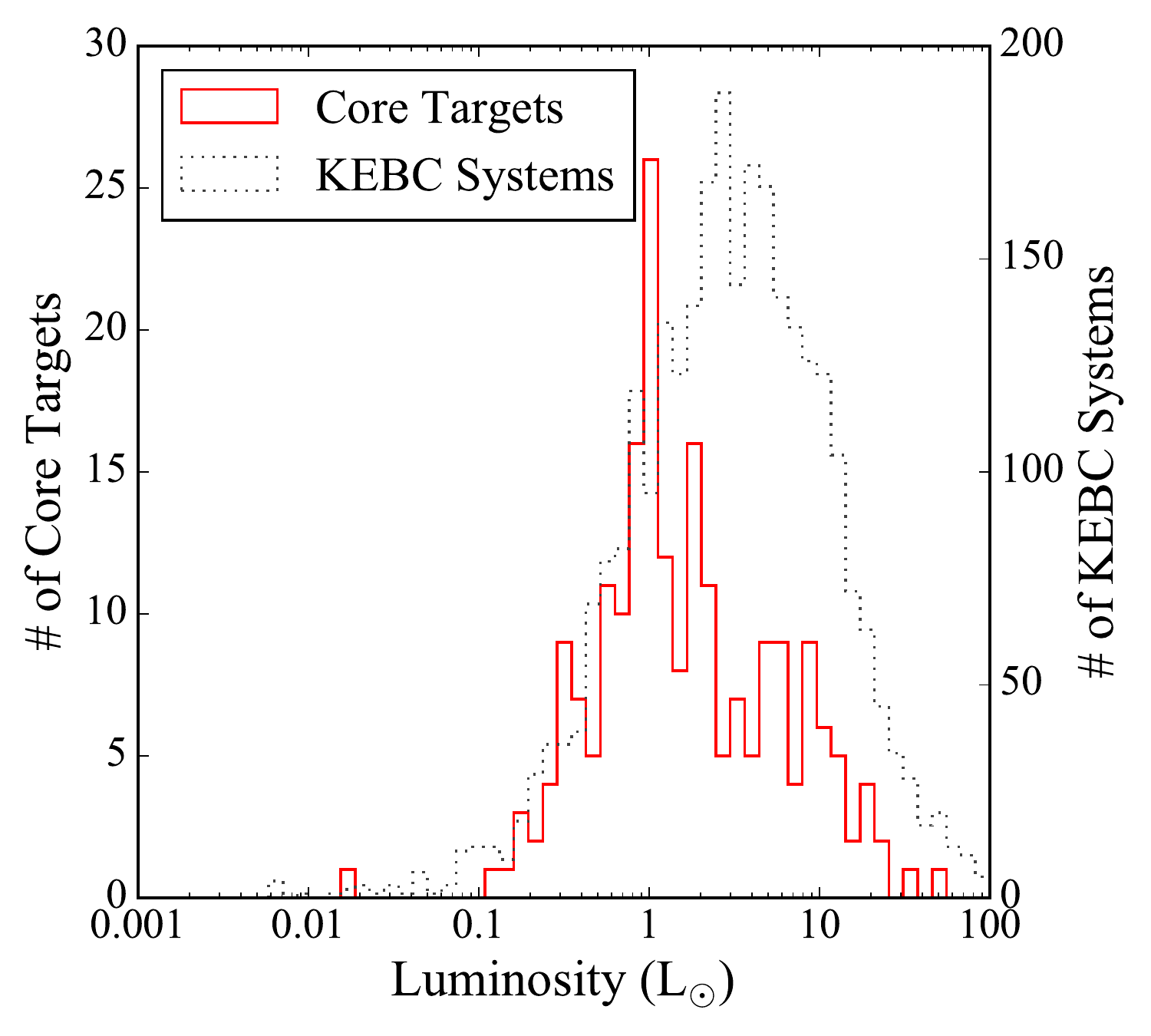}
\par\end{centering}
\caption{Histogram comparing the luminosities for the 212 core sample targets (solid red) and all 2,862 entries in the KEBC (dashed grey) with \emph{Gaia} EDR3 parallaxes. Note the logarithmic scale of the \emph{x}-axis. Our core sample's calculated luminosities range from 0.02~L\textsubscript{$\odot$} (KIC~7671594) to 50.25~L\textsubscript{$\odot$} (KIC~5820209). Systems in our sample are less luminous on average than other KEBC systems.
\label{fig:Luminosity-Histogram}}
\end{figure}

\begin{figure}
\begin{centering}
\includegraphics[width=\columnwidth]{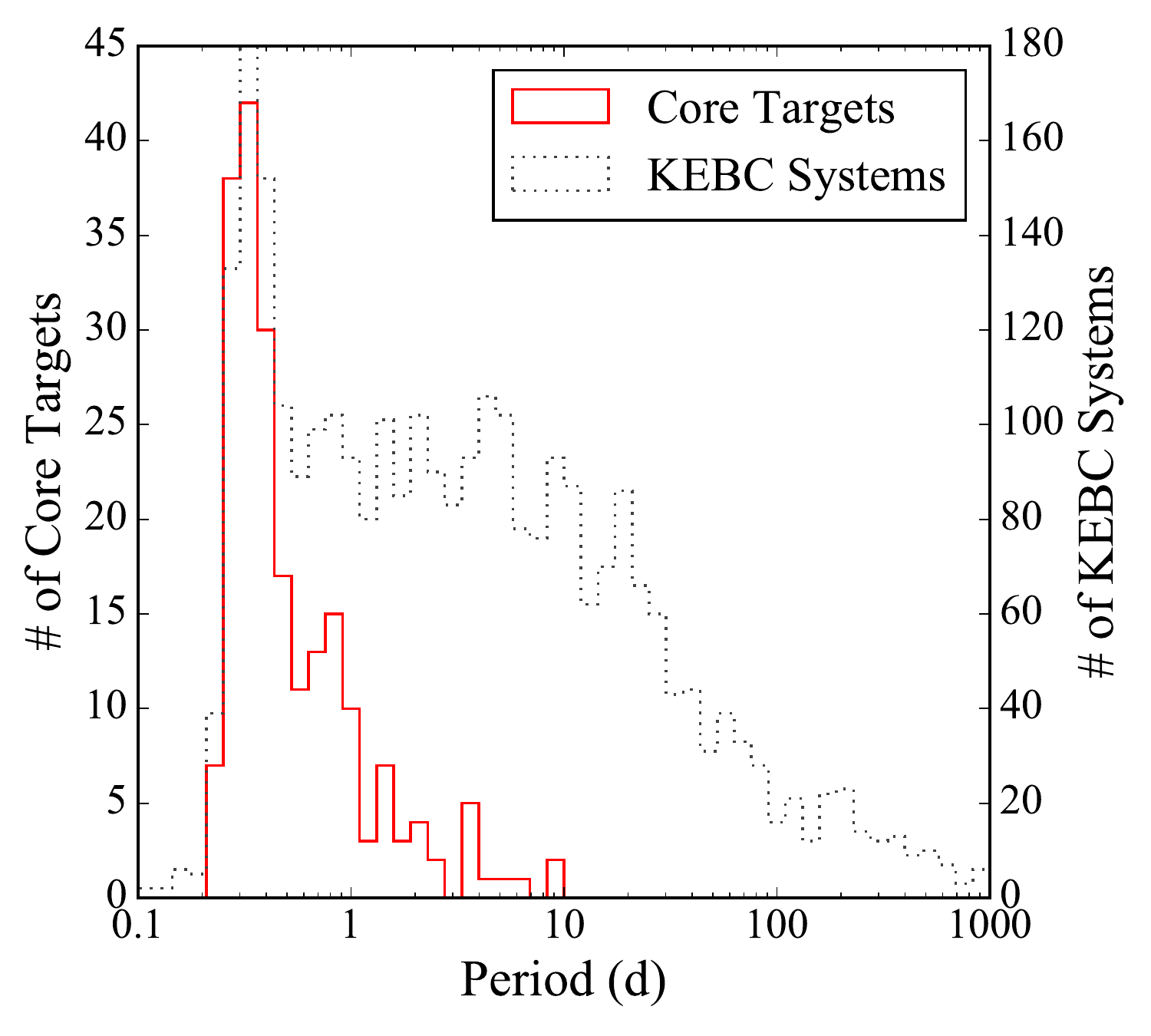}
\par\end{centering}
\caption{Histogram comparing the periods for all 212 core sample targets (solid red) and all 2,915 entries in the KEBC (dashed grey) with a period below 1,000~d. Note the logarithmic scale of the \emph{x}-axis. Our core sample's orbital periods range from 0.234~d (KIC~6050116) to 9.752~d (KIC~6197038). The distributions are similar under 0.5~d, but our sample's long-period population decreases much more quickly than the KEBC's.
\label{fig:Period-Histogram}}
\end{figure}

We determined several physical characteristics from the published information on the KEBC: distance, luminosity, period, temperature, and spectral type. This section focuses on these characteristics' distributions for our core sample and the KEBC\@. We also discuss our sample's color-magnitude diagram and references that indicate the presence of flares, spots, or mass transfer in our systems. Figure~\ref{fig:Distance-Histogram} shows a histogram of the distances derived from \emph{Gaia} parallaxes \citep{Bailer-Jones2021} for our sample, along with the distance distribution for the KEBC\@. These two distributions appear similar. Figure~\ref{fig:Luminosity-Histogram} shows a histogram of the luminosity distribution in our sample and the KEBC, and we note that our sample's systems are less luminous on average than those in the KEBC\@. Figure~\ref{fig:Period-Histogram} shows a histogram of the period distribution for our sample and the KEBC\@. We obtained these periods directly from the KEBC, and Figure~\ref{fig:Period-Histogram} has a logarithmic scale to capture the KEBC's wide range of periods. Figure~\ref{fig:Period-Histogram} shows that our sample has a similar distribution for $P \leq 0.5$~d, but the proportion of systems with longer periods drops precipitously.

Both the KEBC and \emph{Gaia} DR2 \citep{Gaia2018} provide temperatures for most of our targets based on their colors. The KEBC temperatures range from 3,717~K (KIC~7671594) to 8,540~K (KIC~10857342), while the \emph{Gaia} temperatures range from 3,808~K (KIC~7671594) to 8,540~K (KIC~10857342). The \emph{Kepler} and \emph{Gaia} temperatures are similar for the systems in our sample. Nine target systems do not have a \emph{Kepler} temperature, while seven do not have a \emph{Gaia} temperature. We chose to use \emph{Gaia} as our primary source of temperature in light of the more complete coverage \emph{Gaia} provides.

\citet{Frasca2016} provides spectral types for a handful of target systems determined using optical spectra (although they do not state if they assumed solitary stars during their data processing), and the main-sequence spectral types range from K5 (KIC~12109575) to A2 (KIC~8904448). There are eleven systems with an explicit non-main-sequence spectral type listed in Table~\ref{tab:Target-List}. \citet{Ramsay2014} spectrally classified KIC~7667885 and KIC~9786165 as ``mid G'' systems. We used the online table\footnote{\url{http://www.pas.rochester.edu/~emamajek/EEM\_dwarf\_UBVIJHK\_colors\_Teff.txt}} of photometric colors periodically updated by Dr.\ Eric Mamajek \citep{Pecaut2013} in combination with the \emph{Gaia} colors provided by EDR3 to expand the spectral classification to the entire sample. While this does not account for interstellar reddening, evolutionary stage, or color blending of the binary components, it provides a first approximation estimate of the spectral class for all systems in our sample, which Table~\ref{tab:Target-List} also lists. By this estimate, the spectral types of our sample range from M2.5~V (KIC~7671594) to A7~V (KIC~10857342).

\begin{figure}
\begin{centering}
\includegraphics[width=\columnwidth]{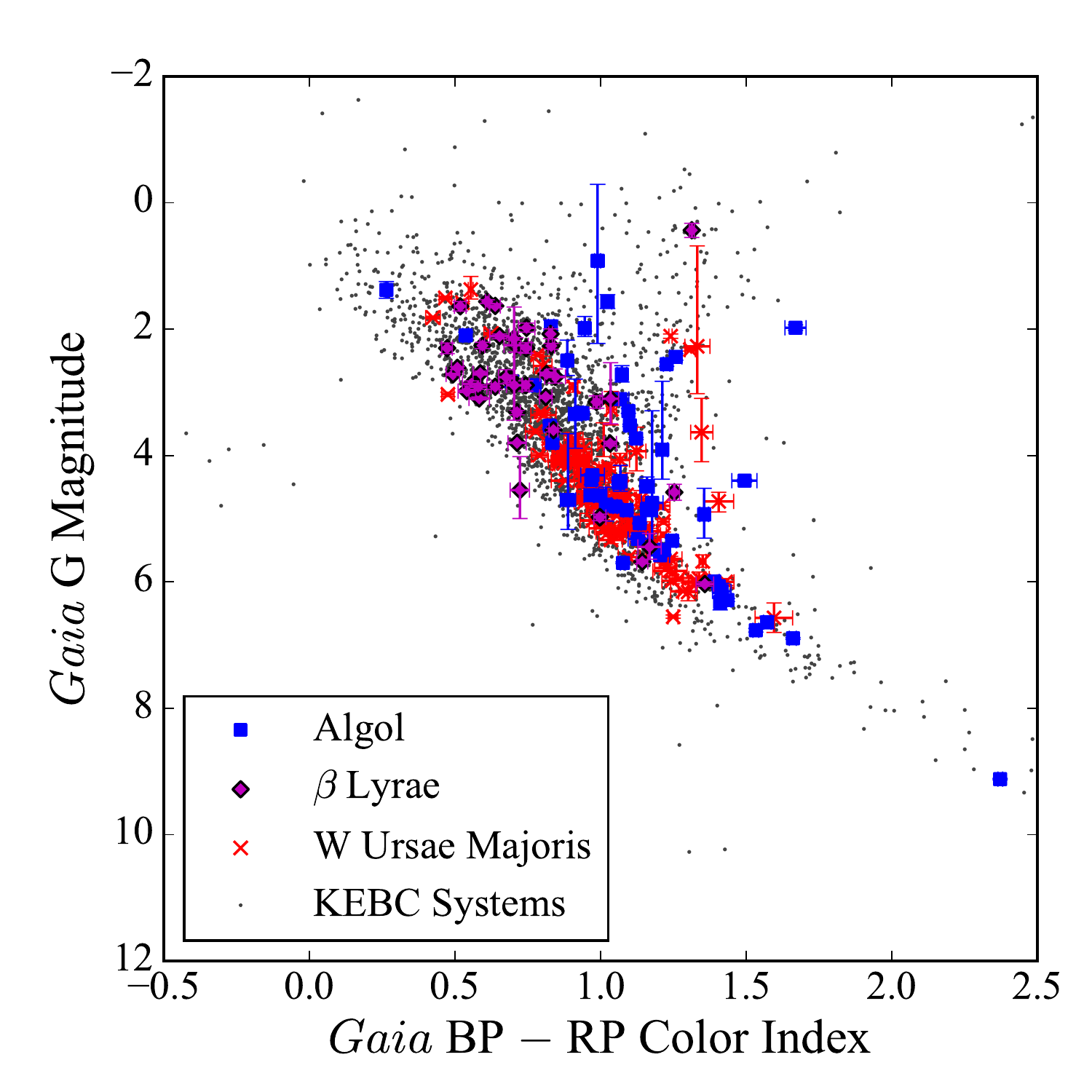}
\par\end{centering}
\caption{Color-magnitude diagram showing the 212 systems in the core sample overplotting the rest of the KEBC systems. The \emph{Gaia} color is plotted against the absolute magnitude in \emph{Gaia}'s G band. The systems are color coded according to their light curve class. Most of our sample are solar-type main-sequence stars.
\label{fig:C-M-Diagram}}
\end{figure}

Figure~\ref{fig:C-M-Diagram} shows a color-magnitude diagram using \emph{Gaia} EDR3 data. The marker color and shape show the light curve classification for each system determined using the criteria given in Section~\ref{subsec:Morphology-Parameter}. Figure~\ref{fig:C-M-Diagram} shows the \emph{Kepler} selection bias discussed in \citet{Batalha2010} and Section~\ref{subsec:Kepler}, as there are few systems with a \emph{Gaia} color index $\text{BP} - \text{RP} \geq 1.5$ or $\text{BP} - \text{RP} \leq 0.5$. Our sample's distribution in Figure~\ref{fig:C-M-Diagram} is similar to the KEBC's along most of the main sequence, suggesting that our sample is not strongly biased beyond \emph{Kepler}'s selection function. Additionally, only two clear giants exist in our sample (KICs~5820209 and 9489411, although the former's 0.656-day period is incongruent with containing a giant because such a period with two 1~M\textsubscript{$\odot$} stars implies a maximum star size of only $\sim$2.5~R\textsubscript{$\odot$}).

Numerous systems in our sample are thought to have observed flares \citep[57\% of the sample;][]{Balona2015, Gao2016, Davenport2016} or be spotted \citep[42\%;][]{Tran2013, Balaji2015}. Table~\ref{tab:Target-List} lists these systems with reference numbers 1-3 and 4-5, respectively, and the flags F and S, respectively. \citet{Kouzuma2018} found evidence for mass transfer in 22 (10\%) systems in our sample. Table~\ref{tab:Target-List} lists these systems with reference number 6 and the flag M. Therefore, our sample consists mainly of solar-type main-sequence stars with evidence for flares, spots, or mass transfer in 148 (70\%) sample systems.

\subsection{Eclipse Timing Variations\label{subsec:ETV}}

\begin{figure*}
\begin{centering}
\includegraphics[width=\columnwidth]{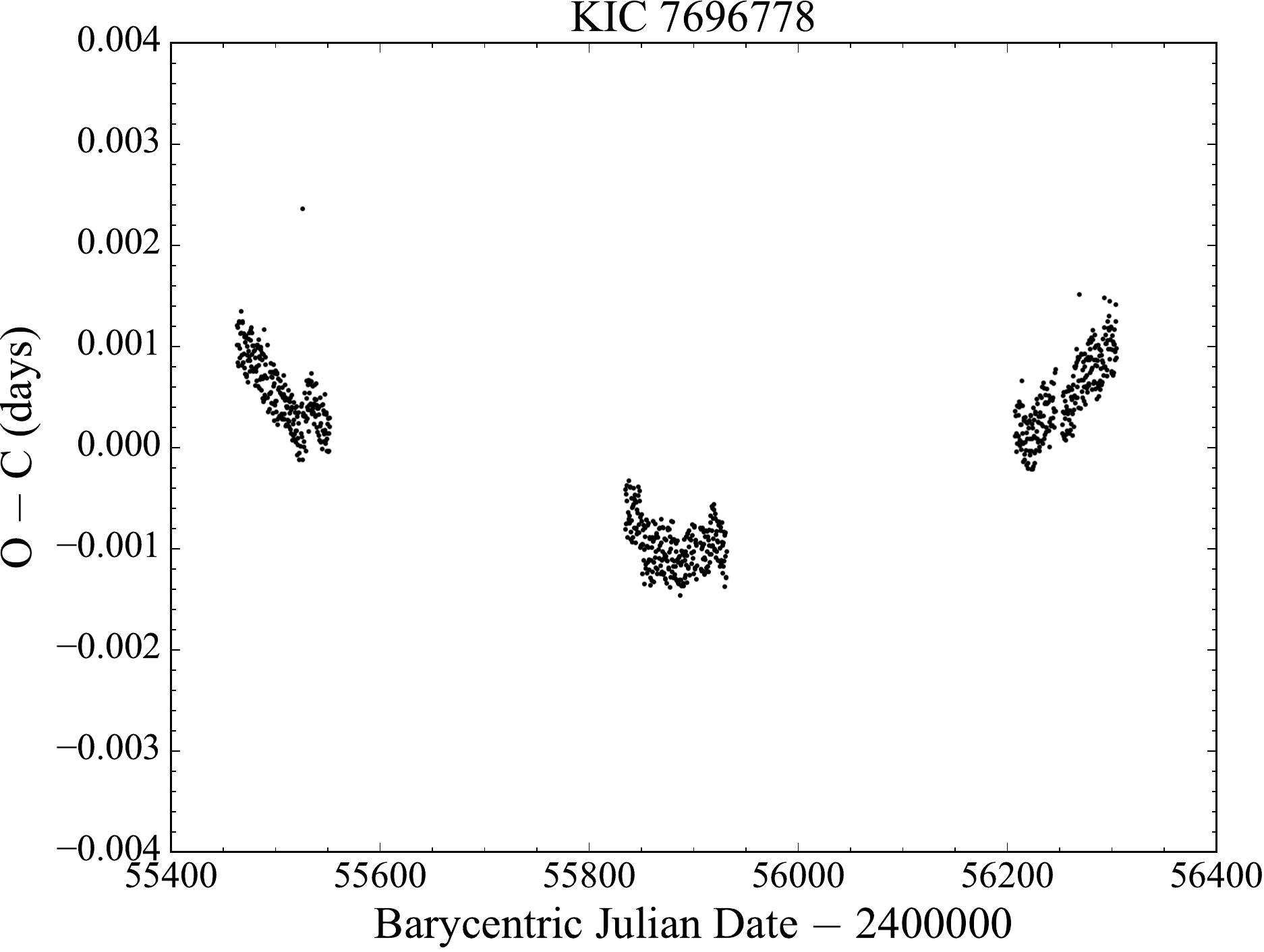}\unskip
\includegraphics[width=\columnwidth]{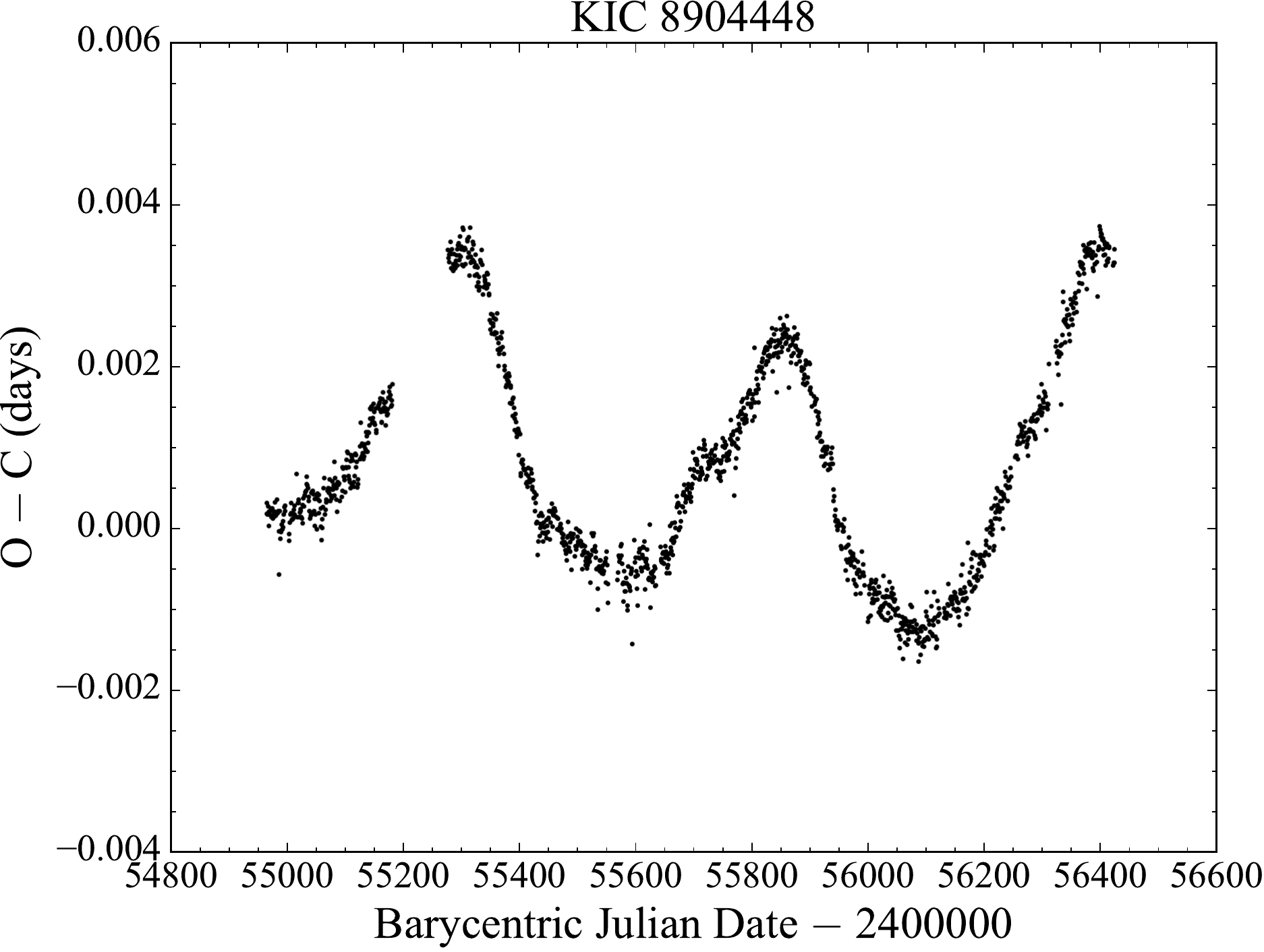}
\par\end{centering}
\caption{Eclipse timing variation of KIC~7696778 (left) and KIC~8904448 (right). KIC~7696778's ETV has a parabolic shape, consistent with a linear change in period caused by mass transfer between components, while KIC~8904448's ETV has a sinusoidal shape, consistent with LTTEs caused by a third body orbiting the binary.
\label{fig:KIC-7696778-&-8904448-ETV}}
\end{figure*}

Eclipse timing variation (ETV) measures how much the primary and secondary eclipses times vary from their theoretical values calculated assuming a constant period. ETV patterns reveal apparent period changes due to light travel time effects (LTTEs) caused by a third body, along with actual period changes due to effects such as mass transfer. Mass transfer creates a parabolic signal (as Figure~\ref{fig:KIC-7696778-&-8904448-ETV}'s left panel shows), while an LTTE creates a sinusoidal signal with a period equal to the third body's orbital period (as Figure~\ref{fig:KIC-7696778-&-8904448-ETV}'s right panel shows). For this paper, we use the ETV data published by \citet{Conroy2014}, a paper describing the ETVs of KEBC systems with morphology parameters between 0.5 and 1.0. \citet{Conroy2014} distinguished a parabolic signal from a long-period sinusoidal signal using the Bayesian Information Criterion \citep{Schwarz1978}. Only three core sample systems (KICs~5020034, 6044064, and 7696778) show a parabolic ETV signal, while none from our marginal sample do. By contrast, forty core sample systems and seven marginal sample systems show a sinusoidal signal. Furthermore, forty-seven core sample systems and seven marginal sample systems show an ``interesting'' ETV, meaning it has a pattern that does not clearly fit into the other two categories.

\begin{figure*}
\begin{centering}
\includegraphics[width=\columnwidth]{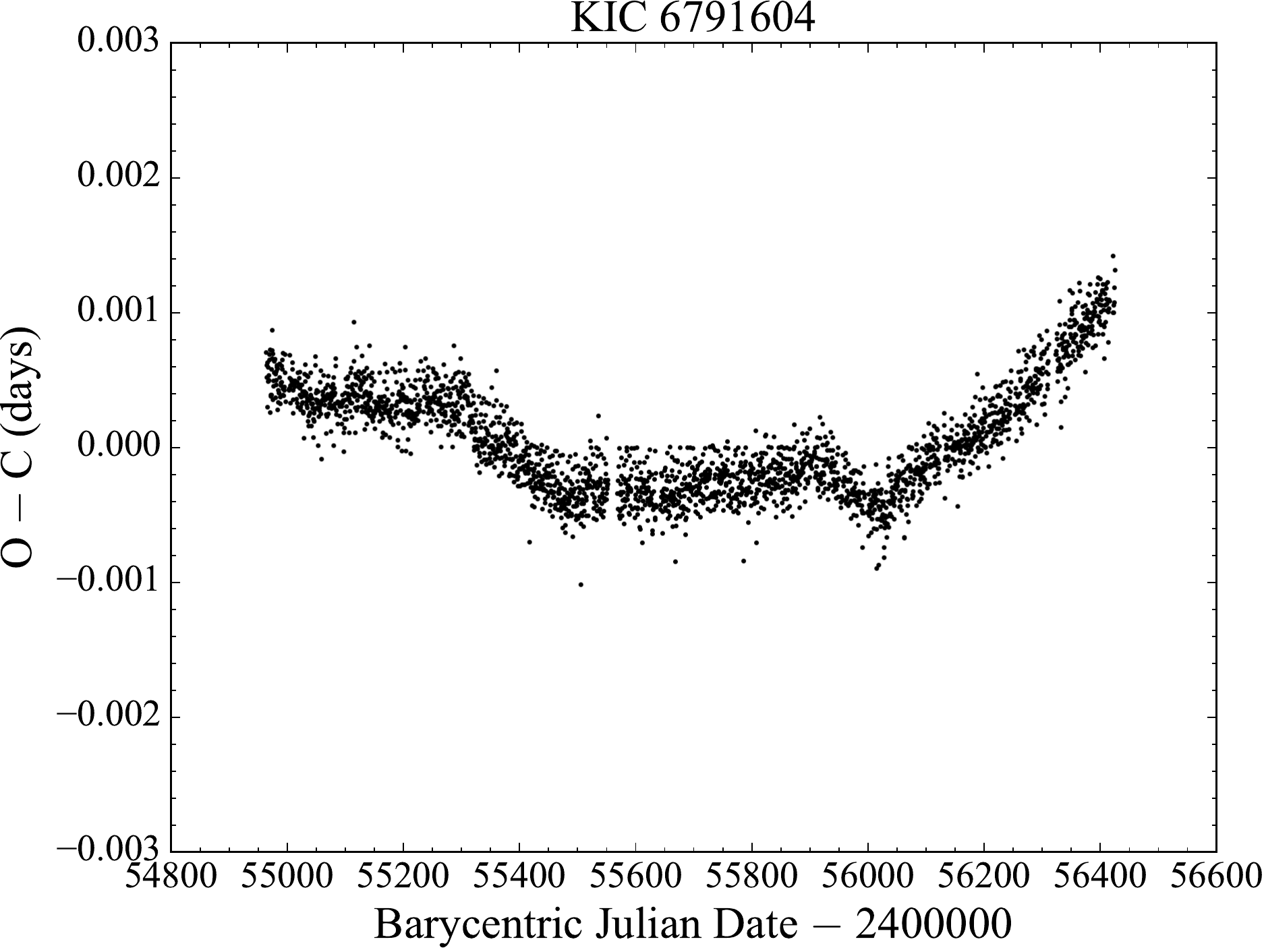}\unskip
\includegraphics[width=\columnwidth]{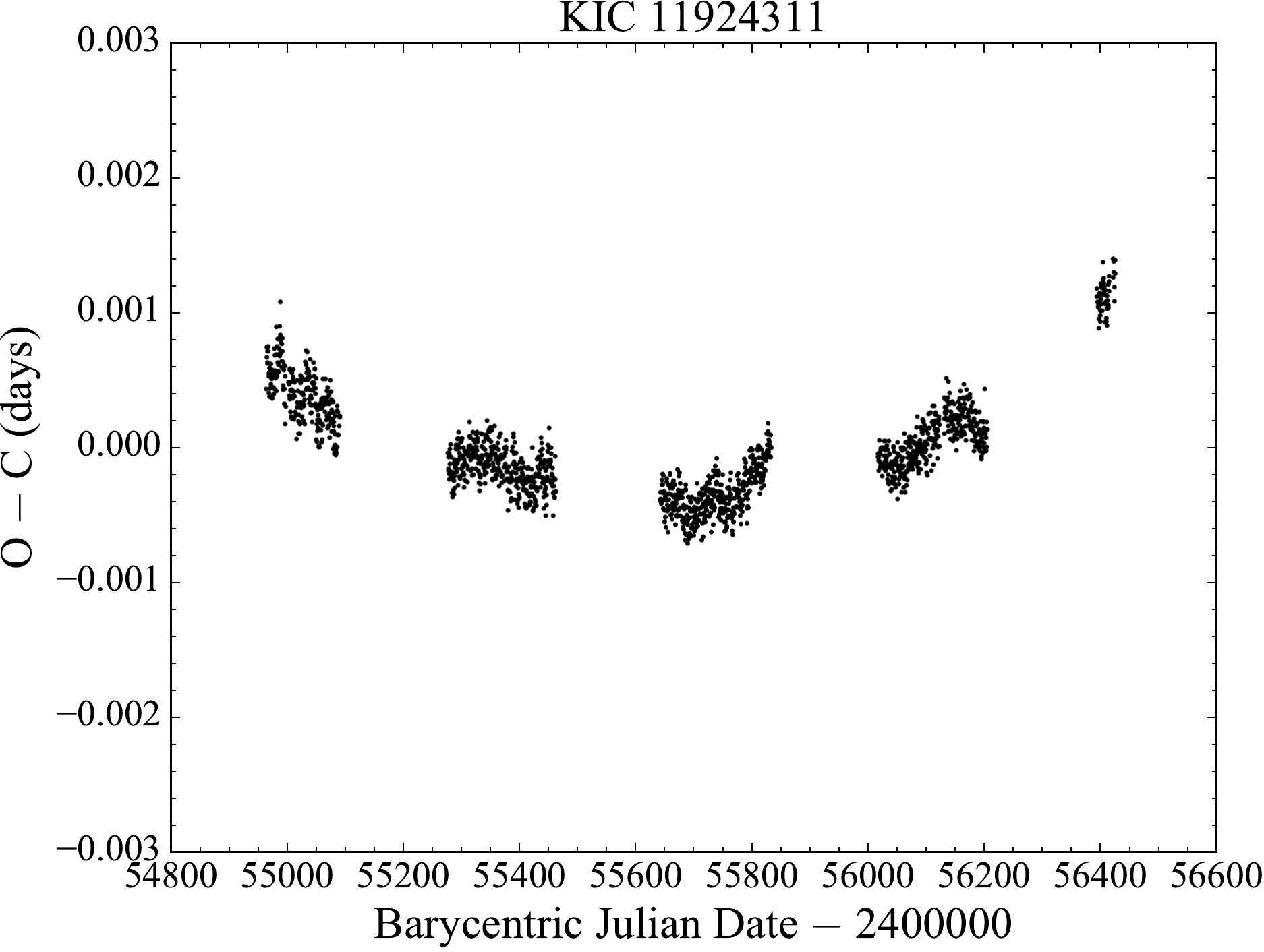}
\par\end{centering}
\caption{Eclipse timing variation of KIC~6791604 (left) and KIC~11924311 (right). KIC~6791604 was flagged with an interesting ETV while KIC~11924311 was flagged as a sinusoidal ETV, but both are consistent with a roughly parabolic ETV as well.
\label{fig:KIC-6791604-&-11924311-ETV}}
\end{figure*}

The twenty-two core sample systems \citet{Kouzuma2018} cited as undergoing mass transfer do not include the three systems \citet{Conroy2014} flagged with a parabolic ETV\@. However, the KEBC flagged sixteen of these twenty-two systems with a sinusoidal or interesting ETV, and visual inspection of their ETVs suggests that, aside from KIC~2437038, their signals are roughly parabolic. Figure~\ref{fig:KIC-6791604-&-11924311-ETV} shows the ETV of two systems \citet{Kouzuma2018} references: KIC~6791604 (flagged as interesting) and KIC~11924311 (flagged as sinusoidal). To further investigate our core sample's ETVs, we applied a digital Butterworth filter \citep{Butterworth1930} of order one to the ETV data using SciPy's \texttt{butter} function to reduce noise from effects such as starspots. The sampling frequency is the orbital frequency, and we set the cutoff frequency to five cycles over the system's observation span, thus removing signals such as starspot modulation that recur more than five times during \emph{Kepler}'s observations. Due to the need for continuous sampling, we removed any system with an observation gap longer than 20~d, leaving a sample of 122 core sample systems. We fit a parabola to each system's filtered data using SciPy's \texttt{curve\_fit} function and calculated the $R^2$ value of the fit. Thirteen systems (11\% of the 122 investigated systems) have $R^2 \geq 0.9$, while 25 systems (20\%) have a less-stringent $R^2 \geq 0.8$ (for reference, KIC~6791604's ETV -- shown in Figure~\ref{fig:KIC-6791604-&-11924311-ETV}'s left panel -- has $R^2 = 0.895$). We regard these 25 systems as having an ETV that is not incompatible with being parabolic. Because these 122 systems form an unbiased subset of our core sample, we expect the same percentage of core sample systems to show a possibly parabolic ETV\@. Therefore, we estimate that up to 43 core sample systems show evidence of mass transfer in their ETV, making mass transfer a plausible, if rare, cause of the O'Connell effect.

\subsection{O'Connell Effect Ratio and Light Curve Asymmetry\label{subsec:OER-and-LCA-Discussion}}

\begin{figure}
\begin{centering}
\includegraphics[width=\columnwidth]{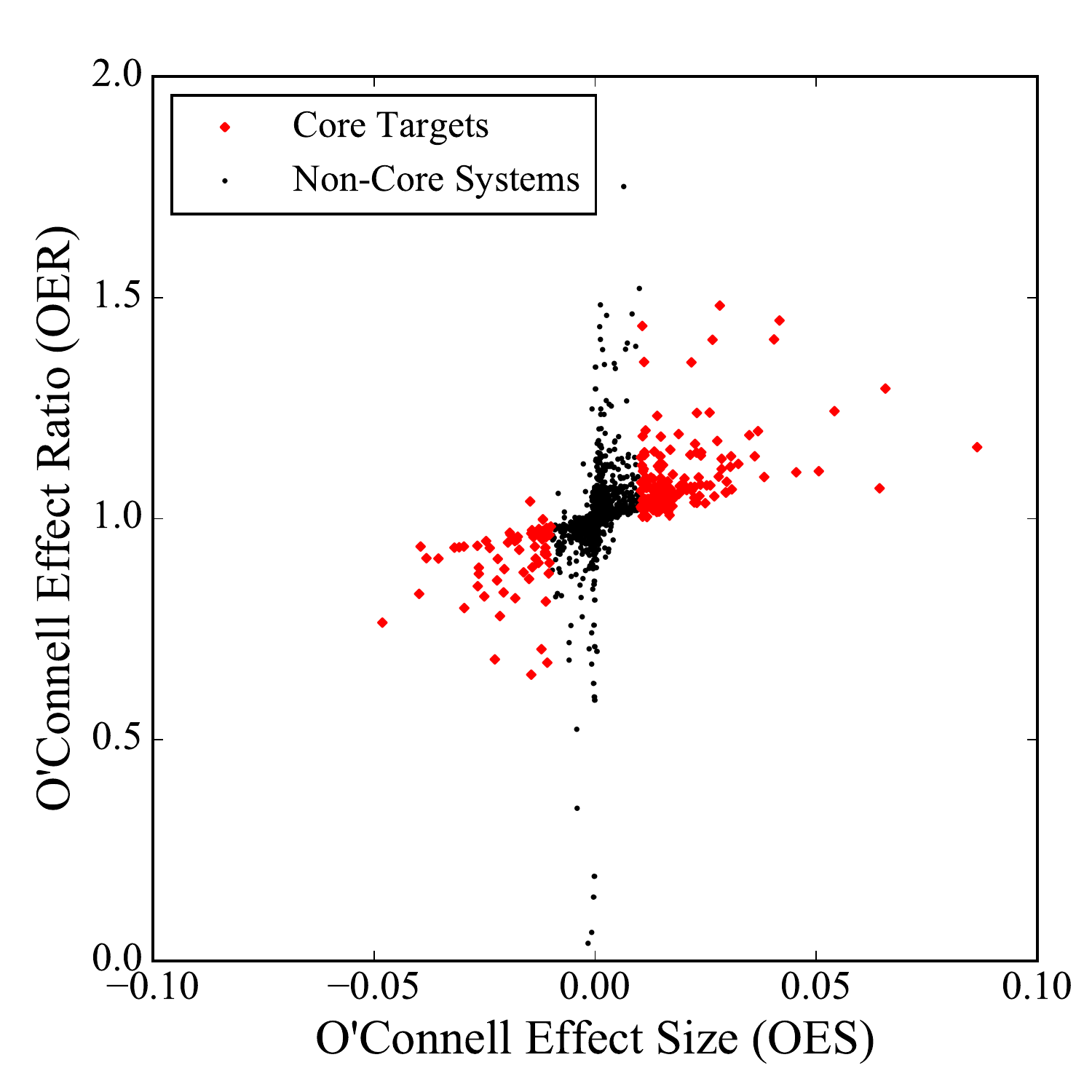}
\par\end{centering}
\caption{Plot comparing the OER to the OES\@. Core sample targets are shown in red diamonds and non-core sample systems in black circles. KICs~9935311 and 11347875 are excluded for clarity.}
\label{fig:OER-Plot}
\end{figure}

\begin{figure}
\begin{centering}
\includegraphics[width=\columnwidth]{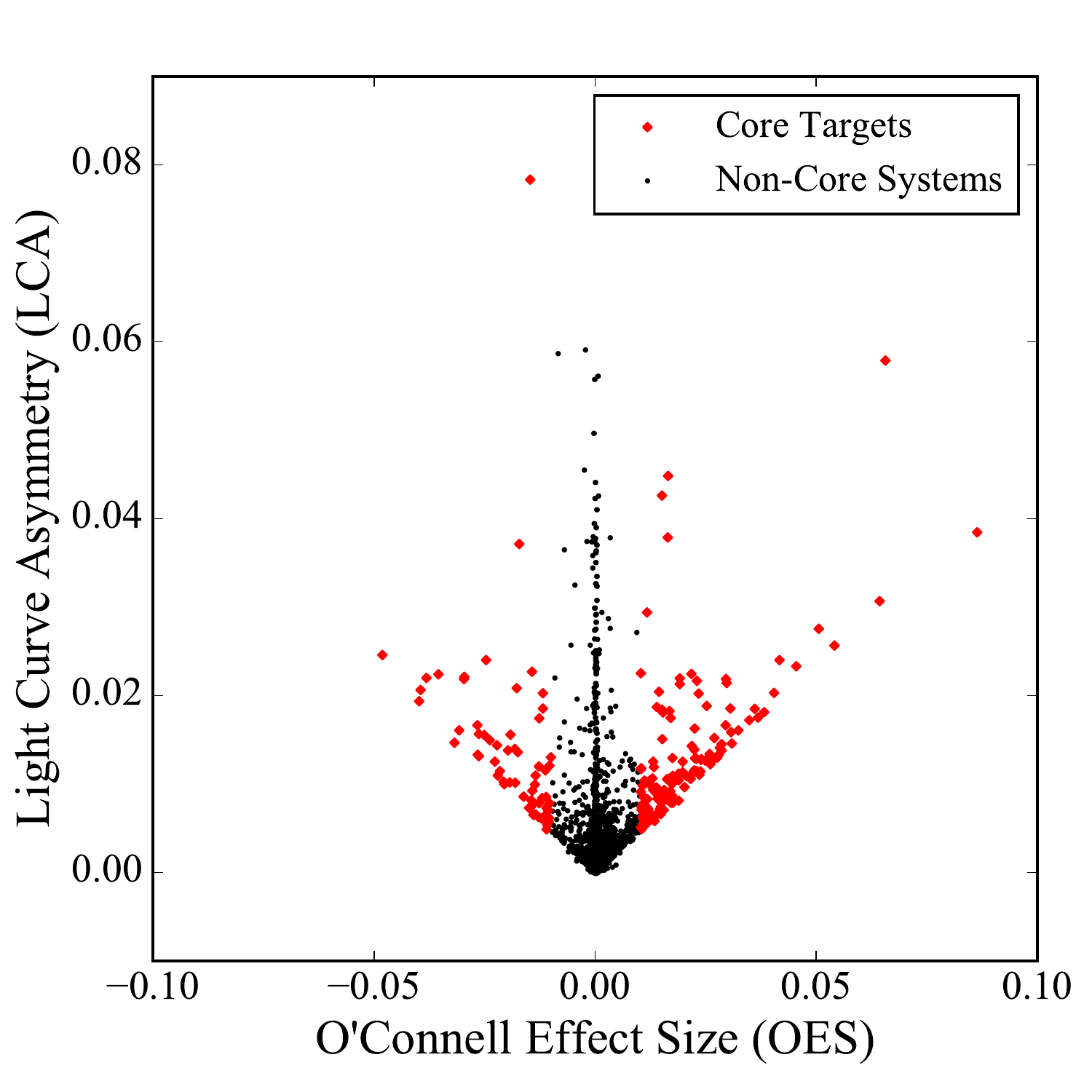}
\par\end{centering}
\caption{Plot comparing the LCA to the OES\@. Core sample targets are shown in red diamonds and non-core sample systems in black circles. KICs~9935311 and 11347875 are excluded for clarity.}
\label{fig:LCA-Plot}
\end{figure}

Figures~\ref{fig:OER-Plot} and \ref{fig:LCA-Plot} compare the OES to the OER (Equation~\ref{eq:OER}) and the LCA (Equation~\ref{eq:LCA}), respectively. The data labeled ``Non-Core Systems'' include all KEBC systems not in the core sample, excluding KIC~5217781. Our analysis found many systems with an OES of $\sim$0 but OER $\neq 1$ or LCA $\neq 0$, producing significant vertical scatter in Figures~\ref{fig:OER-Plot} and \ref{fig:LCA-Plot}. Systems in this scatter region of Figure~\ref{fig:OER-Plot} are low-amplitude (total change in flux $\Delta F \lesssim 0.03$) binaries with an OES that is large relative to their eclipse depth but small on an absolute scale. In Figure~\ref{fig:LCA-Plot}, the scatter region also includes long-period eccentric binaries. We removed a single data point from KICs~9701423 and 10614158 to avoid a nonphysical negative OER, although the low-amplitude system KIC~6948480 retains a negative OER\@. We excluded the two systems with $|$OES$| > 0.1$ (KICs~9935311 and 11347875) from Figures~\ref{fig:OER-Plot} and \ref{fig:LCA-Plot} for clarity. The figures do not plot errors because they are too small to see at this scale for most systems.

Figure~\ref{fig:OER-Plot} shows a strong correlation between the OER and OES\@. This correlation is unsurprising because both measures account for the different amounts of light under both maxima, albeit in different ways. \citet{McCartney1999} notes that the presence of a constant, third source of light contaminating the data will reduce the OES. However, the OER is not affected by third light due to subtracting the minimum flux value of the light curve. Therefore, the OER provides a more consistent measure of the O'Connell effect in this respect. On the other hand, the OER calculation method also means that the OER will be significant for systems like KIC~8912911 and KIC~10905824 with O'Connell effects that are large relative to their eclipse depth but small on an absolute scale. Based on this, we believe that the OER is ill-suited to describe the O'Connell effect in a general population of eclipsing binaries but is applicable to a subset of binaries with amplitudes larger than $\sim$0.01 in normalized flux.

Figure~\ref{fig:LCA-Plot} shows that most of our sample lies along lines with slopes of $\pm$0.5 that meet at the origin. Many systems lie above these lines, but none lie significantly below them, implying that the O'Connell effect produces an LCA at least half as large as the OES\@. The LCA is also clearly sensitive to asymmetries aside from the O'Connell effect, as seen by the vertical scatter mentioned earlier. Many systems in this scatter (like KIC~10909274) are long-period systems with eccentric orbits, implying that the LCA is sensitive to unevenly-spaced eclipses. Therefore, the LCA serves as a good measure of a light curve's asymmetry, but its sensitivity to other asymmetries means that it should not be relied on to detect the O'Connell effect.

\subsection{Systems Showing Peculiar Features\label{subsec:Notable-Systems}}

Four system classes showing peculiar features are present in our sample: those displaying significant temporal variation, those with an asymmetric minimum, those with a concave-up region in their light curves, and a white dwarf. Systems can belong to multiple classes. We have chosen an exemplar system to serve as an identifier for each class for the purpose of categorization. We discuss each class and identify characteristics systems in each class share. Future work will discuss each of these classes more in-depth. In Section~\ref{subsec:Notable-Systems}, our discussion also includes systems from the marginal sample defined in Section~\ref{subsec:Target-Sample}.

\subsubsection{KIC~7433513: Temporally Varying Systems}

\begin{figure}
\begin{centering}
\includegraphics[width=\columnwidth]{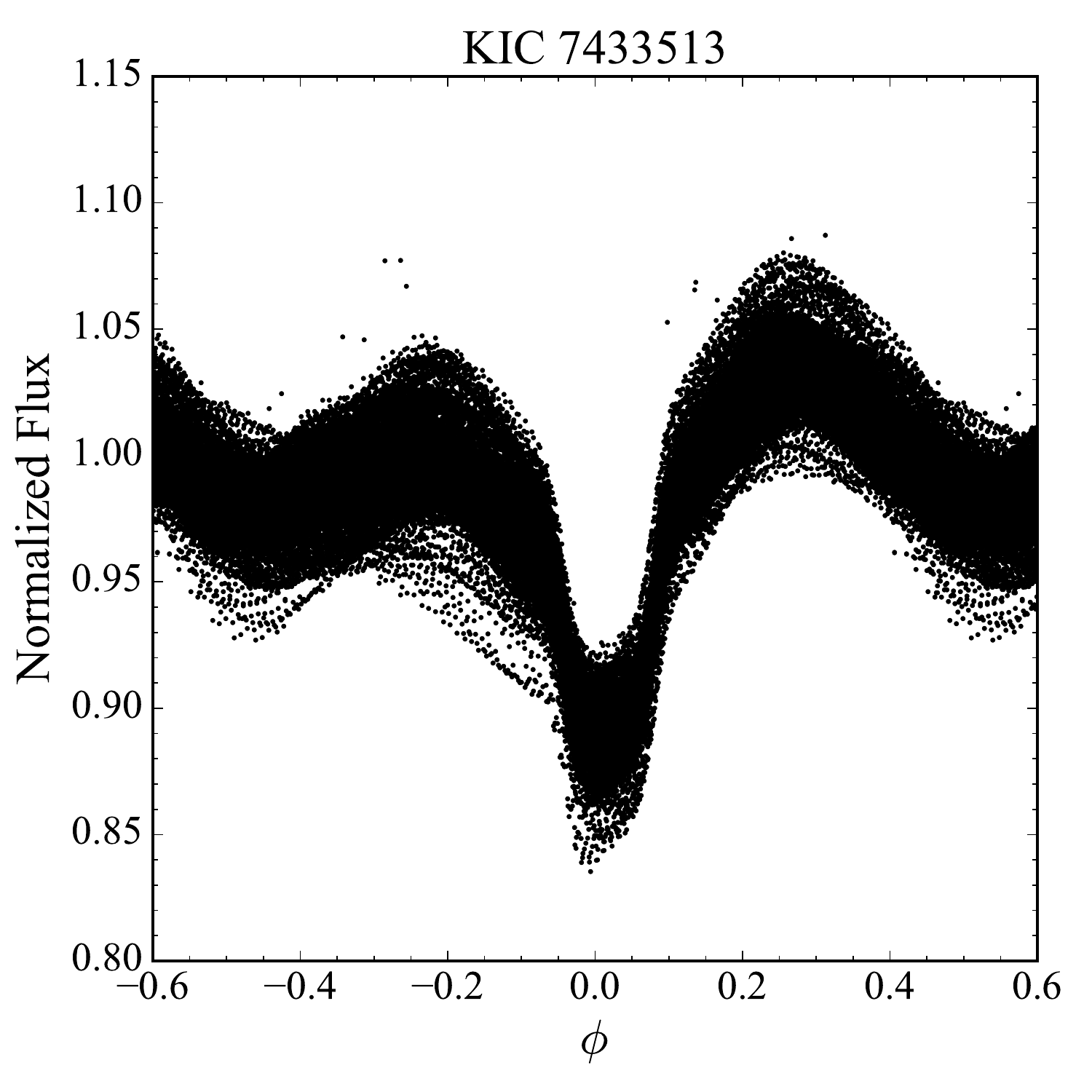}
\par\end{centering}
\caption{\emph{Kepler} light curve of KIC~7433513 showing significant scatter caused by temporal variation.
\label{fig:KIC-7433513-light-curve}}
\end{figure}

\begin{figure}
\begin{centering}
\includegraphics[width=\columnwidth]{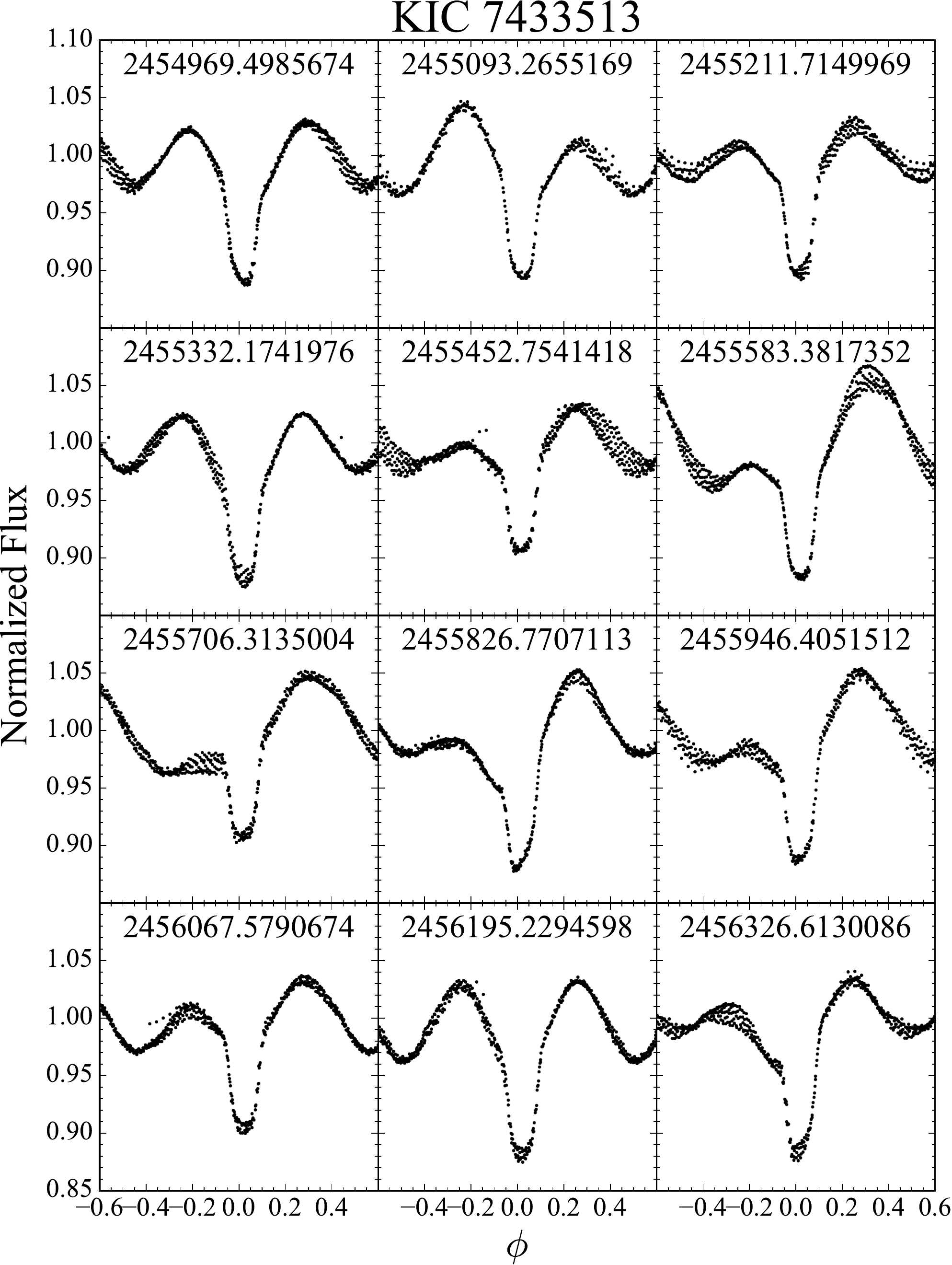}
\par\end{centering}
\caption{Ten-day slices of phased \emph{Kepler} data for KIC~7433513 separated by several months, showing the significant changes in the light curve over time. The number in each time slice's subplot is the BJD for the midpoint of that time slice.
\label{fig:KIC-7433513-multi-epoch-plot}}
\end{figure}

KIC~7433513 and systems like it (labeled with the flag TV in Table~\ref{tab:Target-List}) display strong temporal variation in their light curves. Figure~\ref{fig:KIC-7433513-light-curve} shows KIC~7433513's phased light curve using all \emph{Kepler} data, clearly showing the significant data scatter caused by the temporal variation. Figure~\ref{fig:KIC-7433513-multi-epoch-plot} shows 10-day time slices of KIC~7433513's \emph{Kepler} data separated by a few months, along with the Barycentric Julian Date (BJD) for the midpoint of each time slice. The system changes drastically even over these short timescales. It is difficult to rigorously quantify the number of these systems in our sample because there is a continuum between systems with stable light curves like KIC~5282464 (Figure~\ref{fig:O'Connell-effect-example}) and systems like KIC~7433513. However, we estimate that about 20\% of our core sample and about 65\% of our marginal sample exhibits strong temporal variation. Three further examples from our sample are KICs~2569494, 3659940, and 9137992. Systems displaying this temporal variation are among our sample's cooler systems, with the hottest (KIC~8294484) estimated at 5,760~K\@. \citet{Kunt2017} described KIC~7885570 -- a system from our marginal sample showing strong temporal variation -- as an RS Canum Venaticorum variable. The stars in these systems have active chromospheres that produce large spots, and the evolution of these spots is a plausible source for the observed temporal variation.

\subsubsection{KIC~9164694: Asymmetric Minima Systems\label{subsubsec:Asymmetric-Minima-Systems}}

\begin{figure*}
\begin{centering}
\includegraphics[width=\columnwidth]{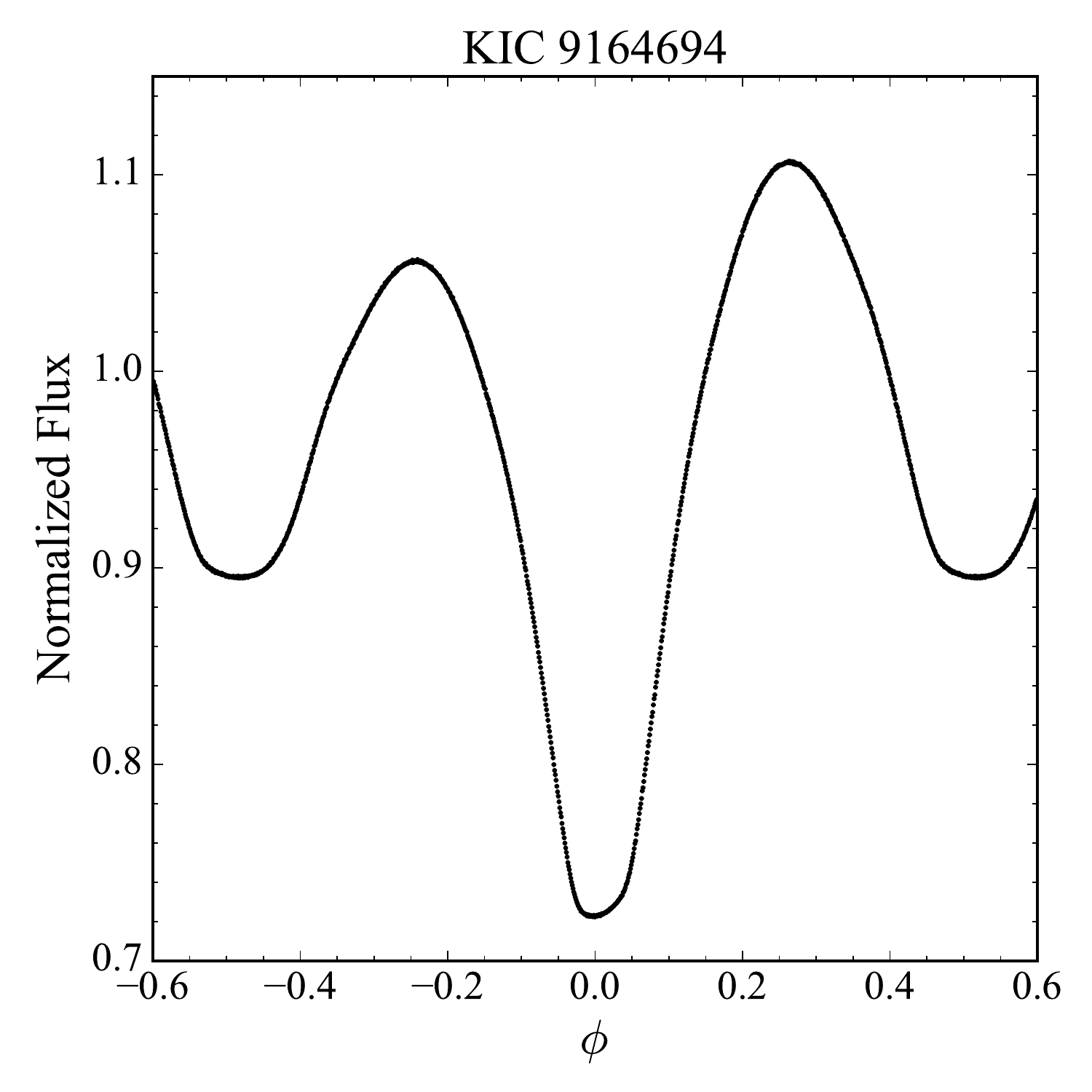}\unskip
\includegraphics[width=\columnwidth]{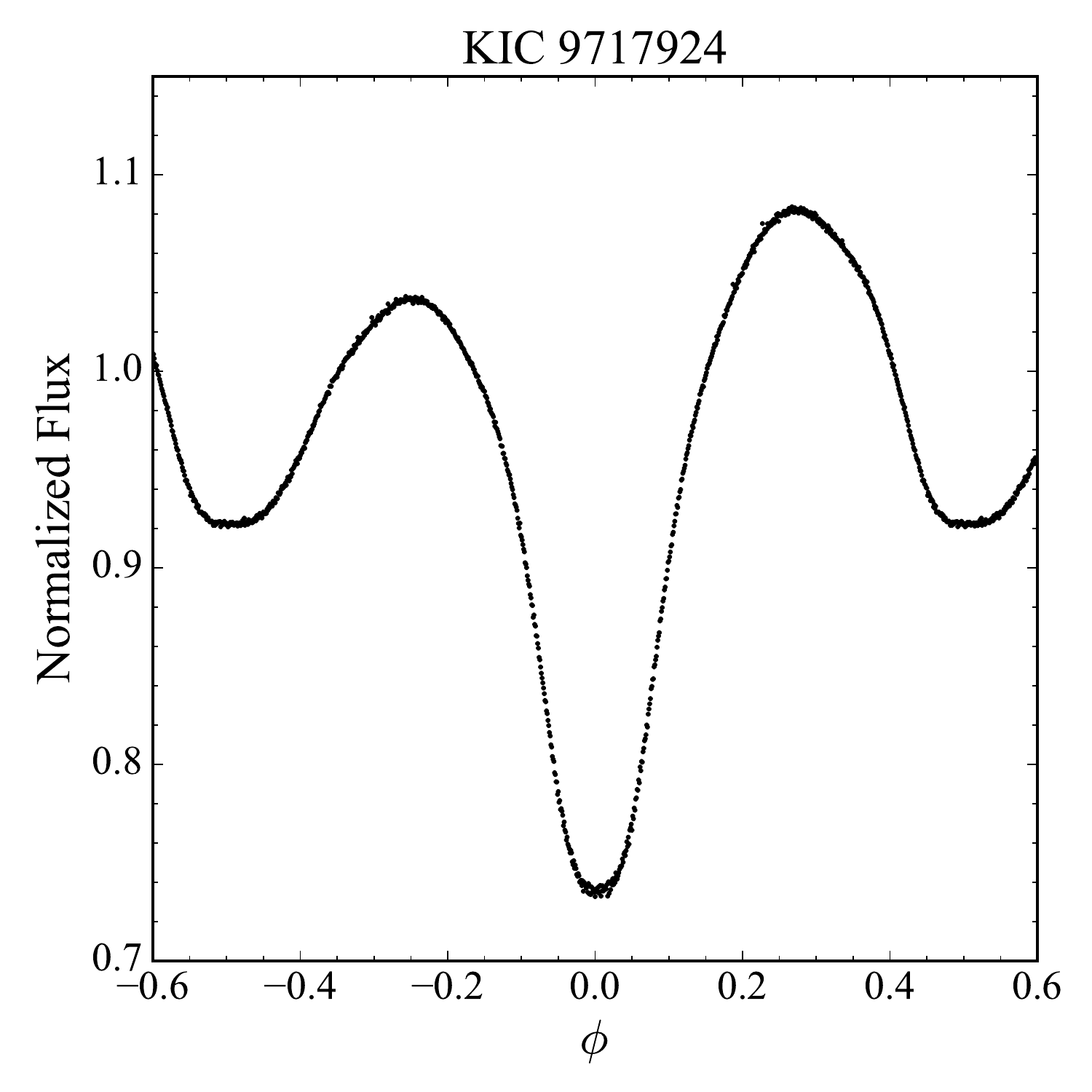}
\par\end{centering}
\caption{Averaged light curves of KIC~9164694 (left) and KIC~9717924 (right) showing an asymmetric primary and secondary minimum, respectively.
\label{fig:KIC-9164694-&-9717924-light-curves}}
\end{figure*}

KIC~9164694 and systems like it (labeled with the flag AM in Table~\ref{tab:Target-List}) contain an asymmetric minimum, as seen in Figure~\ref{fig:KIC-9164694-&-9717924-light-curves}'s left panel showing KIC~9164694's light curve. This asymmetry is rather subtle, and in KIC~9164694's case, is best described as looking like someone has taken their finger and ``pushed'' the light curve upward and leftward in the region just to the right of minimum light. Asymmetric minima are a rarely discussed feature seen in some eclipsing binary systems, including RY Scuti \citep{Djurasevic2008}, GR Tauri \citep{Zhang2002}, AG Virginis \citep{Pribulla2011}, and NSVS~7322420 \citep{Knote2019}. Asymmetries can appear in the primary minimum (as in KIC~9164694) or the secondary minimum (as in KIC~9717924; see Figure~\ref{fig:KIC-9164694-&-9717924-light-curves}'s right panel). About 13\% of our sample displays an asymmetric minimum in either eclipse. There appear to be two sets of asymmetric minima systems: those showing strong temporal variation (like KIC~7433513), and those showing comparatively little (like KICs~9164694 and 9717924). We focus on the latter set for this class, which consists of the marginal sample system KIC~5283839 and 15 core sample systems: KICs~2159783, 2449084, 6205460, 8248967 (according to a 20-term Fourier series), 8696327, 8822555, 8842170, 9164694, 9283826, 9717924, 9786165 (with short-cadence data only), 10528299, 10861842, 11395645, and 11924311. These 16 are among our sample's hotter systems, with nearly half having a temperature over 6,000~K, and most show minimal temporal variation. We distinguish this asymmetry from what older literature like \citet{Brownlee1957} and \citet{van'tVeer} call an asymmetric minimum, as the latter asymmetry occurs closer to the eclipse's beginning and end. By contrast, what we call an asymmetric minimum occurs around the eclipse's center.

The most striking similarity between these systems is that all known asymmetric minima systems exhibit total eclipses. Total eclipses occur when one star is fully occluded during an eclipse and are distinguished by flat-bottomed or nearly flat-bottomed minima. It is unclear why asymmetric minima are only found in totally eclipsing systems and not in partially eclipsing systems. One hypothesis is that spots near one component's poles cause the asymmetry. Another is that the change in flux causing the asymmetry is small enough that it is only detectable during totality, when the flux is nearly constant. We describe these two possibilities in the next two paragraphs. Some systems with total eclipses and a significant O'Connell effect do not show an asymmetric minimum, such as KIC~8386048. Additionally, some totally eclipsing systems with no O'Connell effect show an asymmetric minimum, such as the non-sample system KIC~8265951.

One difference between partially and totally eclipsing binaries is that parts of both stars are always visible for partially eclipsing systems. These regions of persistent visibility occur near each star's visible pole. By contrast, one star is fully occluded in totally eclipsing systems, making it impossible for regions on that star to be persistently visible. Therefore, features near the poles (like spots) can be occluded in totally eclipsing systems but not partially eclipsing systems. BinaryMaker3 (BM3) tests showed that both polar and equatorial spots could produce an asymmetric minimum, however.

\begin{figure}
\begin{centering}
\includegraphics[width=\columnwidth]{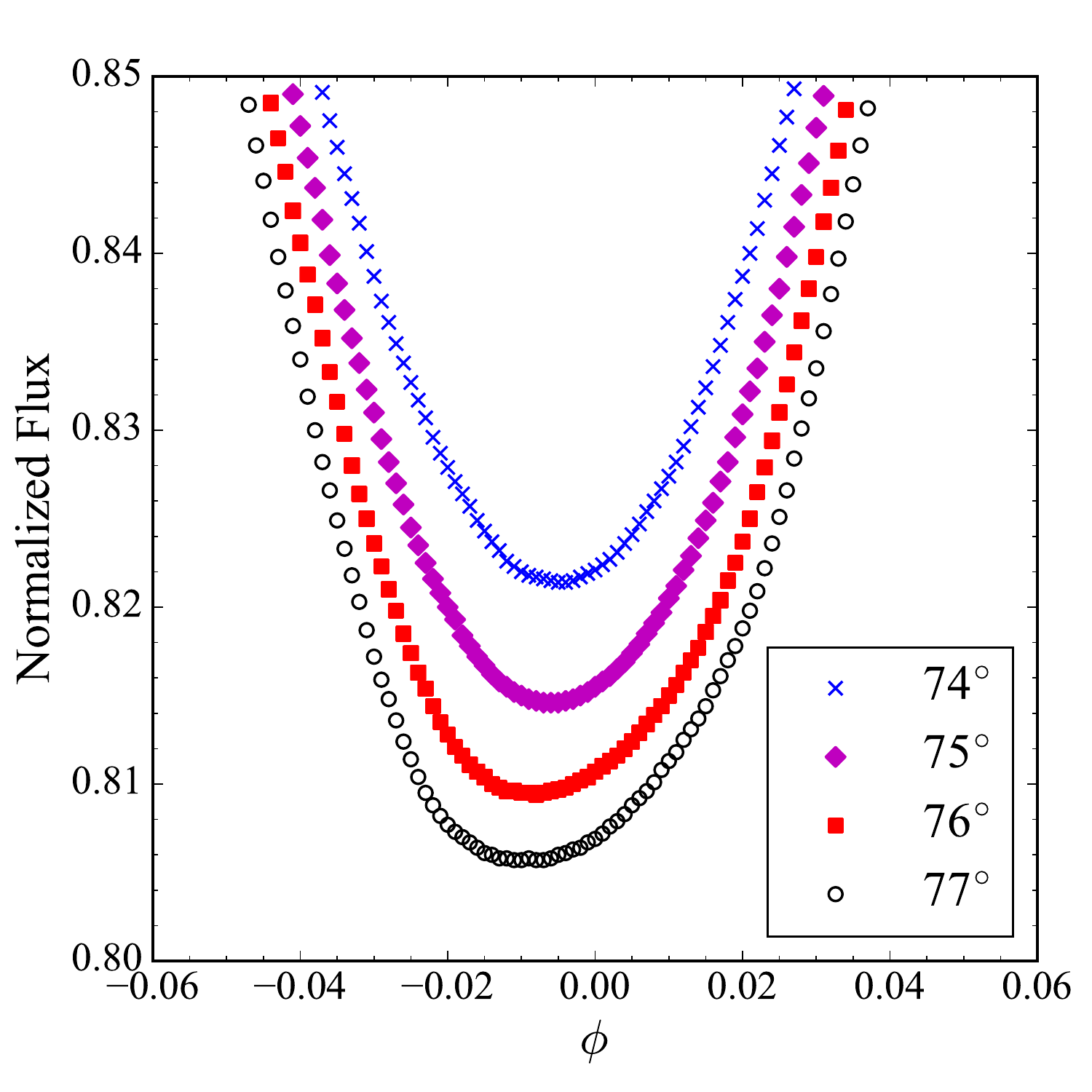}
\par\end{centering}
\caption{Light curves of four eclipsing binary models detailing the area around the primary minimum. Parameters for all four models are identical except for inclination. The asymmetry grows significantly more pronounced as the system changes from partially eclipsing ($i = 74,~75^{\circ}$) to totally eclipsing ($i = 76,~77^{\circ}$), demonstrating the asymmetry's strong dependence on total eclipses.
\label{fig:Asymmetric-Minimum-Example}}
\end{figure}

Our BM3 testing also indicated that the effect is strongly dependent on the presence of total eclipses. Figure~\ref{fig:Asymmetric-Minimum-Example} shows that a one-degree change in inclination is enough to transform a largely symmetrical eclipse into a significantly asymmetric one. Therefore, another hypothesis is that total eclipses are a prerequisite for a significant minimum asymmetry. We posit that a feature on one component, which we will call a hot starspot or spot, causes the asymmetry via its changing aspect. The spot must be azimuthally offset from the plane both 1.)\ perpendicular to the orbital plane and 2.)\ containing both stellar centers to produce an asymmetric minimum. The change in flux caused by the spot's changing aspect is small compared to the change caused by the eclipsed star being covered (or uncovered) during the eclipse's partial phase. Therefore, the large change in flux during the partial phase prevents the spot from creating a significant asymmetry. During the almost flux constant total phase, however, the small change in flux caused by the spot's changing aspect has no other change in flux to compete with and can produce a significant asymmetry. Therefore, under our starspot model, asymmetric minima imply total eclipses.

\newpage
\subsubsection{KIC~10544976: The White Dwarf}

\begin{figure*}
\begin{centering}
\includegraphics[width=\columnwidth]{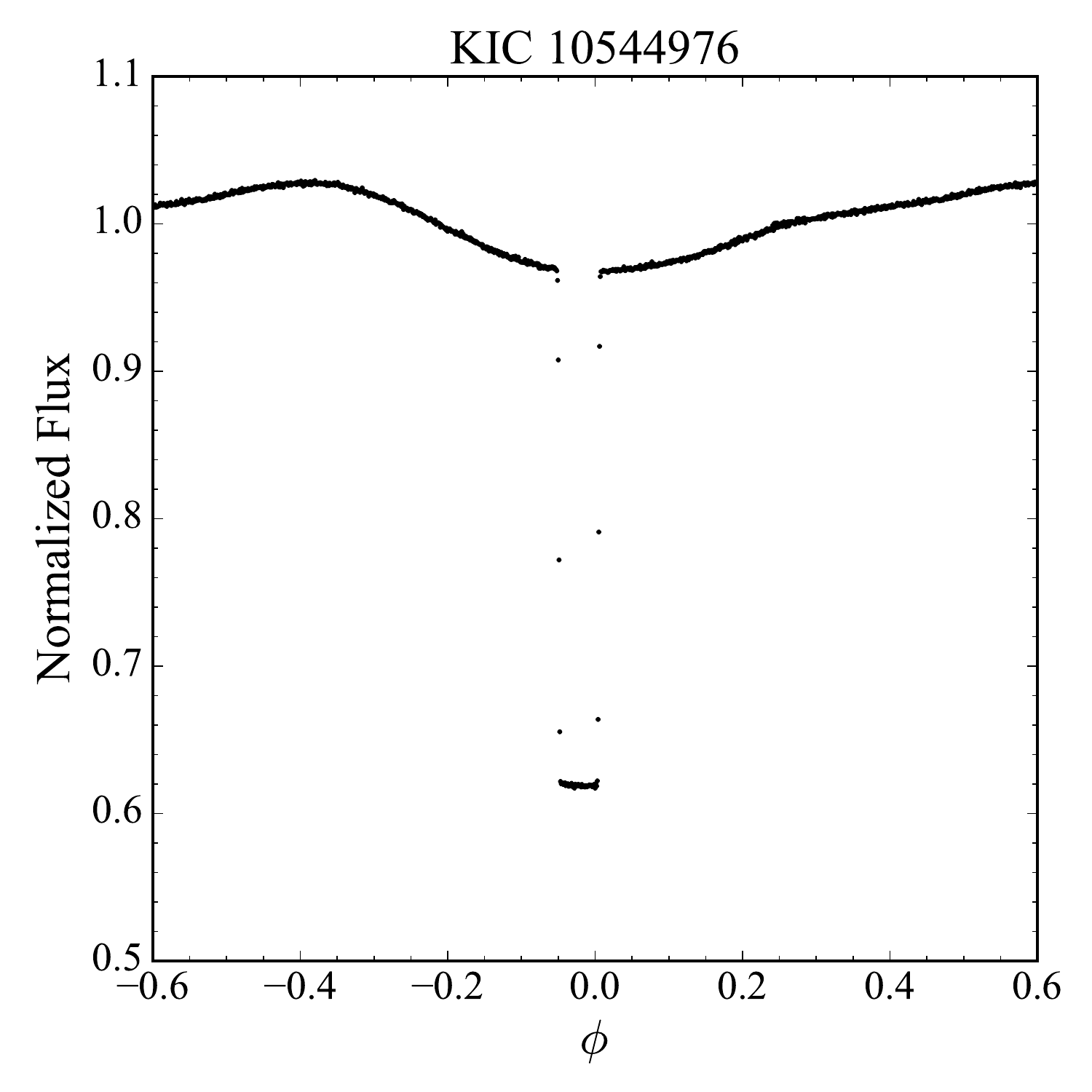}\unskip
\includegraphics[width=\columnwidth]{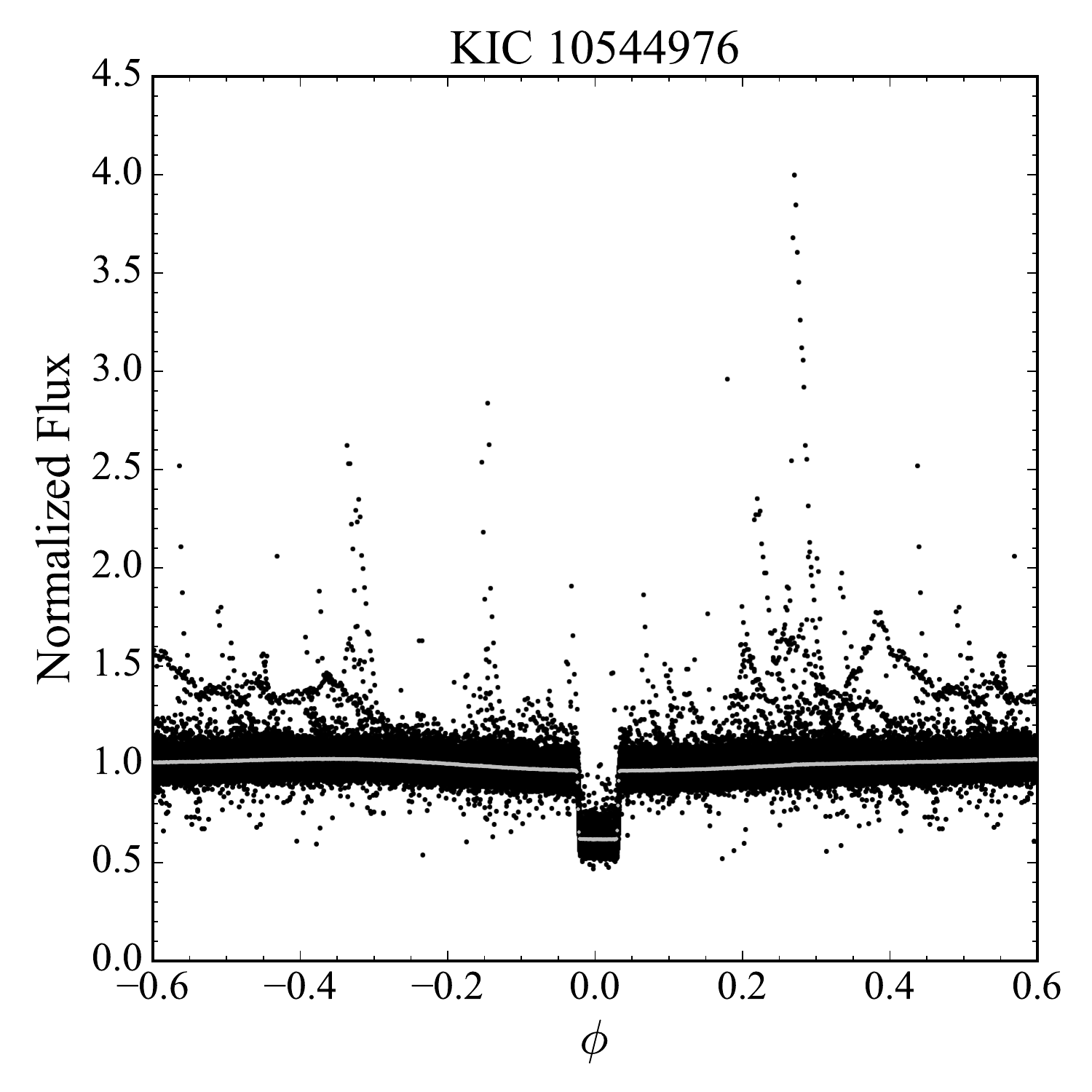}
\par\end{centering}
\caption{Averaged (left) and \emph{Kepler} (right) light curves of KIC~10544976 using short-cadence data showing important features, including the sharp primary eclipse and numerous flares. The right panel also plots the averaged light curve in grey for comparison.
\label{fig:KIC-10544976-light-curve}}
\end{figure*}

KIC~10544976 (labeled with the flag WD in Table~\ref{tab:Target-List}) is unique within our sample as it is the only system containing a degenerate component. The lack of white dwarfs in our sample is unsurprising because white dwarfs were not priority targets for \emph{Kepler}. \citet{Almenara2012} identifies the system as consisting of a DA white dwarf primary and an M4~V secondary in an 8.4-hour orbit. The left panel of Figure~\ref{fig:KIC-10544976-light-curve} displays KIC~10544976's short-cadence average light curve, showing that the system has a very sharp, well-defined primary eclipse and no secondary eclipse. The red dwarf companion produces many flares, as Figure~\ref{fig:KIC-10544976-light-curve}'s right panel shows. We consider this system to have a non-traditional O'Connell effect because the observed flux increases monotonically after the primary eclipse until phase $-0.35$. Therefore, the light curve lacks two inter-eclipse maxima to measure a difference between. KIC~10544976 has an OES of $-0.0096$, placing it within our marginal sample near the core sample cutoff.

\subsubsection{KIC~11347875: Concave-Up Systems\label{subsubsec:Concave-Up-Systems}}

\begin{figure}
\begin{centering}
\includegraphics[width=\columnwidth]{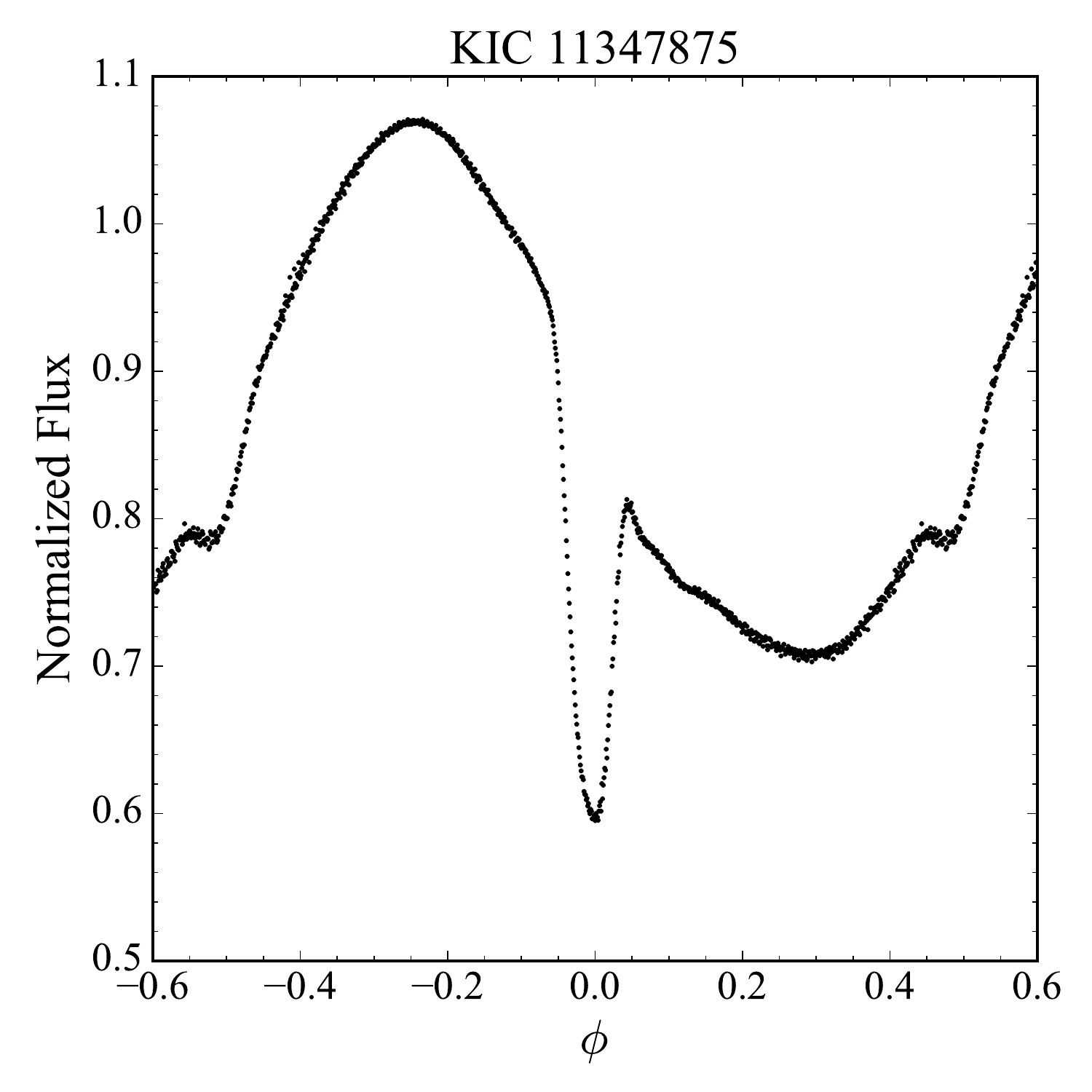}
\par\end{centering}
\caption{Averaged light curve of KIC~11347875 the system's concave-up region following the primary eclipse.
\label{fig:KIC-11347875-light-curve}}
\end{figure}

KIC~11347875 has the largest measured O'Connell effect in our sample, but its light curve has an unusual appearance shared with six systems in the sample (labeled with the flag CU in Table~\ref{tab:Target-List}). Figure~\ref{fig:KIC-11347875-light-curve} shows that KIC~11347875's light curve is concave-up after the primary minimum, giving the system the appearance of an eclipsing signal superimposed on a rough sinusoid. Figure~\ref{fig:KIC-11347875-Fourier-comparison} shows an enlarged view of KIC~11347875's concave-up region. As with KIC~10544976, the lack of inter-eclipse maxima indicates that the system has a non-traditional O'Connell effect. \citet{Gao2016} identifies KIC~11347875 as containing two late-type red giants based on effective temperature and surface gravity estimates, although our luminosity estimate from \emph{Gaia} EDR3 data suggests a total luminosity of only 1.306~L\textsubscript{$\odot$}. Six systems (KICs~5300878, 6044064, 6197038, 6697716, 7671594, and 9119652) show a similar concave-up region, as do some temporally varying systems like KIC~8479107 during some time intervals. However, KIC~11347875 is by far the most extreme example. All systems exhibiting a concave-up region are quite cool ($T < 5,\!100$~K), have a longer orbital period ($P > 1.2$~d), and show flares.

\section{Analysis \& Discussion\label{sec:Analysis-and-Discussion}}

We now present a graphical representation of the correlations between the characteristics we studied (except distance and luminosity). We further discuss several correlations and their implications on the study of the O'Connell effect, with a focus on correlations involving OES, eclipse depth, and the morphology parameter.

\subsection{Characteristic Trends\label{subsec:Characteristic-Trends}}

Figures~\ref{fig:Corner-Plot-Part-One}, \ref{fig:Corner-Plot-Part-Two}, and \ref{fig:Corner-Plot-Part-Three} show corner plots comparing period, OES, \emph{Gaia} temperature, \emph{Gaia} BP -- RP color index, morphology parameter, absolute \emph{Gaia} G magnitude, and primary eclipse depth. We chose the plot limits for clarity and, as such, the OES plots exclude KICs~9777984, 9935311, and 11347875. The banding seen in the temperature plots (most prominent in Figure~\ref{fig:Corner-Plot-Part-Three}'s right column) is a known issue with \emph{Gaia} DR2 data \citep{Andrae2018}. The banding seen in the morphology plots (most prominent in the KEBC data of Figure~\ref{fig:Corner-Plot-Part-Three}'s top-center panel) is a quantization effect due to the morphology parameter being given to only two decimal places.

Figures~\ref{fig:Corner-Plot-Part-One}-\ref{fig:Period-vs-OES} used the trend subset of the KEBC discussed in Section~\ref{subsec:KEBC}, which excluded systems for two reasons: to analyze our sample without raising Python errors and not have results be affected by invalid parameter values. We excluded a given system because either it had no \emph{Gaia} parallax data (43 systems, plus 1 system without a known \emph{Gaia} EDR3 identifier), multiple entries in the KEBC (11 systems, 24 total entries), or had $\mu = -1$, indicating that the KEBC team could not assign a valid morphology parameter to the system (174 systems). We also explicitly excluded KICs~5217781, 7667885, 7950964, and 9137819, as in Section~\ref{sec:Results}. These removals reduced the KEBC's size to 2,678 systems. While we removed these 228 systems (241 entries) from our KEBC analysis in Section~\ref{subsec:Statistical-Analysis-Discussion}, we did not remove any from our core sample analysis.

\subsection{Statistical Analysis\label{subsec:Statistical-Analysis-Discussion}}

Table~\ref{tab:K-S-Test-Results} gives the results of the K--S test described in Section~\ref{subsec:Statistical-Analysis}. The K--S statistic describes the degree of difference between the populations the two samples were drawn from. The \emph{p}-value indicates how compatible the K--S test results are with the null hypothesis that the two samples were drawn from the same population. We expected the significantly different OES distributions as we are looking at the wings of the distribution shown in Figure~\ref{fig:O'Connell-size-histogram}. This difference is intended by design as it was how we defined our sample. We also expected the distance distribution similarity, reinforced by Figure~\ref{fig:Distance-Histogram}, because a phenomenon that should not depend on distance differentiates the samples. There is no reason to expect spatial differences between the two samples, and any biases in spatial distribution are due to biases in the \emph{Kepler} sample. The remaining characteristics in Table~\ref{tab:K-S-Test-Results} are of greater interest as the results point to possible underlying connections between those characteristics and the O'Connell effect.

\setcounter{table}{1}
\begin{deluxetable}{lcD}
\tablecaption{Kolomgorov--Smirnov Test Results}
\label{tab:K-S-Test-Results}
\tablehead{
\colhead{Characteristic} & \colhead{K--S Statistic} & \multicolumn{2}{r}{\emph{p}-Value}}
\decimals
\startdata
O'Connell Effect Size       & 0.604 & <0.001 \\
$|$O'Connell Effect Size$|$ & 0.871 & <0.001 \\
Primary Eclipse Depth       & 0.398 & <0.001 \\
Morphology Parameter        & 0.242 & <0.001 \\
Temperature                 & 0.277 & <0.001 \\
Distance                    & 0.092 &  0.080 \\
Absolute Magnitude          & 0.285 & <0.001 \\
Period                      & 0.274 & <0.001 \\
\enddata
\end{deluxetable}
\vspace{-12pt}

Table~\ref{tab:Spearman-Test-Results} gives the results of our correlation analysis described in Section~\ref{subsec:Statistical-Analysis}. Kendall's $\tau$ coefficients are uniformly smaller than Spearman's $\rho$ coefficients and indicate the same correlations, so Table~\ref{tab:Spearman-Test-Results} presents only Spearman's $\rho$ coefficients for the sake of brevity. The coefficient describes the correlation's strength, while the \emph{p}-value indicates how compatible the results of the Spearman test are with the null hypothesis that the characteristics are uncorrelated. We consider two characteristics correlated if $|\rho| \geq 0.1$ -- consistent with \citetalias{Davidge1984} -- and strongly correlated if $|\rho| \geq 0.2$. Characteristic pairs have their values bolded whenever they are correlated for a given sample. Table~\ref{tab:Spearman-Test-Results} also lists how many of the 20 random subsets produced by our bootstrapping procedure (described in Section~\ref{subsec:Statistical-Analysis}) showed a correlation for each characteristic pair. Several correlations, such as the one between distance and absolute magnitude, are expected and serve as validations of our correlation analysis. We do not discuss such expected correlations despite their robustness as they are present for reasons unrelated to the O'Connell effect.

\begin{deluxetable*}{llDDCDDC}
\tablecaption{Spearman's $\rho$ Test Results}
\label{tab:Spearman-Test-Results}
\tablehead{
\colhead{Characteristic} & \colhead{Characteristic} & \multicolumn{5}{c}{~~~Core Sample\tablenotemark{a}} & \multicolumn{5}{c}{~~KEBC\tablenotemark{a}}\\
\colhead{One} & \colhead{Two} & \multicolumn{2}{r}{Coeff.} & \multicolumn{2}{r}{\emph{p}-Value} & \colhead{Surv.\tablenotemark{b}} &  \multicolumn{2}{r}{Coeff.} & \multicolumn{2}{r}{\emph{p}-Value} & \colhead{Surv.\tablenotemark{b}}}
\decimals
\startdata
\multicolumn{12}{c}{Informative O'Connell Effect Results}\\
\hline
OES           & Temperature   &  \mathbf{0}.\mathbf{385} & \mathbf{<0}.\mathbf{001} & \mathbf{20} &                    0.020 &                    0.423 &           0 \\
OES           & Morphology    & \mathbf{-0}.\mathbf{148} &  \mathbf{0}.\mathbf{032} & \mathbf{16} &                    0.024 &                    0.326 &           0 \\
OES           & Absolute Mag. & \mathbf{-0}.\mathbf{286} & \mathbf{<0}.\mathbf{001} & \mathbf{19} &                    0.033 &                    0.183 &           2 \\
OES           & Distance      &  \mathbf{0}.\mathbf{218} &  \mathbf{0}.\mathbf{001} & \mathbf{17} &                   -0.012 &                    0.641 &           0 \\
OES           & Period        &  \mathbf{0}.\mathbf{212} &  \mathbf{0}.\mathbf{002} & \mathbf{17} &                   -0.069 &                    0.005 &           4 \\
Eclipse Depth & Morphology    & \mathbf{-0}.\mathbf{489} & \mathbf{<0}.\mathbf{001} & \mathbf{20} & \mathbf{-0}.\mathbf{158} & \mathbf{<0}.\mathbf{001} & \mathbf{18} \\
Absolute Mag. & Period        & \mathbf{-0}.\mathbf{609} & \mathbf{<0}.\mathbf{001} & \mathbf{20} & \mathbf{-0}.\mathbf{370} & \mathbf{<0}.\mathbf{001} & \mathbf{20} \\
\hline
\multicolumn{12}{c}{Other Results}\\
\hline
OES           & Eclipse Depth &                    0.033 &                    0.628 &          13 &  \mathbf{0}.\mathbf{124} & \mathbf{<0}.\mathbf{001} & \mathbf{12} \\
$|$OES$|$     & Eclipse Depth &  \mathbf{0}.\mathbf{136} &  \mathbf{0}.\mathbf{048} & \mathbf{10} &  \mathbf{0}.\mathbf{573} & \mathbf{<0}.\mathbf{001} & \mathbf{20} \\
$|$OES$|$     & Morphology    &                   -0.055 &                    0.428 &          10 &  \mathbf{0}.\mathbf{210} & \mathbf{<0}.\mathbf{001} & \mathbf{20} \\
$|$OES$|$     & Temperature   &                   -0.025 &                    0.726 &           9 & \mathbf{-0}.\mathbf{213} & \mathbf{<0}.\mathbf{001} & \mathbf{19} \\
$|$OES$|$     & Distance      &                    0.044 &                    0.524 &           9 &                    0.010 &                    0.687 &           0 \\
$|$OES$|$     & Absolute Mag. &                   -0.027 &                    0.696 &          11 &  \mathbf{0}.\mathbf{236} & \mathbf{<0}.\mathbf{001} & \mathbf{20} \\
$|$OES$|$     & Period        &                    0.046 &                    0.506 &           8 & \mathbf{-0}.\mathbf{366} & \mathbf{<0}.\mathbf{001} & \mathbf{20} \\
Eclipse Depth & Temperature   &                   -0.039 &                    0.575 &           9 &                   -0.091 &                   <0.001 &          10 \\
Eclipse Depth & Distance      &                    0.057 &                    0.405 &          10 &                    0.071 &                    0.004 &           6 \\
Eclipse Depth & Absolute Mag. &                   -0.004 &                    0.955 &          12 &  \mathbf{0}.\mathbf{110} & \mathbf{<0}.\mathbf{001} & \mathbf{13} \\
Eclipse Depth & Period        &                    0.054 &                    0.432 &          12 &                   -0.084 &                   <0.001 &          10 \\
Morphology    & Temperature   &  \mathbf{0}.\mathbf{113} &  \mathbf{0}.\mathbf{109} & \mathbf{12} &  \mathbf{0}.\mathbf{196} & \mathbf{<0}.\mathbf{001} & \mathbf{20} \\
Morphology    & Distance      & \mathbf{-0}.\mathbf{148} &  \mathbf{0}.\mathbf{032} & \mathbf{13} &                    0.025 &                    0.317 &           1 \\
Morphology    & Absolute Mag. &                    0.065 &                    0.350 &          12 & \mathbf{-0}.\mathbf{156} & \mathbf{<0}.\mathbf{001} & \mathbf{19} \\
Morphology    & Period        & \mathbf{-0}.\mathbf{591} & \mathbf{<0}.\mathbf{001} & \mathbf{20} & \mathbf{-0}.\mathbf{562} & \mathbf{<0}.\mathbf{001} & \mathbf{20} \\
Temperature   & Distance      &  \mathbf{0}.\mathbf{245} & \mathbf{<0}.\mathbf{001} & \mathbf{17} &  \mathbf{0}.\mathbf{139} & \mathbf{<0}.\mathbf{001} & \mathbf{19} \\
Temperature   & Absolute Mag. & \mathbf{-0}.\mathbf{705} & \mathbf{<0}.\mathbf{001} & \mathbf{20} & \mathbf{-0}.\mathbf{695} & \mathbf{<0}.\mathbf{001} & \mathbf{20} \\
Temperature   & Period        &  \mathbf{0}.\mathbf{248} & \mathbf{<0}.\mathbf{001} & \mathbf{16} &  \mathbf{0}.\mathbf{236} & \mathbf{<0}.\mathbf{001} & \mathbf{20} \\
Distance      & Absolute Mag. & \mathbf{-0}.\mathbf{512} & \mathbf{<0}.\mathbf{001} & \mathbf{20} & \mathbf{-0}.\mathbf{364} & \mathbf{<0}.\mathbf{001} & \mathbf{20} \\
Distance      & Period        &  \mathbf{0}.\mathbf{348} & \mathbf{<0}.\mathbf{001} & \mathbf{20} &  \mathbf{0}.\mathbf{100} & \mathbf{<0}.\mathbf{001} &  \mathbf{8} \\
\enddata
\tablenotetext{a}{Entries in bold indicate correlated characteristics (i.e.\ having $|\rho| \geq 0.1$)}
\tablenotetext{b}{Number of ``surviving'' (i.e.\ having $|\rho| \geq 0.1$) random subsets out of 20; see Section~\ref{subsec:Statistical-Analysis}}
\end{deluxetable*}
\vspace{-24pt}

\begin{figure*}
\begin{centering}
\includegraphics[width=\textwidth]{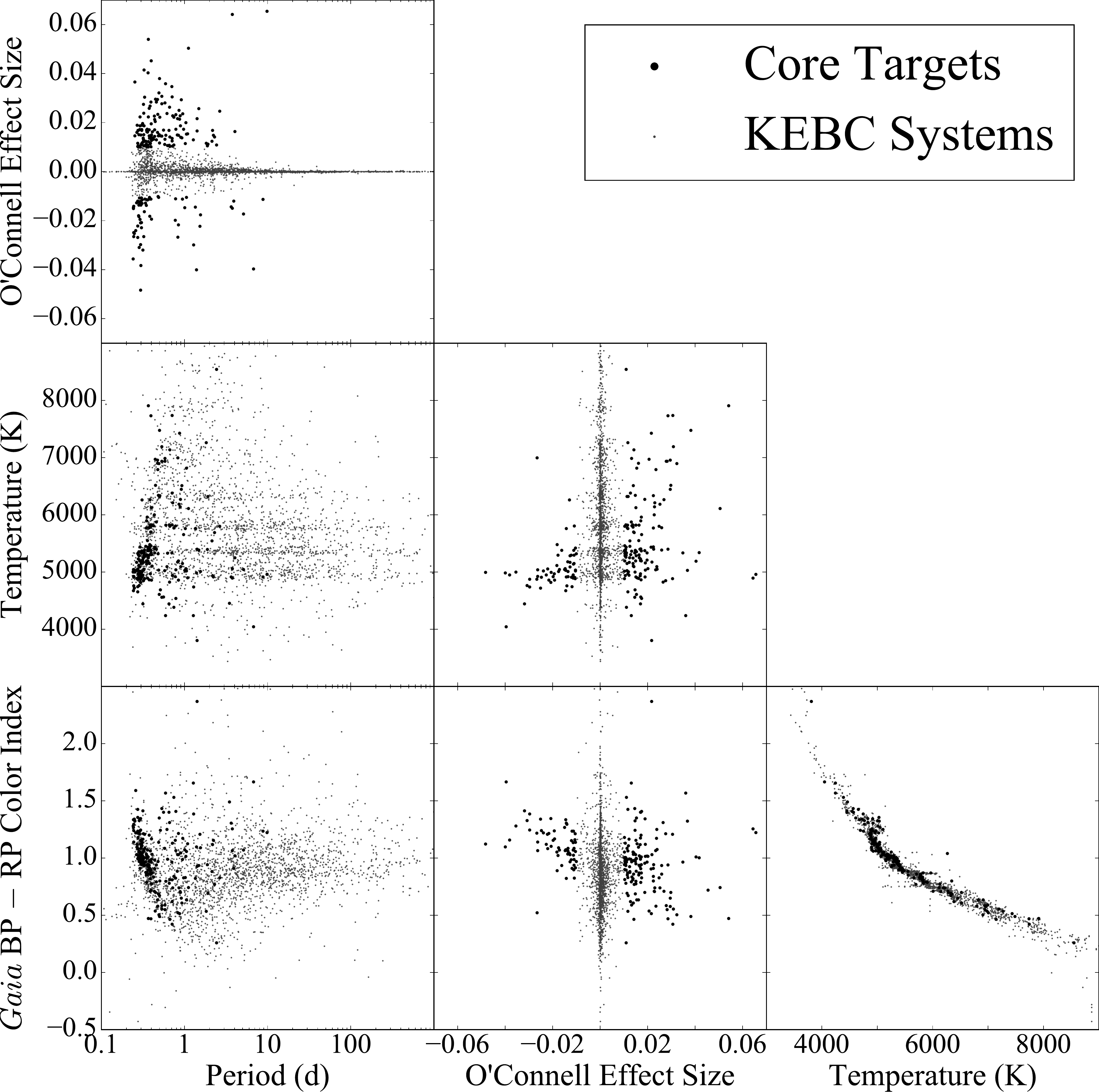}
\par\end{centering}
\caption{Corner plot showing the correlations between three characteristics of interest (period, OES, and temperature) and OES, temperature, and \emph{Gaia} color. Core sample targets are shown in black while non-core KEBC systems are shown in grey. A few outlier systems have been removed from the OES plots for clarity (see discussion in text). Note the logarithmic \emph{x}-axis of the period plots.
\label{fig:Corner-Plot-Part-One}}
\end{figure*}

\begin{figure*}
\begin{centering}
\includegraphics[width=\textwidth]{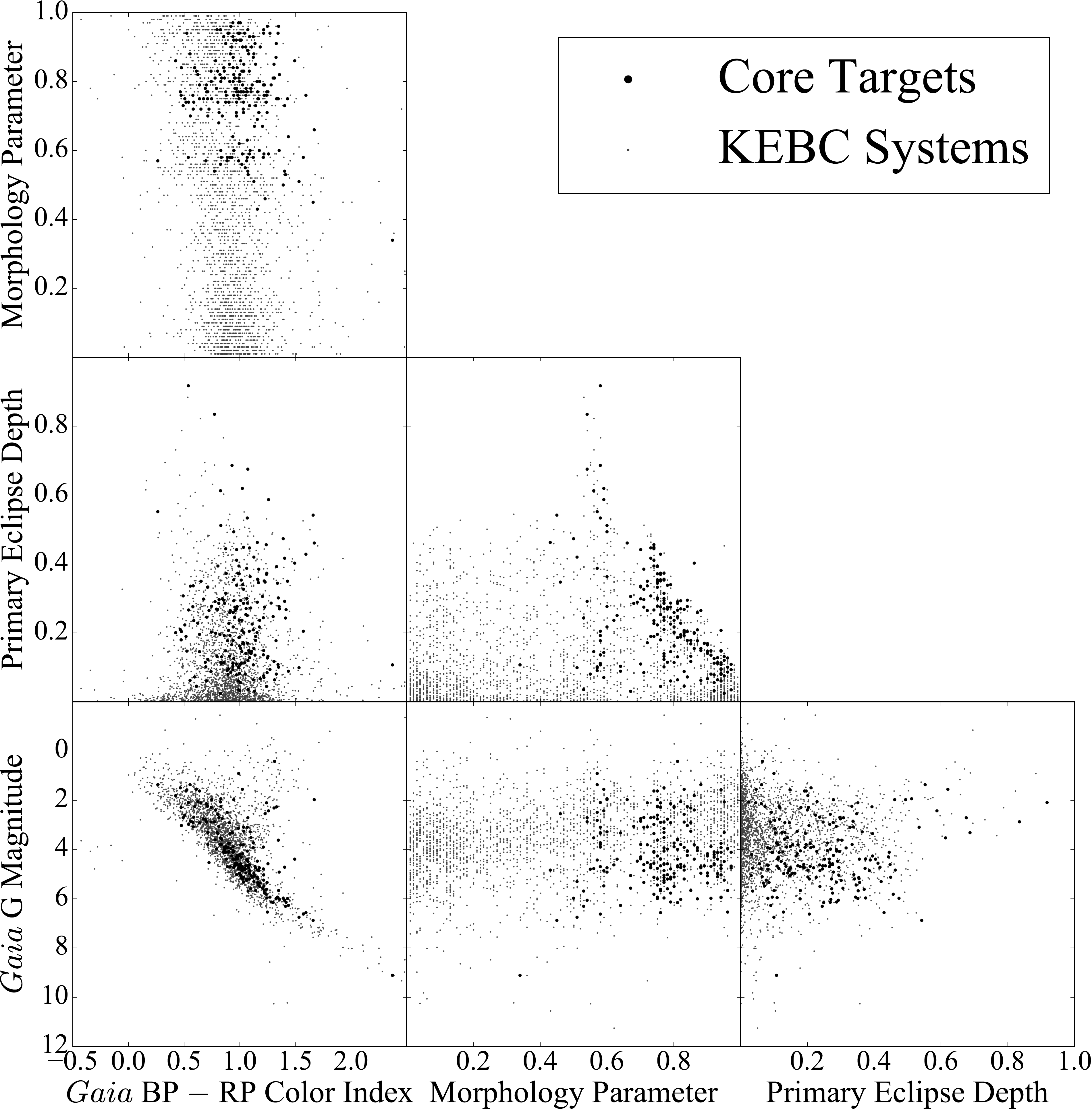}
\par\end{centering}
\caption{Corner plot showing the correlations between three characteristics of interest (\emph{Gaia} color, morphology parameter, and primary eclipse depth) and morphology parameter, primary eclipse depth, and absolute \emph{Gaia} G magnitude. Core sample targets are shown in black while non-core KEBC systems are shown in grey.
\label{fig:Corner-Plot-Part-Two}}
\end{figure*}

\begin{figure*}
\begin{centering}
\includegraphics[width=\textwidth]{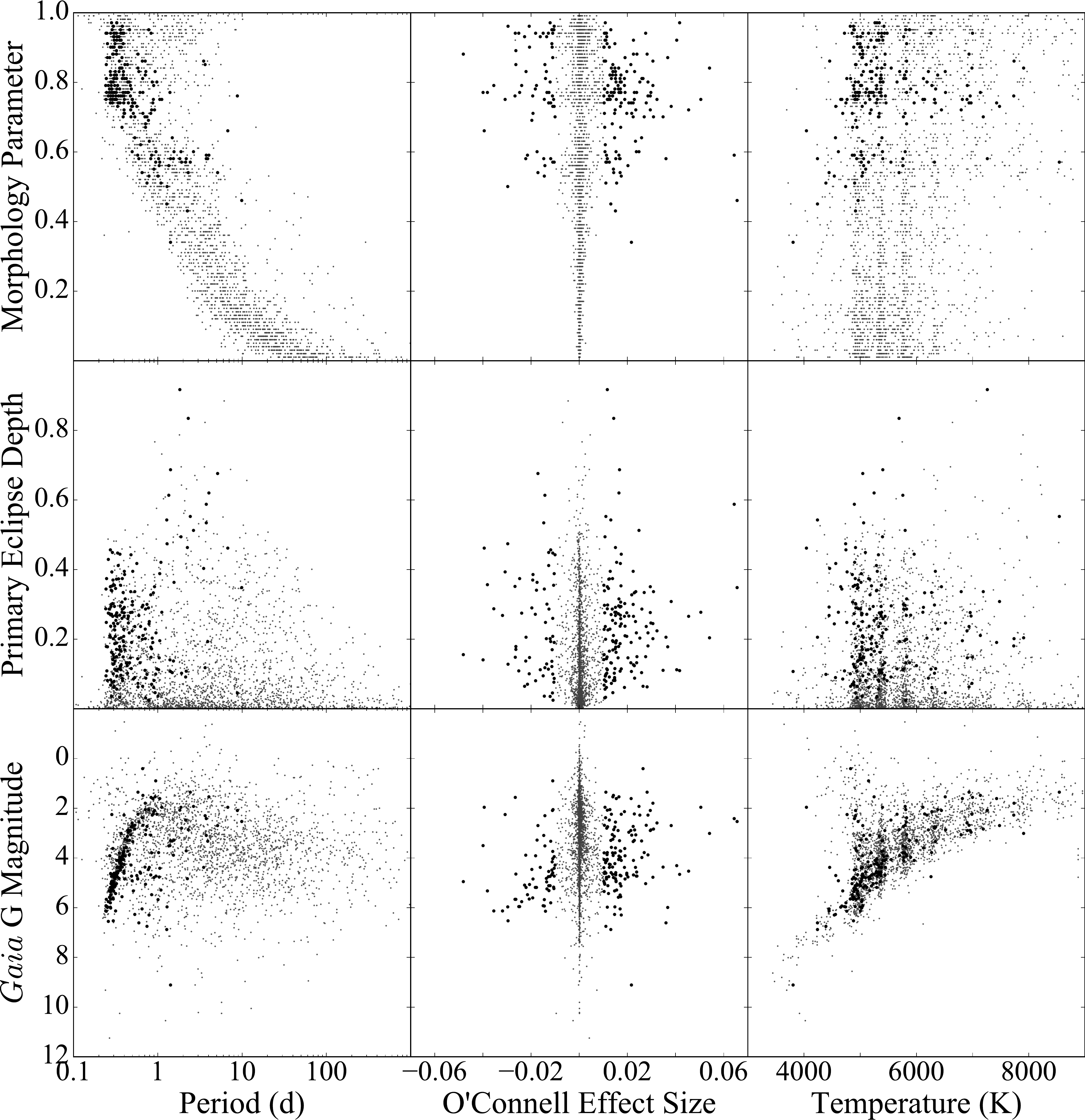}
\par\end{centering}
\caption{Corner plot showing the correlations between three characteristics of interest (period, OES, and temperature) and morphology parameter, primary eclipse depth, and absolute \emph{Gaia} G magnitude. Core sample targets are shown in black while non-core KEBC systems are shown in grey. A few outlier systems have been removed from the OES plots for clarity (see discussion in text). Note the logarithmic \emph{x}-axis of the period plots.
\label{fig:Corner-Plot-Part-Three}}
\end{figure*}

We found that seven of the characteristic pairs we studied provided more insight into the O'Connell effect than the remaining ones. Table~\ref{tab:Spearman-Test-Results} highlights these seven pairs (OES and temperature, OES and morphology parameter, OES and absolute magnitude, OES and distance, OES and period, eclipse depth and the morphology parameter, and absolute magnitude and period). The bootstrapping procedure discussed in Section~\ref{subsec:Statistical-Analysis} guided our choice in which characteristic pairs were important: five of these pairs are robust and the other two are nearly so. These levels of robustness further raises our confidence in the results we present. Our discussion in Sections~\ref{subsubsec:O'Connell-Effect-Size-Correlations}-\ref{subsubsec:Period-Correlations} focuses primarily on these characteristic pairs, along with several trends visible in Figures~\ref{fig:Corner-Plot-Part-One}, \ref{fig:Corner-Plot-Part-Two}, and \ref{fig:Corner-Plot-Part-Three}. These trends include the lack of core systems with morphology parameters between 0.6 and 0.7 in the morphology panels, the clustering of systems along the right edge of Figure~\ref{fig:Corner-Plot-Part-Two}'s center panel, and the stark contrast between the positive and negative O'Connell effect systems' temperature distributions in Figure~\ref{fig:Corner-Plot-Part-One}'s center panel.

Our initial analysis of the KEBC found strong correlations between the OES and both period and the morphology parameter. Our inclusion of many long-period and well-detached systems that are fundamentally different from our sample's systems strengthened these correlations. As a result of these fundamental differences, we considered these initial correlations less relevant to our study. Therefore, our final KEBC analysis used the analysis subset of the KEBC discussed in Section~\ref{subsec:KEBC}, which, in addition to the systems excluded in the earlier subsets, excluded systems with $P \geq 10$~d (690 systems) or $\mu \leq 0.3$ (1,018 systems), leaving only 1,639 KEBC systems. As in Sections~\ref{sec:Results} and \ref{subsec:Characteristic-Trends}, we removed no systems from our core sample analysis, and the KEBC population includes the core sample (except KIC~7667885). Our analysis considers both OES and $|$OES$|$.

\subsubsection{O'Connell Effect Size\label{subsubsec:O'Connell-Effect-Size-Correlations}}

The OES (as defined in Section~\ref{subsec:O'Connell-Effect-Size-Determination}) reflects the difference in brightness at the two maxima, which in turn depends on the brightness difference between the leading and trailing hemispheres of each star. We gave particular focus to correlations and trends involving the OES because it defined our sample and is the characteristic most directly related to the O'Connell effect. This section discusses OES and $|$OES$|$ correlations with the other characteristics we studied. It also discusses the differences between the OES and $|$OES$|$ correlations.

Figure~\ref{fig:Corner-Plot-Part-One}'s center panel comparing OES and temperature shows that a negative O'Connell effect, wherein the brighter maximum occurs before the primary minimum, mainly occurs in systems with $T < 6,\!000$~K\@. Thirty-four (23\% of 143 systems with a \emph{Gaia} temperature) positive O'Connell effect systems have a temperature above 6,000~K, while only two (3\% of 62 systems) negative O'Connell effect systems do. Furthermore, one of the two hot negative O'Connell effect systems, KIC~7773380, has a \emph{Kepler} temperature and \emph{Gaia} color implying that it is likely significantly cooler than its \emph{Gaia} temperature indicates. The other system, KIC~7950962, has primary and secondary eclipse depths that differ by only $\sim$0.001 in normalized flux. Furthermore, our light curves of the system have a phase offset of 0.5 from the KEBC light curve. These two facts indicate an ambiguity regarding which of KIC~7950962's eclipses is the primary, and thus if its O'Connell effect is positive or negative. The dearth of hot systems displaying a negative O'Connell effect was unknown before now and is consistent with the idea that starspots are the predominant cause of a negative O'Connell effect. Starspots are expected to exist in the convective envelopes of cooler stars but not in the radiative envelopes of hotter stars.

\begin{figure}
\begin{centering}
\includegraphics[width=\columnwidth]{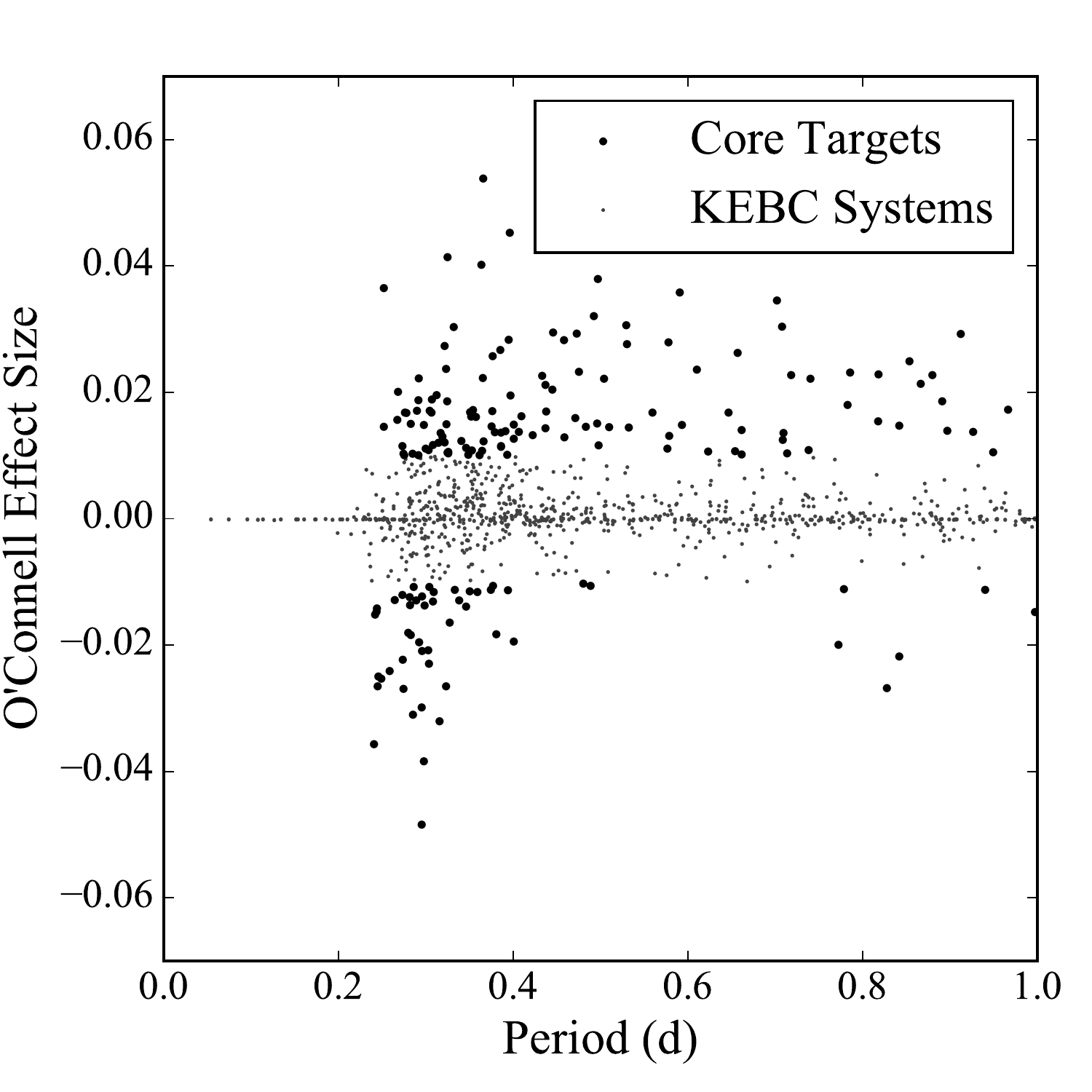}
\par\end{centering}
\caption{Plot comparing the OES to the period. Core sample targets are shown in black while non-target KEBC systems are shown in grey. The \emph{y}-axis is in units of normalized flux. KICs~9777984, 9935311, and 11347875 are excluded for clarity. Note the positive correlation in core sample systems with $P < 0.5$~d.
\label{fig:Period-vs-OES}}
\end{figure}

Figure~\ref{fig:Corner-Plot-Part-One}'s top-left panel comparing OES and period shows that the OES tends towards zero at longer periods. This trend is consistent with the idea that binary interaction ultimately causes the O'Connell effect. The lack of any systems in our sample with a period greater than ten days strengthens this idea. Table~\ref{tab:Spearman-Test-Results} shows that this correlation is not robust. Of greater concern is that Table~\ref{tab:Spearman-Test-Results} indicates that the correlation between OES and period is positive in our sample, contradicting the trend observed in Figure~\ref{fig:Corner-Plot-Part-One}. Our explanation for this contradiction is a positive trend between OES and period for the shortest period systems in our sample.

To better display this short-period system trend, Figure~\ref{fig:Period-vs-OES} shows a rescaled view of Figure~\ref{fig:Corner-Plot-Part-One}'s top-left panel focusing on systems with periods under 1~d. Systems with $P \leq 0.5$~d show this positive correlation clearly. Since Figure~\ref{fig:Period-Histogram} shows that most systems in our sample have $P \leq 0.5$~d, this positive correlation dominates the $\rho$ coefficient. We found that the $\rho$ coefficient for OES and period more than doubled ($\rho = 0.425$) when analyzing only core sample systems with $P \leq 0.5$~d rather than the entire core sample, supporting our explanation. Furthermore, the short-period system correlation is robust (20 of 20 subsets show a correlation). When we analyzed the core sample systems with $P > 0.5$~d, we found no correlation ($\rho = -0.069$). We interpret the short-period system trend as a result of the following: hotter systems ($T \geq 6,\!000$~K) almost always have OES~$> 0$ (as discussed in the previous paragraph), temperature is positively correlated with luminosity, and luminosity is positively correlated with orbital period (as discussed in Section~\ref{subsubsec:Period-Correlations}). Therefore, the shortest period systems are more likely to have OES~$< 0$ than longer period systems, producing the observed trend.

\begin{figure*}
\begin{centering}
\includegraphics[width=\columnwidth]{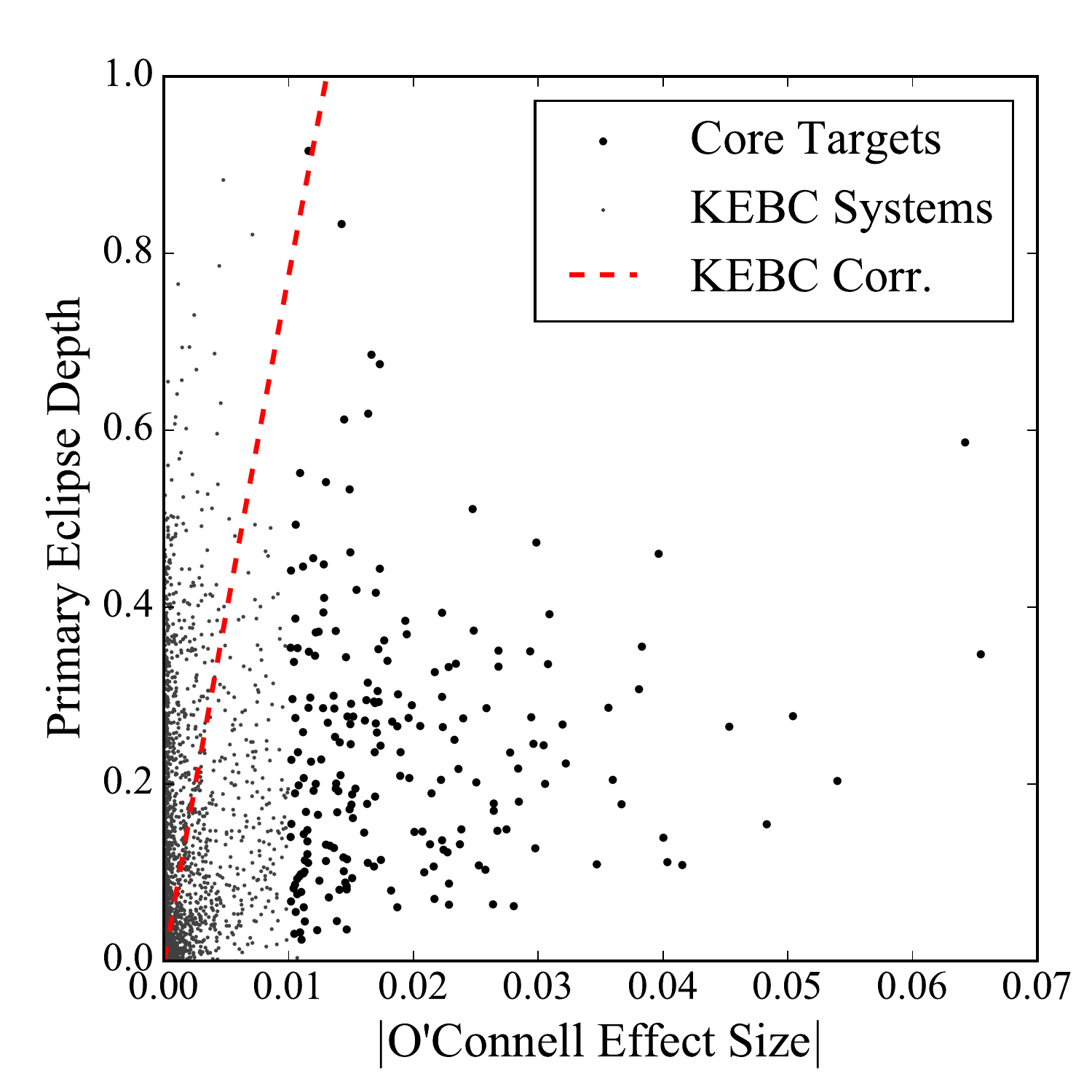}\unskip
\includegraphics[width=\columnwidth]{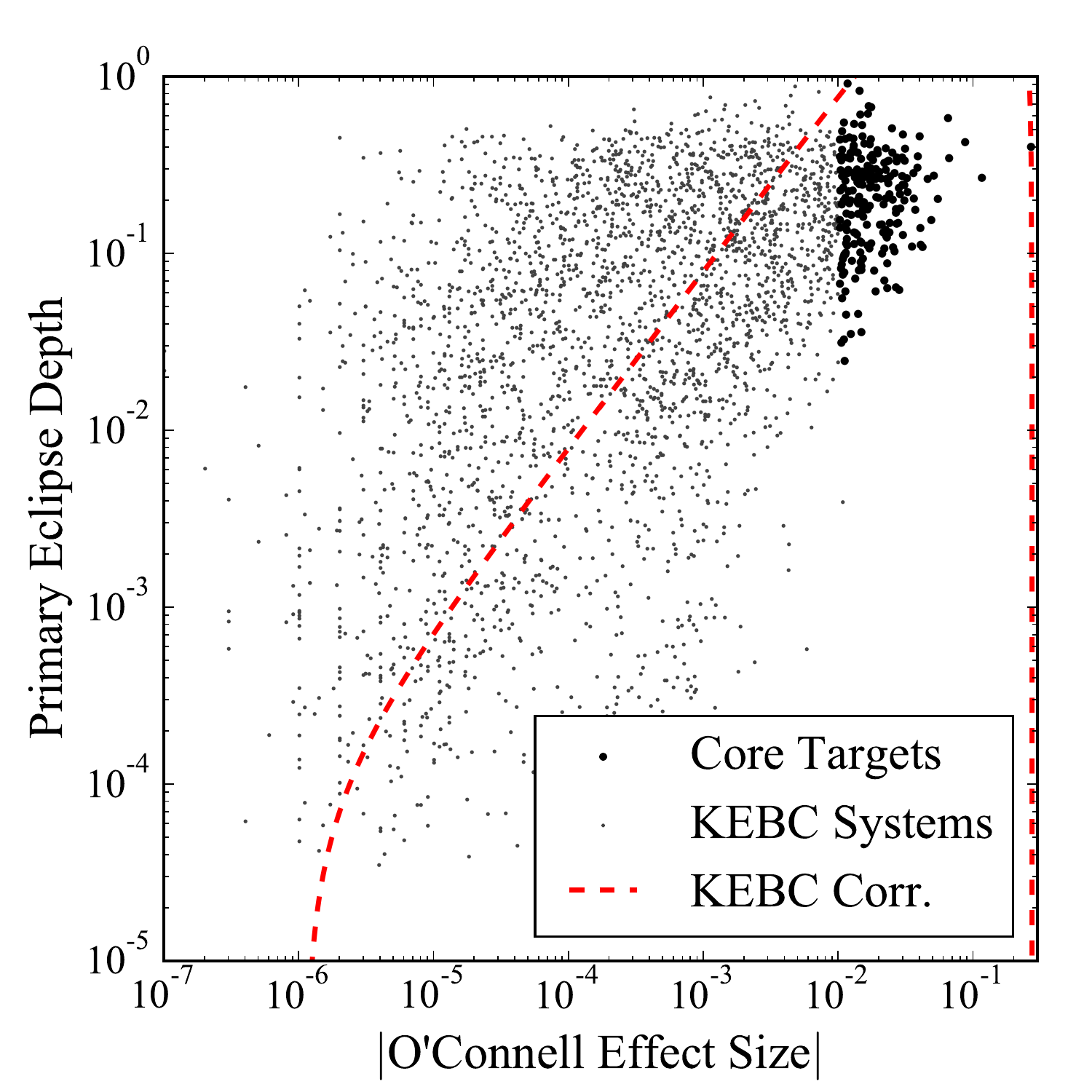}
\par\end{centering}
\caption{Plots comparing the primary eclipse depth to $|$OES$|$. Core sample targets are shown in black while non-target KEBC systems are shown in grey. The red dashed line shows the ODR correlation for the KEBC\@. The left panel's axes are scaled linearly, while the right panel's axes are scaled logarithmically. Both axes are in units of normalized flux. KICs~9777984, 9935311, and 11347875 are excluded from the left panel for clarity. No correlation is apparent in the left panel, but a weak correlation can be seen in the right panel.
\label{fig:Depth-vs-|OES|}}
\end{figure*}

We expected strong correlations between $|$OES$|$ and other characteristics under the premise that the positive and negative O'Connell effects are fundamentally similar. This premise implies that the positive and negative OES correlations should strengthen when neglecting the OES's sign. Therefore, we were surprised by the \emph{lack} of such correlations in our core sample. The only $|$OES$|$ correlation we found was between $|$OES$|$ and the primary eclipse depth, and the correlation is both weak and not robust. Furthermore, Figure~\ref{fig:Depth-vs-|OES|}'s left panel shows no visible correlation between these two characteristics. The lack of $|$OES$|$ correlations implies that our premise that the positive and negative O'Connell effects are fundamentally similar is incorrect.

Note that every characteristic is correlated with $|$OES$|$ in the KEBC, the strongest of which is the correlation with primary eclipse depth. We determined a quadratic fit of $D = -298(19)|\text{OES}|^2 + 81(4)|\text{OES}|$ for this correlation using ODR, where $D$ is the primary eclipse depth and the parentheticals are the uncertainties. While we do not see this correlation in Figure~\ref{fig:Depth-vs-|OES|}'s left panel, either, Figure~\ref{fig:Depth-vs-|OES|}'s right panel showing a log-log plot of these characteristics displays a visible, if weak, correlation between them. We suspect that the $\rho$ coefficient detected this trend, resulting in the large coefficient in Table~\ref{tab:Spearman-Test-Results}. Figure~\ref{fig:Depth-vs-|OES|}'s right panel indicates that this trend extends from our sample down to much smaller values of $|$OES$|$ and eclipse depth. Therefore, some systems that are not in our sample may be fundamentally similar to our sample's systems, only with smaller values for $|$OES$|$ and eclipse depth. As such, this correlation may be more related to the O'Connell effect than its strength in our sample would indicate.

Both panels of Figure~\ref{fig:Depth-vs-|OES|} also show a lack of systems displaying a significant O'Connell effect and a small primary eclipse depth. Our selection criterion ($|\text{OES}| \geq 0.01$) does not exclude such systems, and their absence is conspicuous since most KEBC systems have small eclipse depths. A bias in the \emph{Kepler} selection function against these systems may explain their absence, which would imply a fundamental difference between such systems and the systems \emph{Kepler} observed. Their absence may also be a true representation of O'Connell effect binaries, although we cannot identify a plausible reason why systems with a significant O'Connell effect cannot have a small primary eclipse depth. As a final note, Figure~\ref{fig:Depth-vs-|OES|}'s left panel shows that all four KEBC systems with an eclipse depth above 0.8 have a non-negligible $|$OES$|$.

We found different correlations for the OES (the morphology parameter, temperature, distance, absolute magnitude, and period) as compared to $|$OES$|$ (eclipse depth). Because the OES distinguishes between the positive and negative O'Connell effects but $|$OES$|$ does not, this suggests a more fundamental difference between the positive and negative O'Connell effects than previously thought. Assuming that spots are the O'Connell effect's primary cause, a larger OES implies that spots cover a greater area, have a more extreme temperature factor, or are further offset from the line connecting the stellar centers. \citet{Kouzuma2019} found a weak positive correlation between stellar temperature and spot temperature factor for cool starspots in W-type overcontact systems, wherein the smaller star is hotter (in contrast to A-type overcontact systems; \citealt{McCartney1999}). The same correlation is much stronger in semi-detached systems. \citet{Kouzuma2019} found weak positive correlations in overcontact systems between stellar temperature and spot size and between orbital period and spot size. The correlations for cool spots are similar but generally stronger than for hot spots. His results imply positive correlations between temperature and $|$OES$|$ and between orbital period and $|$OES$|$. Table~\ref{tab:Spearman-Test-Results} shows both correlations for OES but neither for $|$OES$|$.

To further test this correlation, we found the $\rho$ coefficient for the positive O'Connell effect systems and for the negative O'Connell effect systems. We found that the correlations for the positive O'Connell effect systems ($\rho_{\text{temp}} = 0.122$ and $\rho_{\text{per}} = 0.083$) are weak. Furthermore, while the correlations are much stronger for the negative O'Connell effect systems ($\rho_{\text{temp}} = 0.402$ and $\rho_{\text{per}} = 0.142$), the former has the wrong sign (i.e.\ the OES gets closer to zero as the temperature increases) while the latter is still weak. These correlations are therefore inconsistent with the results of \citetp{Kouzuma2019} starspot study. However, such inconsistencies may not indicate that starspots do not cause the O'Connell effect. Instead, they may result from our sample's mixture of systems with different O'Connell effect causes, or perhaps from differences between our sample and his. Determining the cause of these inconsistencies is beyond the scope of this paper.

The three other robust correlations that we have not discussed are between OES and the morphology parameter, distance, and absolute magnitude. We interpret these three correlations as arising from unrelated correlations with other characteristics. For instance, the correlation between the OES and absolute magnitude arises from the discussed correlation between OES and temperature combined with the strong correlation between temperature and absolute magnitude that is a well-known feature of main-sequence stars. We do not discuss these secondary correlations due to their dependence on other correlations that have a more fundamental explanation.

The different correlations we find among positive and negative O'Connell effect systems, combined with the fact that a much larger number of systems display a positive O'Connell effect, leads us to conclude that the preference for a brighter maximum following the primary eclipse is not an observational bias. It is instead fundamental to the O'Connell effect. Additionally, we conclude that the positive and negative O'Connell effects have different causes, or that one cause is common in one case and rare in the other. We consider these findings important results of our study.

\subsubsection{Eclipse Depth\label{subsubsec:Eclipse-Depth-Correlations}}

The primary eclipse depth (as defined in Section~\ref{subsec:Eclipse-Depth-Determination}) is influenced by four parameters: temperature ratio, relative radii, component shapes, and orbital inclination. This section discusses the correlation between the eclipse depth and the morphology parameter. It also discusses a couple of trends between these two characteristics seen in Figure~\ref{fig:Corner-Plot-Part-Two}'s center panel.

The eclipse depth strongly correlates with the morphology parameter in our sample, while the KEBC shows a weaker correlation. We expect such a correlation because, as \citet{Wilson2006} states, the eclipse depth is a monotonic function of the stellar radii ratio for overcontact systems (i.e.\ systems with a large morphology parameter). Meanwhile, \citet{Matijevic2012} notes that the primary eclipse width increases with the morphology parameter, and the eclipse widths measure the sum of the relative stellar radii. Thus, in overcontact systems, the primary eclipse depth decreases as the morphology parameter increases. The best fit to this correlation is the linear function $\mu = -2.13(13)D + 1.26(9)$. The KEBC's correlation is weaker as it includes more small $\mu$ systems, for which the eclipse depth and stellar radii ratio relation does not hold.

Figure~\ref{fig:Corner-Plot-Part-Two}'s center panel comparing the primary eclipse depth to the morphology parameter shows a sharp edge toward the panel's right side in both our sample and the KEBC. This trend indicates that eclipses get shallower as $\mu$ increases above 0.7, a consequence of the relation discussed in the previous paragraph. The systems in our sample appear to be clustered along this edge in Figure~\ref{fig:Corner-Plot-Part-Two}'s center panel. This clustering indicates that O'Connell effect systems tend to have the deepest eclipses of systems with similar light curves (as measured by the morphology parameter), hinting at a connection between the OES and the eclipse depth. The clustering around the right edge is also seen with the non-core systems, although it is not as pronounced.

Figure~\ref{fig:Corner-Plot-Part-Two}'s center panel shows another trend: all KEBC systems with $D \geq 0.6$ have $\mu \approx 0.6$. \citet{Soderhjelm2005} states that the maximum eclipse depth for two main-sequence stars is 0.75 magnitudes (equivalent to a relative flux of 0.5), implying that systems with deeper eclipses must have an evolved component. \citet{Matijevic2012} says that a $\mu$ of 0.6 indicates a semi-detached system, which occurs when an evolving star fills its Roche lobe. This trend therefore make sense because close binaries with evolved components would be expected to have $\mu \approx 0.6$ by \citet{Matijevic2012}. However, we cannot discount the possibility that this trend is a statistical artifact caused by the rarity of systems with such deep eclipses in the KEBC\@. Our eclipse depth determination method discussed in Section~\ref{subsec:Eclipse-Depth-Determination} may also influence this trend, as it can underestimate the eclipse depth of Algol-type systems.

\subsubsection{Morphology Parameter\label{subsubsec:Morphological-Correlations}}

\begin{figure}
\begin{centering}
\includegraphics[width=\columnwidth]{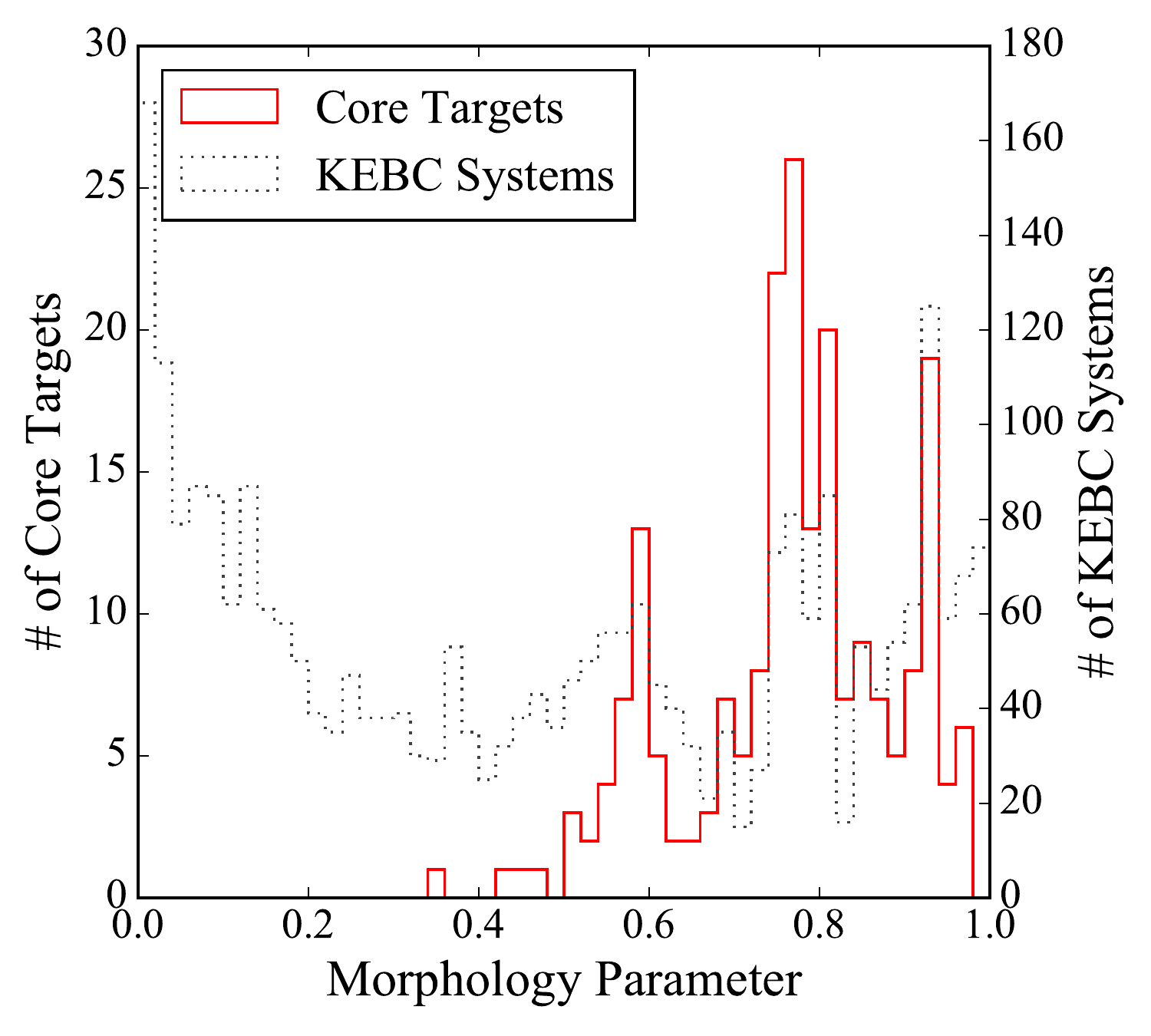}
\par\end{centering}
\caption{Histogram comparing the morphology parameters for 211 of the 212 core sample targets (solid red) and all 2,745 entries in the KEBC (dashed grey) that have $\mu \neq -1$. KIC~7667885 is not included because it lacks a value for $\mu$ (see discussion in Section~\ref{subsec:Target-Sample}). The two distributions differ significantly, particularly for $\mu < 0.5$. The core sample also has few systems with $0.6 \leq \mu \leq 0.7$.
\label{fig:Morphology-Histogram}}
\end{figure}

The morphology parameter (described in Section~\ref{subsec:Morphology-Parameter}) is primarily a measure of eclipse widths \citep{Matijevic2012}, but it correlates well with the morphology class of a given system. We wish to reemphasize the point from \citet{Matijevic2012} that the morphology parameter only provides a ``best-guess'' estimate of the morphology class. This section focuses on the morphology parameter distribution shown in Figure~\ref{fig:Morphology-Histogram}. It also discusses the initially surprising correlation between the morphology parameter and temperature.

Figure~\ref{fig:Morphology-Histogram} shows a lower abundance of systems with $0.6 \leq \mu \leq 0.7$. The same region is depleted in the KEBC, but not to the extent of our sample, suggesting that this underabundance is not wholly due to the parent population. \citet{Matijevic2012} states that systems with $\mu$ in this range are semi-detached. Our interpretation of systems in this range is that the accreting star grows larger with increasing morphology parameter, with the system becoming contact around $\mu = 0.7$. The larger the accretor is, the less deeply a matter stream will penetrate into its potential well. The matter stream will therefore impact the surface with less energy and cause less dramatic heating. Assuming the accretor is on the main sequence, a larger star will be hotter, further reducing the degree of heating. These factors could cause the paucity of systems in this region of parameter space showing a significant O'Connell effect.

Figure~\ref{fig:Morphology-Histogram} also shows that systems with $\mu \lesssim 0.5$ rarely show a significant O'Connell effect, and none with $\mu \leq 0.3$ do. This is significant because our selection criterion should not be biased with respect to $\mu$. Stars in systems with a small $\mu$ are generally farther apart and are less likely to significantly affect each other, so this result reinforces the link between binary interaction and the O'Connell effect discussed in Section~\ref{subsubsec:O'Connell-Effect-Size-Correlations}.

We were surprised by the morphology parameter and temperature correlation because we assumed they would be uncorrelated. We now believe that this correlation results from a selection effect. Two processes work in tandem to cause this selection effect: the influence of geometry on the morphology parameter and the influence of the initial mass function (IMF) on temperature. Regarding the first process, eclipses only occur if $\Delta < R_1 + R_2$ at $\phi = 0$ (conjunction), where $\Delta = a\cos{i}$ \citep[Equations~3.37 and 3.38]{Prsa2006} and the $R_n$ are the component radii. Therefore, closer binaries (smaller $a$) exhibit eclipses over a broader range of inclinations. Since the components of large morphology parameter systems are generally closer than in small morphology parameter systems, we are more likely to observe a given system eclipse in the former case. The IMF, meanwhile, makes hot stars rarer than cool stars, a fact compounded by \emph{Kepler}'s selection biases discussed in Section~\ref{subsec:Kepler}. Acting together, these processes mean that most hot eclipsing binaries observed by \emph{Kepler} should have large morphology parameters. The resulting dearth of hot, small morphology parameter systems creates an apparent positive correlation between the characteristics.

\subsubsection{Period\label{subsubsec:Period-Correlations}}

The period is a function of component masses and orbital separation due to Kepler's third law. This section discusses the correlation between the period and the absolute magnitude/luminosity in the core sample. Figure~\ref{fig:Corner-Plot-Part-Three}'s lower-left panel demonstrates this correlation in the dense clustering seen near the panel's left side.

A correlation between period and absolute magnitude initially seems odd. Low luminosity stars can have very wide orbits, after all, while very luminous stars can be short-period contact binaries. However, there is a lower period limit for stars of a given luminosity. For example, on the main sequence, a star's size and surface temperature (and thus absolute magnitude) are a function of its mass. If binary components are too close, however, they will merge into a single star. Kepler's third law states that $P \propto a^{1/3}$, while the Stefan-Boltzmann law states that $L_* \propto R^2$. Therefore, a lower limit on component separation based on stellar radius implies a lower limit on the period for a given absolute magnitude for main-sequence stars. This limit increases for evolved stars since they become cooler and larger as they evolve. The correlation between period and absolute magnitude is weaker in the KEBC because of the KEBC's numerous long-period systems, which do not have a relationship between component separation and stellar radius.

\begin{figure}
\begin{centering}
\includegraphics[width=\columnwidth]{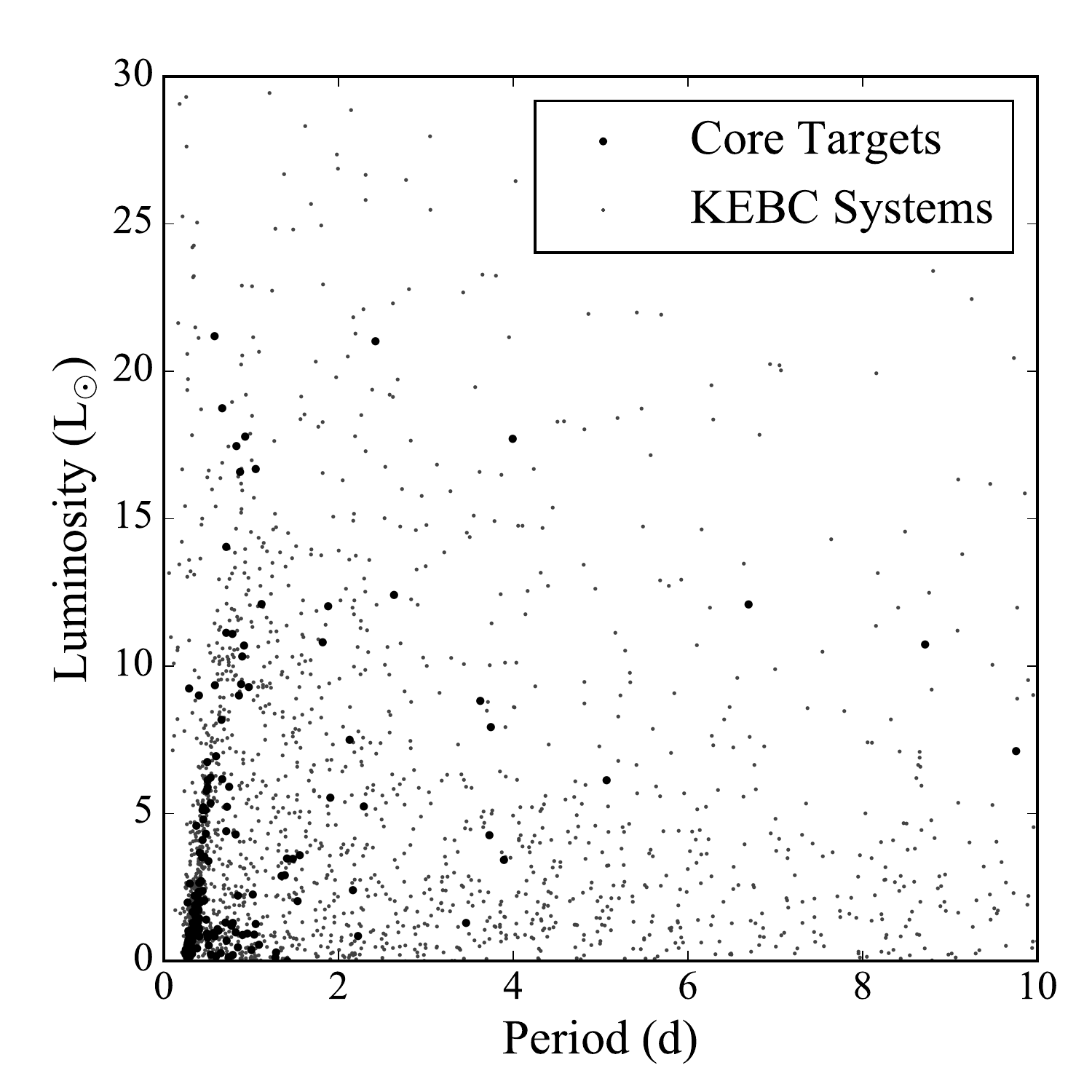}
\par\end{centering}
\caption{Plot comparing the luminosity to the period. Core sample targets are shown in black while non-target KEBC systems are shown in grey. KICs~3662635 and 5820209 are excluded for clarity. The sharp edge near the lower-left corner denotes the short-period limit for a given luminosity. Most of the core sample is clustered near or along this edge.
\label{fig:Luminosity-vs-Period}}
\end{figure}

Both Figure~\ref{fig:Corner-Plot-Part-Three}'s lower-left panel and Figure~\ref{fig:Luminosity-vs-Period} show a tight clustering of systems along their left edge. Our interpretation of this clustering is that it represents the lower limit on the period for a given absolute magnitude or luminosity. Systems to the left of this edge are likely subdwarf systems or have poorly determined distances resulting in large luminosity errors. The fact that such a large percentage of our sample lies on or near this edge -- particularly systems with $L_* \geq 1~\text{L}_{\odot}$ in Figure~\ref{fig:Luminosity-vs-Period} -- suggests that the O'Connell effect is more common in contact or near-contact binaries. This trend is therefore strong evidence that binary interaction is a critical factor in causing the O'Connell effect.

\section{Conclusion\label{sec:Conclusion}}

Our study of the characteristics of systems displaying a significant O'Connell effect takes advantage of the unprecedented amount of data available from the \emph{Kepler} and \emph{Gaia} missions. It represents a significant step forward in our understanding of this poorly understood phenomenon. We studied a significantly different parameter space than the earlier works of \citetalias{O'Connell1951} and \citetalias{Davidge1984} out of necessity: we lack detailed information on the bulk characteristics of the component stars that were the basis of those works. We instead focused on those observational properties we could directly determine from the data currently available.

We found a significant O'Connell effect in 212 eclipsing binaries, with the 211 KEBC systems representing 7.3\% of the 2,907 systems cataloged in the KEBC\@. This shows that a significant O'Connell effect is present in a non-negligible fraction of eclipsing binaries. Our core sample systems range in period from a quarter of a day to nearly ten days, temperature from around 3,800~K to 8,500~K, and luminosity from two-hundredths of a solar luminosity to over fifty solar luminosities. Clearly, a wide range of system types shows this phenomenon.

The O'Connell effect also manifests itself quite differently in different systems. Many systems have light curves that vary wildly with time, and the O'Connell effect varies along with it. Other systems have extremely stable light curves and a stable value of the OES\@. The existence of the latter type of system would indicate a cause of the O'Connell effect that is more stable in time than starspots, such as a hotspot caused by mass transfer. Furthermore, up to 20\% of our sample displays the ETV signal associated with mass transfer. We conclude that there are multiple causes of the O'Connell effect.

The dearth of hot ($T \geq 6,\!000$~K) systems displaying a negative O'Connell effect is a new result and can be interpreted as evidence that starspots cause the O'Connell effect in negative O'Connell effect systems. Additionally, we found a moderate correlation between OES and temperature and a weaker correlation between OES and period, both predicted by \citetp{Kouzuma2019} study on starspots, which also suggests that starspots predominately cause the O'Connell effect. However, looking at positive and negative O'Connell effects separately does not yield the same correlations. Instead, we found correlations that are inconsistent with \citetp{Kouzuma2019} results. Therefore, we draw no conclusion as to whether starspots are a predominant cause of the O'Connell effect, but note that the idea is plausible. At the same time, it should not automatically be assumed that starspots cause the O'Connell effect in a given system without further supporting evidence for starspots. Many more systems display a positive O'Connell effect than a negative one, suggesting that which maximum is brighter is not chance but fundamentally tied to the O'Connell effect. We conclude that, whatever the physical causes of the O'Connell effect are, they preferentially cause the maximum following the primary eclipse to be brighter than the one preceding it.

The overarching trend found in this study is that the O'Connell effect and binary interaction are strongly correlated. We do not see this phenomenon in binaries with widely separated stars, and we almost always see evidence (ellipticity effects in light curves) of non-spherical stars in our sample. Looking beyond our sample, W Crucis \citep{Zola1996, Pavlovski2006} exhibits the O'Connell effect, and its light curve indicates an interacting, semi-detached system similar to other systems in our sample. However, W Crucis' 198-day orbital period -- the longest of an O'Connell effect binary that we are aware of -- is over 20 times longer than the longest period in our sample. Nevertheless, W Crucis' ellipsoidal variations shows that it has interacting components despite the long period, reinforcing the correlation between binary interaction and the O'Connell effect. We conclude that this correlation and the physical causes of the O'Connell effect are related: the interaction (whether gravitational, magnetic, or physical) between close binary stars causes the O'Connell effect instead of a phenomenon an isolated star (including non-interacting components in wider binaries) would display. If the phenomenon occurred on isolated stars, we would expect to see the O'Connell effect in \emph{all} types of binaries, not just those with interacting components.

There are several ways to expand upon our work in the future. The recently published TESS Eclipsing Binary Catalog \citep[TEBC;][]{Prsa2022} provides an opportunity to expand our analysis to a much larger data set. \emph{Gaia} Data Release 3 (DR3) -- currently slated for release in 2022 -- will include \emph{Gaia} time-series photometry for eclipsing binaries \citep{Gaia2019}. \emph{Gaia}'s tricolor photometry allows temperature measurements for both stars and an exploration of the O'Connell effect's temperature dependence. This data greatly increases the parameter space available for study. We will further discuss the system classes Section~\ref{subsec:Notable-Systems} described in future work. This discussion will include searching for periodicities and quasi-periodicities in temporally varying systems, finding the second and third derivatives of concave-up systems to quantify their properties, and determining the degree of asymmetry in asymmetric minima systems. We will also discuss the subset of our sample brighter than $K_p = 14$ in future work due to \citet{Wolniewicz2021} finding such systems to be unbiased. The new \emph{Kepler} eclipsing binaries \citet{Bienias2021} found require further study to ensure our sample's completeness. Future work will also discuss our observations of ten sample systems and the PHOEBE \citep{Prsa2005, Prsa2016} models we create using our data. Analysis of the sub-groupings seen in Figures~\ref{fig:Corner-Plot-Part-One}-\ref{fig:Corner-Plot-Part-Three} is left for a future study. Finally, we intend to further study the differences between the positive and negative O'Connell effects our study found.

\begin{acknowledgments}This material is based upon work supported by the National Aeronautics and Space Administration under Grant No. 80NSSC19K1021 issued through the NNH18ZDA001N Astrophysics Data Analysis Program (ADAP)\@. This paper includes data collected by the \emph{Kepler} mission. Funding for the \emph{Kepler} mission is provided by the NASA Science Mission directorate. This research has made use of the SIMBAD database, operated at CDS, Strasbourg, France. This research has made use of the VizieR catalogue access tool, CDS, Strasbourg, France (DOI : 10.26093/cds/vizier). The original description of the VizieR service was published in 2000, A\&AS 143, 23. We would like to thank Dr.\ Ron Kaitchuck for his helpful feedback during the writing of this paper. We would also like to thank Drs.\ Andrej Pr\v{s}a and Kyle Conroy for their help in understanding the creation and contents of the KEBC\@. Finally, we would like to thank AAVSO astronomer Chuck Cynamon for observing several systems in our sample.
\end{acknowledgments}

\facilities{\emph{Gaia}, \emph{Kepler}}

\software{AstroPy \citep{AstroPy2013, AstroPy2018}, BinaryMaker3 \citep{Bradstreet2002}, IPython/Jupyter \citep{Perez2007}, Matplotlib \citep{Hunter2007}, NumPy \citep{Harris2020}, Peranso \citep{Paunzen2016}, PHOEBE v0.31a \citep{Prsa2005}, SciPy \citep{Virtanen2020}}

\clearpage

\bibliographystyle{aasjournal}
\bibliography{MasterReferenceList}

\setcounter{table}{0}

\begin{longrotatetable}
\begin{deluxetable}{lccccccccccccccc}
\tablecaption{Complete \emph{Kepler} O'Connell Effect Target List
\label{tab:Target-List}}
\tabletypesize{\tiny}
\tablehead{
\colhead{\textit{Kepler}} & \colhead{R.A. (J2000)} & \colhead{Dec. (J2000)} & \colhead{Period} & \colhead{LC} & \colhead{Morph.} & \colhead{BP -- RP} & \colhead{\textit{Gaia}} & \colhead{\textit{Gaia}} & \colhead{Dist.} & \colhead{Lum.} & \multicolumn{2}{c}{Spectral Type} & \colhead{Ref.\tablenotemark{f}\tablenotemark{g}\tablenotemark{h}} & \colhead{Samp.\tablenotemark{i}} & \colhead{Notes\tablenotemark{j}}\\
\colhead{ID\tablenotemark{a}} & \colhead{hh:mm:ss.s} & \colhead{+dd:mm:ss.ss} & \colhead{(days)} & \colhead{Class\tablenotemark{b}} & \colhead{Param.} & \colhead{Color} & \colhead{BP Mag.} & \colhead{G Mag.} & \colhead{(pc)} & \colhead{(L$_\odot$)\tablenotemark{c}} & \colhead{Lit.\tablenotemark{d}} & \colhead{Color\tablenotemark{e}} & \colhead{} & \colhead{} & \colhead{}}
\startdata
\href{http://keplerebs.villanova.edu/overview/?k=2159783}{2159783}   & 19:25:26.2 & +37:33:11.90 & 0.3739 & WU &    0.87 & 0.938 & 15.365 & 14.982 &    1546 &   1.810 &     F6 V\textsuperscript{{1}} &   G9 &                                           4, 5, 6, 22\textsuperscript{{*}} & C &      AM, S, M\\
\href{http://keplerebs.villanova.edu/overview/?k=2161623}{2161623}   & 19:26:58.7 & +37:35:27.60 & 2.2835 & Al &    0.54 & 0.771 & 14.571 & 14.273 &    1901 &   5.262 &                       \nodata & F9.5 &                                                                    \nodata & C &       \nodata\\
\href{http://keplerebs.villanova.edu/overview/?k=2302092}{2302092}   & 19:25:02.9 & +37:41:52.49 & 0.2947 & WU &    0.88 & 1.124 & 14.982 & 14.503 &     811 &   0.775 &                       \nodata &   K2 &                                                                       5, 7 & C &             S\\
\href{http://keplerebs.villanova.edu/overview/?k=2305372}{2305372}   & 19:27:57.7 & +37:40:21.83 & 1.4047 & Al &    0.58 & 0.929 & 14.214 & 13.846 &    1273 &   3.498 &    F8 Vs\textsuperscript{{1}} &   G9 &                                  2, 7, 13, 14\textsuperscript{{*}}, 16, 19 & C &             F\\
\href{http://keplerebs.villanova.edu/overview/?k=2437038}{2437038}   & 19:20:47.6 & +37:44:31.83 & 0.2677 & WU &    0.87 & 1.120 & 16.501 & 16.149 &    2784 &   2.004 &                       \nodata &   K2 &                                                                    3, 5, 6 & C &       F, S, M\\
\href{http://keplerebs.villanova.edu/overview/?k=2449084}{2449084}   & 19:30:46.9 & +37:46:19.13 & 0.7398 & BL &    0.70 & 0.842 & 15.319 & 14.983 &    2798 &   5.926 &                       \nodata &   G4 &                                                                          3 & C &         AM, F\\
\enddata
\end{deluxetable}
\end{longrotatetable}

\listofchanges

\end{document}